\documentclass[12pt,aps,tightenlines,nofootinbib]{revtex4}

\pdfoutput=1 

\usepackage{amsmath}
\usepackage{amsfonts}
\usepackage{bm}
\usepackage{wasysym}
\usepackage{graphicx}
\usepackage{subfigure}
\usepackage{color}
\usepackage{lscape}
\usepackage{rotating}
\usepackage{natbib}

\bibpunct{(}{)}{;}{a}{,}{,}

\newcommand{\Ttensor}{\mathsf{T}}

\newcommand{\rtensor}{\mathsf{r}}

\newcommand{\mathd}{\mathrm{d}}

\newcommand{\Diff}{\mathrm{D}}

\begin{document}
\title{Interfacial instability in turbulent flow over a liquid film in a channel}
\author{Lennon \'O N\'araigh$\,^1$\footnote{Present address: School of Mathematical Sciences, University College Dublin, Belfield, Dublin 4, Ireland.}, P. D. M. Spelt$\,^1$, O. K. Matar$\,^1$, and T. A. Zaki$\,^2$}
\affiliation{Departments of Chemical$\,^1$ and Mechanical$\,^2$
Engineering, Imperial College London, SW7 2AZ, United Kingdom}
\date{\today}
%
%

%
%
\begin{abstract}
We revisit here the stability of a deformable interface that
separates a fully-developed turbulent  gas flow from a thin layer of
laminar liquid. Although this problem has been investigated in many
previous studies, a model that requires no parameters that must be
chosen {\it a posteriori}, and that uses a base state profile that
is validated against experiments is as yet unavailable. Furthermore,
the significance of wave-induced perturbations in turbulent stresses
remains unclear. Much emphasis in the oceanographic literature has
been on the distortion of turbulence by the presence of interfacial
waves, but the relevance of this has not been tested for turbulent
flow over a thin liquid layer in industrial channel or pipe flows.
Therefore, unlike previous work, the turbulent base state velocity
profile proposed here requires only a specification of a flowrate or
pressure drop, and no {\it a posteriori} choice of parameters.
Moreover, the base state contains sufficient detail such that it
allows for instability due to a viscosity-contrast mechanism (which
turns out to be dominant) as well as instability due to a
critical-layer-type mechanism, and it is validated against the
experimental and numerical data available in the literature.
Furthermore, the effect of perturbations in the turbulent stress
distributions is investigated, and demonstrated, for the first time,
to be small for cases wherein the liquid layer is thin. The detailed
modelling of the liquid layer elicits two unstable modes, and mode
competition can occur, although in most cases the instability is due
to the viscosity-contrast mechanism.  In particular, there is the
possibility that surface roughness can reduce the growth rate of the
interfacial mode, and promote a liquid-layer instability to the
status of most dangerous mode. Our base-state model facilitates a
new definition of `slow' and `fast' waves. We use our linear
stability analysis to determine the factors that affect the wave
speed and demonstrate that the waves are `slow' according to the
definition proposed here. Finally, we compare our results with
experimental data, and good agreement is obtained.
\end{abstract}
\maketitle
\section{Introduction}
\label{sec:intro}

A linear stability analysis of small-amplitude waves on an otherwise
flat liquid film would provide a powerful tool in understanding and
modelling the onset of droplet entrainment from a liquid layer by a
shearing superposed turbulent gas flow, which has numerous
industrial applications (e.g. \citet{Hewitt1970}). Furthermore, this
would serve as a benchmark for direct numerical simulations of
two-layer flows, as in
\citet{Boeck2007,Valluri2008,Fuster2009,Valluri2010} for laminar
flows. Although early work in the area focussed on a
`divide-and-attack' approach~\citep{Benjamin1958,Miles1962b},
wherein first the perturbation in the shear stress exerted on a wavy
film would be estimated from measurements and models in turbulent
flow over a wavy solid wall, which would then feed into a stability
analysis of the liquid layer, the advance of computational methods
and facilities has enabled one to solve stability problems wherein
perturbations in the gas and liquid are fully coupled. Various
studies have pursued this for liquid films sheared by turbulent flow
of a gas~\citep{Miesen1995,Kuru1995}. However, several difficulties
have arisen in these studies, which prevent such previous work to be
of direct use as benchmark tests for direct numerical simulations of
turbulent stratified channel flows and industrial applications.
First, a robust model for a base-state velocity profile that has
been tested against experiments and numerical simulations is
not available, and its detailed modelling turns out to be
important. Next, the base state should not require specification of
any ad-hoc parameters or parameters that must be chosen a
posteriori: merely the flow rate or imposed pressure drop (along
with the physical and geometrical properties) should suffice.
Previous models lack at least one of these
aspects; the base state model proposed here satisfies all of these
criteria.  
A second motivation for this study is to ascertain 
the role of perturbations in
turbulent stresses, which are caused by the presence of waves. 
These
could feed back to the growth rate and speed of the wave even in a
linear analysis. A further objective of this study is to confirm the
physical mechanism that leads to instability, as various mechanisms
have been proposed for general two-layer flows, and to relate this
to a classification of slow and fast waves. In many cases, previous
work was conducted in the context of wind-driven waves in
oceanographic applications, and it remains unclear whether, for
instance, a Miles-type instability could be approached in stratified
channel flows. In addressing this issue, the present work
leads to a new, naturally classification of slow and fast waves. We briefly
expand on these issues here.

In previous work, a `lin-log' base-state profile has been used in a
boundary-layer setting, and the friction velocity $U_*$ was
guessed~\citep{Miesen1995}.  Others have adopted an empirical
profile, the validity of which is unclear~\citep{Kuru1995}. Here, we
derive a base-state model that contains no free parameters: the
friction velocity is determined as a function of Reynolds number. We
develop our base state starting from a rigorously validated model
(that of \citet{Biberg2007}), and generalize it in order to take
account of the near-interfacial zone -- a significant region of the
flow for determining the viscosity-contrast instability.  We compare
the resulting model with direct numerical simulations and
experiments. Furthermore, our model contains no logarithmic
singularities: full modelling of the viscous sublayers is provided,
in contrast to the model of \citet{Biberg2007}. The base state
(i.e., the flat interface profiles) is presented in
Sec.~\ref{sec:flat}.

At least four mechanisms have been reported that relate to the
instability of a laminar liquid layer sheared by an external turbulent  gas flow.  The
first kind of instability was identified by \citet{Miles1957},
and is called the \textit{critical-layer instability}. The transfer
of energy from the mean flow into the wave perturbations is governed
by the sign of the second derivative of the base-state flow at the
critical layer -- the height where the wave speed and the
base-state velocity match.  Another kind of instability was identified by \citet{Yih1967}, and is called the
\textit{viscosity-contrast mechanism}.  Here the instability arises
due to the jump in the viscosity across the interface. In addition,
instability can arise due to direct forcing by turbulent pressure
oscillations in the gas~\citep{Phillips1957}.  The so-called
\textit{internal mode}~\citep{Miesen1995,Boomkamp1996}, is observed
when the bottom layer is laminar.  This mode derives its energy both
from the interface, and from conditions in the bulk of the bottom
layer.  Thus, this mode persists even when the upper layer is
void~\citep{Miesen1995}.  It has the characteristics of a
Tollmien--Schlichting wave, and depends sensitively on the viscosity
contrast.  This sensitivity is amplified when the bottom layer has
non-Newtonian rheology (\'O~\citet{ONaraigh2010}). \citet{Ozgen1998} studied such a scenario for power-law fluids,
where they find that the internal mode dominates over the
interfacial (viscosity-contrast) mode, a reversal of the situation
observed by Miesen and
co-workers~\citep{Miesen1995,Boomkamp1996,Boomkamp1997} for the
Newtonian case.  This effect is especially visible at high Reynolds
numbers.  This \textit{mode competition} is also possible for
Newtonian fluids: \citet{YeckoZaleski2002} have observed it for two-phase
mixing layers.
\citet{Benjamin1958} described an instability mechanism
called \textit{non-separated sheltering},  wherein the viscous
stresses near the interface give rise to a faster velocity
downstream of the wave crest, and thus cause a pressure asymmetry.
These effects are also observed in the viscosity-contrast
mechanism~\citep{Boomkamp1997}, of which the Benjamin description can
be regarded as a limiting case.  \citet{Belcher1993} proposed
that the turbulent stresses could produce the same effect, although
in this paper, we find that the viscous stresses are dominant in the
creation of the instability, at least under the thin-film parameter
regime studied here.
Finally, we note here that approximate mechanisms such as those
identified in Kelvin-Helmholtz-type theories,  although perhaps
relevant in large-amplitude waves, do not arise in the full
linearized problem wherein viscous effects are fully accounted for~\citep{Boomkamp1996}.

In order to assess which of these mechanisms dominates in sheared
liquid films by turbulent gas flow,  it is useful to note that the
critical-layer waves are typically fast, and the viscosity-contrast
waves are typically slow (relative to an appropriate scale); thus,
we propose in this paper to classify waves as fast or slow depending
on the kind of instability mechanism at work.  This is not only a
semantic choice: our definition leads to a rule-of-thumb for
deciding when waves are slow or fast, a rule wherein the wave speed
is measured relative to the interfacial gas friction-velocity, which
in turn is based on the pressure gradient in the channel. For fast
waves, the liquid layer has been treated like a moving wavy wall,
while for slow waves, detailed understanding of the liquid layer and
interfacial viscous sublayer layer is required.

\citet{Boomkamp1996} classify interfacial
instabilities according to an energy budget: an energy law is
associated to the dynamical equations, and positive inputs of energy
are identified with various physical mechanisms of instability.
Thus, they place the critical-layer and viscosity-contrast
instability into a rigorous framework.  They comment on the
equivalence between slow waves and the viscosity-contrast
instability, and the equivalence  between fast waves and the
critical-layer instability; we take this analysis further by
deriving a prediction for the character of the wave based on the
pressure gradient, and several other base-state parameters.  The
classification is presented in Sec.~\ref{subsec:slowandfast}.

In order to explain a further motivation for the present study, we
first summarize the averaging  approach presumed herein. Consider a
large ensemble of realizations of a (three-dimensional)
pressure-driven turbulent channel flow. The velocity field contains
perturbations due to turbulence, and due to the presence of
small-amplitude waves. At any time, a Fourier decomposition can be
taken of the interface height and, simultaneously, of the velocity
and pressure fields. These Fourier-decomposed velocity and pressure
fields can be averaged over the ensemble of realizations (as well as
over the spanwise direction). These ensemble-averaged velocity and
pressure fields are not uni-directional, but are distorted due to
the presence of the corresponding (normal mode) interfacial wave.
Example fields that have been obtained in conceptually the same
manner from DNS  (albeit for turbulent flow over a wavy wall) can be
found in the paper of~\citet{Sullivan2000}. In the present study, results are
presented (and compared) from several RANS models.
Now, a significant issue here is that such wave-induced perturbation
stresses may, in principle, affect the growth rate and speed of
waves.  Questions concerning the importance of these stresses have
been much debated in the
literature~\citep{Miles1957,Belcher1993,Belcher1994,Kuru1995,OceanwavesBook2004}.
Previous studies on sheared liquid films have not accounted for
these effects, hence the significance of these stresses are not
known at present. The present study does at last provide convincing
evidence that these are indeed not important in the determination of
wave speed and growth rate for sheared thin films. The linear
stability analysis and assessment of the significance of
perturbation turbulent stresses (PTS) are the subject of
Sec.~\ref{sec:model_perturb}.

In Sec.~\ref{sec:linear_stability}, we carry out a detailed linear
stability analysis to investigate the character of the  interfacial
instability, and to find out which parameters affect the wave speed.
We also study there mode competition, which can be achieved by
devising a roughened interface, this being the averaged effect of
instantaneous pressure fluctuations there due to the Phillips
mechanism. Finally, we compare the model predictions with
experimental data and simplified (Kelvin-Helmholtz-based) theories
in Sec.~\ref{sec:experiments}.

\section{The flat-interface model and its properties}
\label{sec:flat}

\begin{figure}
\begin{center}
\includegraphics[width=0.7\textwidth]{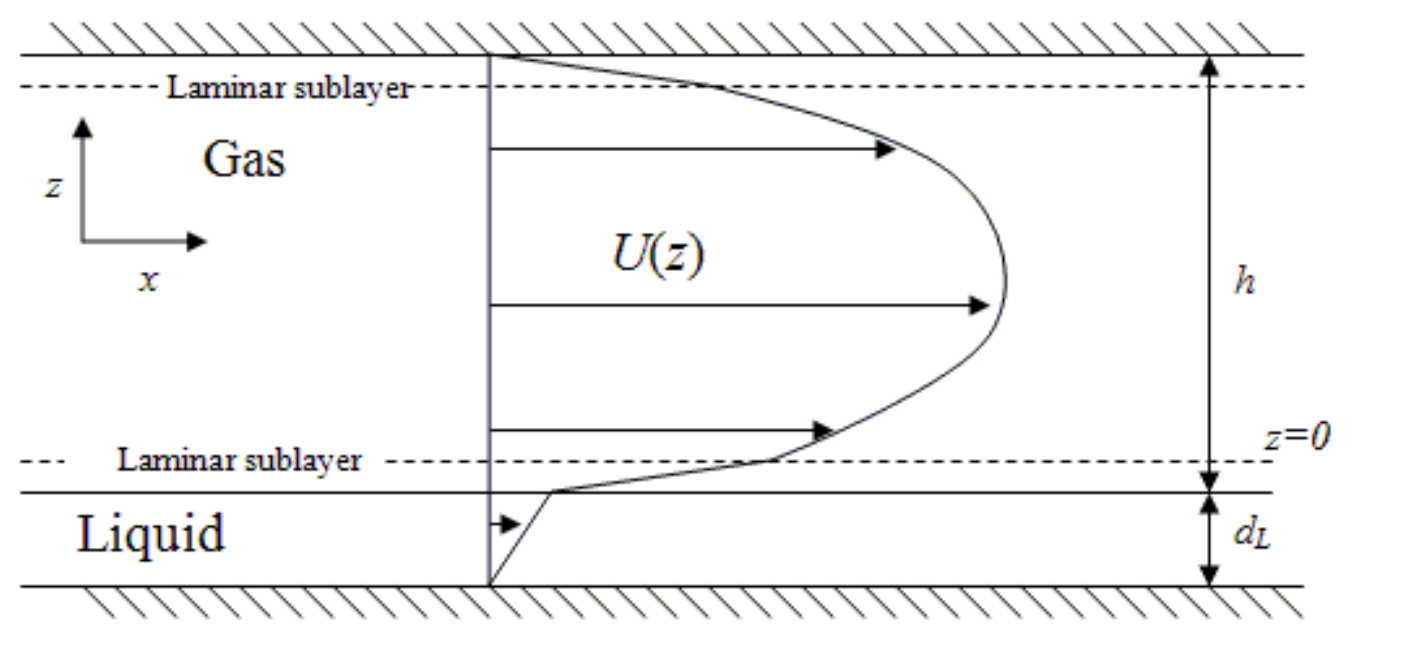}
\end{center}
\caption{A schematic diagram of the base flow.  The liquid layer is laminar,
while the gas layer exhibits fully-developed turbulence, described here by
a Reynolds-averaged velocity profile.  A pressure gradient in the $x$-direction
drives the flow.}
\label{fig:schematic}
\end{figure}

In this section we derive a base state appropriate for a two-layer system
in a channel, described schematically in Fig.~\ref{fig:schematic}.  This
is an equilibrium state of the system, in the sense that the mean velocities
are independent of time, and the mean interfacial height that demarcates
the phases is flat.
The bottom layer is a thin, laminar, liquid layer, while the top
layer is gaseous, turbulent and fully-developed. A pressure gradient
is applied along the channel. The mean profile of the system is a
uni-directional flow in the horizontal, $x$-direction. In the gas
layer, near the gas-liquid interface and the gas-wall boundary, the
flow profile is linear, and the viscous scale exceeds the
characteristic length scale of the
turbulence~\citep{TurbulenceMonin,TurbulencePope}. In the bulk of the
gas region, the flow possesses a logarithmic
profile~\citep{TurbulenceMonin,TurbulencePope}. We assume that the
gas-liquid interface is smooth, although we do take account of
surface roughness  in Sec.~\ref{sec:flat_roughness}, and again in
Sec.~\ref{subsec:roughness}.


The growth rate of the wave amplitude depends sensitively on the
choice of mean flow. Therefore, it is necessary to derive a
mean flow-profile that incorporates the characteristics of the flow
observed in experiments. In this section, we generalize the model of \citet{Biberg2007}
and accurately model the viscous sublayers
found in two-phase turbulent flows.  This is a non-trivial improvement, since
accurate descriptions of near-interfacial conditions are important in understanding
the stability of the interface.
 The functional form of the derived
velocity profile enables us to express the wall and
interfacial shear stresses as functions of the mean pressure
gradient. This treatment also enables us to write down a turbulent
closure scheme using an eddy viscosity, which is constituted as a
simple function of the vertical coordinate, $z$.
Using the above-mentioned assumptions and approximations, 
we derive the mean flow in each layer.

\subsection{The mean flow}
\label{subsec:mean_flow}

\textit{The liquid film:}  The Reynolds-averaged Navier--Stokes (RANS)
equation appropriate for the liquid film is the following:
\begin{equation}
\mu_L\frac{\partial^2 U_L}{\partial z^2}-\frac{\partial p}{\partial x}=0,
\label{eq:base_liquid}
\end{equation}
where $U_L$ and $\mu_L$ denote the liquid mean flow velocity and
viscosity, respectively.  Although the pressure gradient $\partial
p/\partial x$ is written using partial  derivatives, the assumption
of hydrostatic balance means that this pressure gradient is merely a
function of $x$.  For, the hydrostatic balance assumption amounts to
the equation $-\partial p_j/\partial z=\rho_j g$, in which $j=L,G$
labels the phase.  Integrating once gives $p_j=-\rho_j g
z+\tilde{p}_j\left(x\right)$, hence $\partial p_j/\partial
x=\mathd\tilde{p}_j/\mathd x$.  Since this derivative is independent
of the $z$-coordinate, it is phase-independent too, and we therefore
drop the phase-label from this pressure gradient in the rest of the
work.
We integrate
Eq.~\eqref{eq:base_liquid} and apply the following boundary
conditions, which correspond, respectively, to continuity of
tangential stress at the interface and no-slip at the channel bottom
wall:
\begin{equation} \mu_L\frac{\partial U_L}{\partial
z}\bigg|_{z=0}=\tau_{\mathrm{i}}=\rho_G U_{*\mathrm{i}}^2,\qquad U_L\left(-d_L\right)=0,
\label{eq:int_cond}
\end{equation}
where $-d_L\leq z\leq 0$ is the domain of the liquid film. The
quantity $\tau_{\mathrm{i}}$ is the interfacial stress and
$U_{*\mathrm{i}}$ is the interfacial friction velocity on the gas
side.  We are going to close the model by finding an expression for
this friction velocity in terms of the pressure gradient  $\partial
p/\partial x$. The boundary conditions~\eqref{eq:int_cond} yield the
following relation for the mean-flow velocity profile in the film,
\begin{equation}
U_L\left(z\right)=\frac{1}{2\mu_L}\frac{\partial p}{\partial x}
\left(z^2-d_L^2\right)+\frac{\tau_{\mathrm{i}}}{\mu_L}\left(z+d_L\right),\qquad
-d_L\leq z\leq 0.
\end{equation}
Nondimensionalizing on the scale $U_0$, where
\[
\rho_G U_0^2=h\left|\frac{\partial p}{\partial x}\right|,
\]
gives
\[
\tilde{U}_L=\frac{\mu_G}{\mu_L}\left[-\tfrac{1}{2}Re_0\left(\tilde{z}^2-\delta^2\right)+\frac{Re_*^2}{Re_0}\left(\tilde{z}+\delta\right)\right],\qquad
\delta=\frac{d_L}{h},
\]
where the tildes denote dimensionless quantities: $\tilde{U}=U/U_0$, and
$\tilde{z}=z/h$, and where
\[
Re_0=\frac{\rho_GU_0h}{\mu_G},\qquad
Re_*=\frac{\rho_GU_{*\mathrm{i}}h}{\mu_G},
\]
are the Reynolds numbers.  The definition of $Re_0$ differs from the definition
of the Reynolds number in single-phase channel flow, $Re_P=\rho_G h^3|\partial
p/\partial x|/2\mu_G^2$.  They are related, however, by the formula $Re_0=\sqrt{2Re_P}$.
 Note furthermore that
\begin{equation}
\tilde{U}\left(0\right)=\frac{\mu_G}{\mu_L}\left(\tfrac{1}{2}Re_0\delta^2+\frac{Re_*^2}{Re_0}\delta\right).
\label{eq:u0}
\end{equation}

\paragraph*{The gas layer:}  The RANS equation in the gas is
\begin{equation} \mu_G\frac{\partial U_G}{\partial
z}+\tau_{\mathrm{TSS}}=\tau_{\mathrm{i}}+\frac{\partial p}{\partial
x}z,
\label{eq:rans0}
\end{equation}
where $\tau_{TSS}=-\rho\langle u' w'\rangle$ is the turbulent shear stress
due to the averaged effect of the turbulent fluctuating velocities.  In channel
flows, it is appropriate to model this term using an eddy-viscosity model~\citep{TurbulenceMonin}.
 In mixing-length theory, the eddy viscosity depends on the local
 rate of strain~\citep{Bradshaw1974}, which means that the turbulent shear
 stress depends on the square of the rate of strain.  Instead of this standard
 mixing-length theory, we introduce an interpolation function for the eddy
 viscosity, which mimics the ordinary mixing-length theory near the interface
 and near the wall, and transitions smoothly from having a positive slope
 near the interface, to having a negative slope near the wall.  Thus, the
 turbulent shear stress is linear in the rate of strain, and
\begin{equation}
\tau_{\mathrm{TSS}}=\mu_T\frac{\partial U_G}{\partial z},
\qquad\mu_T=\kappa\rho_G h U_{*\mathrm{w}}G\left(\tilde{z}\right)\psi_{\mathrm{i}}\left(\tilde{z}\right)\psi_{\mathrm{w}}\left(1-\tilde{z}\right),
\label{eq:tss0}
\end{equation}
where
$\mu_T$ is the eddy viscosity, $U_{*\mathrm{w}}$ is the friction
velocity at the upper wall $\tilde{z}=1$,  and where
$G\left(\tilde{z}\right)$,
$\psi_{\mathrm{i}}\left(\tilde{z}\right)$, and
$\psi_{\mathrm{w}}\left(1-\tilde{z}\right)$ are functions to be
determined.  Here $\psi_{\mathrm{i}}$ and $\psi_{\mathrm{w}}$ are
interface and wall functions respectively, which damp the effects of
turbulence to zero rapidly near the interface and the wall, while
$G$ is an interpolation function designed to reproduce the law of
the wall near the interface and the upper wall.  This interpolation
scheme is based on the work of \citet{Biberg2007}.  The
precise choice of $G$ and the wall functions is given below, choices
that are confirmed by the agreement between our predictions of the
base state and experiments and numerical simulation.
Substituting Eq.~\eqref{eq:tss0} into Eq.~\eqref{eq:rans0} gives the formula
\begin{eqnarray}
U_G\left(z\right)&=&U_G\left(0\right)+\tau_{\mathrm{i}}h\int_0^{z/h}\frac{\left(1+\frac{h}{\tau_{\mathrm{i}}}\frac{\partial
p}{\partial x}s\right)ds}{\mu_G+\kappa\rho_Gh U_{*\mathrm{w}}G\left(s\right)\psi_{\mathrm{i}}\left(s\right)\psi_{\mathrm{w}}\left(1-s\right)},\nonumber\\
&=&U_G\left(0\right)+\tau_{\mathrm{i}}h\int_0^{z/h}\frac{\left(1+\frac{h}{\tau_{\mathrm{i}}}\frac{\partial
p}{\partial x}s\right)ds}{\mu_G+\frac{\kappa\rho_Gh U_{*\mathrm{i}}}{\sqrt{|R|}}G\left(s\right)\psi_{\mathrm{i}}\left(s\right)\psi_{\mathrm{w}}\left(1-s\right)},
\end{eqnarray}
where $R=\tau_{\mathrm{i}}/\tau_{\mathrm{w}}$.  Non-dimensionalizing and using Eq.~\eqref{eq:u0}, this is
\begin{equation}
\tilde{U}_G\left(\tilde{z}\right)=\frac{\mu_G}{\mu_L}\left(\tfrac{1}{2}Re_0\delta^2+\frac{Re_*^2}{Re_0}\delta\right)
+\frac{Re_*^2}{Re_0}\int_0^{\tilde{z}}\frac{\left(1-\frac{Re_0^2}{Re_*^2}s\right)ds}{1+\frac{\kappa Re_{*}}{\sqrt{|R|}}G\left(s\right)\psi_{\mathrm{i}}\left(s\right)\psi_{\mathrm{w}}\left(1-s\right)}.
\end{equation}
The ratio $R$ can be obtained in closed form as follows.  Since
\begin{eqnarray*}
\tau\left(z\right)&=&\tau_{\mathrm{i}}+\frac{\partial p}{\partial x}z,\\
&=&-\tau_{\mathrm{w}}+\frac{\partial p}{\partial x}\left(z-h\right),
\end{eqnarray*}
these formulas can be equated to give
\[
\tau_{\mathrm{i}}=-\tau_{\mathrm{w}}-\frac{\partial p}{\partial x}h,
\]
or,
\[
-\frac{\tau_{\mathrm{w}}}{\tau_{\mathrm{i}}}=1+\frac{\partial p}{\partial
x}\frac{h}{\tau_{\mathrm{i}}}=1-\left(\frac{Re_0}{Re_*}\right)^2,
\]
hence,
\[
|R|=\left|1-\left(\frac{Re_0}{Re_*}\right)^2\right|^{-1}.
\]
We use the following form for the $G$-function.  This function is
designed to reproduce the logarithmic  profile near the interface
and near the upper wall, a result that we demonstrate below
(P.~\pageref{tag:ulog}); further evidence of the correctness of this
choice is provided when we compare our predictions for the
base-state velocity with experiments (Sec.~\ref{sec:compare}).
Thus,
\begin{equation}
G\left(s\right)=s\left(1-s\right)\underbrace{\left[\frac{s^3+|R|^{5/2}\left(1-s\right)^3}{R^2\left(1-s\right)^2+Rs\left(1-s\right)+s^2}\right]}_{=\mathcal{V}\left(s\right)},\qquad
0\leq s\leq 1.
\label{eq:G}
\end{equation}
We use a Van-Driest type of formalism~\citep{TurbulencePope} for the wall functions $\psi_{\mathrm{i}}$
and $\psi_{\mathrm{w}}$:
\begin{equation}
\psi_{\mathrm{i}}\left(s\right)=1-e^{-s^n/A_{\mathrm{i}}},\qquad
\psi_{\mathrm{w}}\left(1-s\right)=1-e^{-\left(1-s\right)^n/A_{\mathrm{w}}},
\label{eq:psi}
\end{equation}
where $n$, $A_{\mathrm{i}}$, and $A_{\mathrm{w}}$ are input parameters.  We are now in a position to determine
$Re_*$: it is obtained as the zero of the function $\tilde{U}\left(1;Re_*\right)=0$, or
\begin{equation}
\frac{\mu_G}{\mu_L}\left(\tfrac{1}{2}Re_0\delta^2+\frac{Re_*^2}{Re_0}\delta\right)
+\Bigg\{\frac{Re_*^2}{Re_0}\int_0^{1}\frac{\left(1-\frac{Re_0^2}{Re_*^2}s\right)ds}{1+\frac{\kappa Re_{*}}{\sqrt{|R|}}G\left(s\right)\psi_{\mathrm{i}}\left(s\right)\psi_{\mathrm{w}}\left(1-s\right)}\Bigg\}_{|R|=\left|1-\left(\frac{Re_0}{Re_*}\right)^2\right|^{-1}}=0.
\end{equation}
In summary, we have the following velocity profile in the base state:
\begin{equation}
\tilde{U}\left(\tilde{z}\right)=\begin{cases}\frac{\mu_G}{\mu_L}\left[-\tfrac{1}{2}Re_0\left(\tilde{z}^2-\delta^2\right)+\frac{Re_*^2}{Re_0}\left(\tilde{z}+\delta\right)\right],&-\delta\leq\tilde{z}\leq0,\\
\frac{\mu_G}{\mu_L}\left(\tfrac{1}{2}Re_0\delta^2+\frac{Re_*^2}{Re_0}\delta\right)
+\frac{Re_*^2}{Re_0}\int_0^{\tilde{z}}\frac{\left(1-\frac{Re_0^2}{Re_*^2}s\right)ds}{1+\frac{\kappa Re_{*}}{\sqrt{|R|}}G\left(s\right)\psi\left(s\right)\psi\left(1-s\right)},&0\leq\tilde{z}\leq1.\end{cases}
\label{eq:U_base}
\end{equation}
\begin{figure}
  \begin{center}
\subfigure[]{
\includegraphics[width=0.3\textwidth]{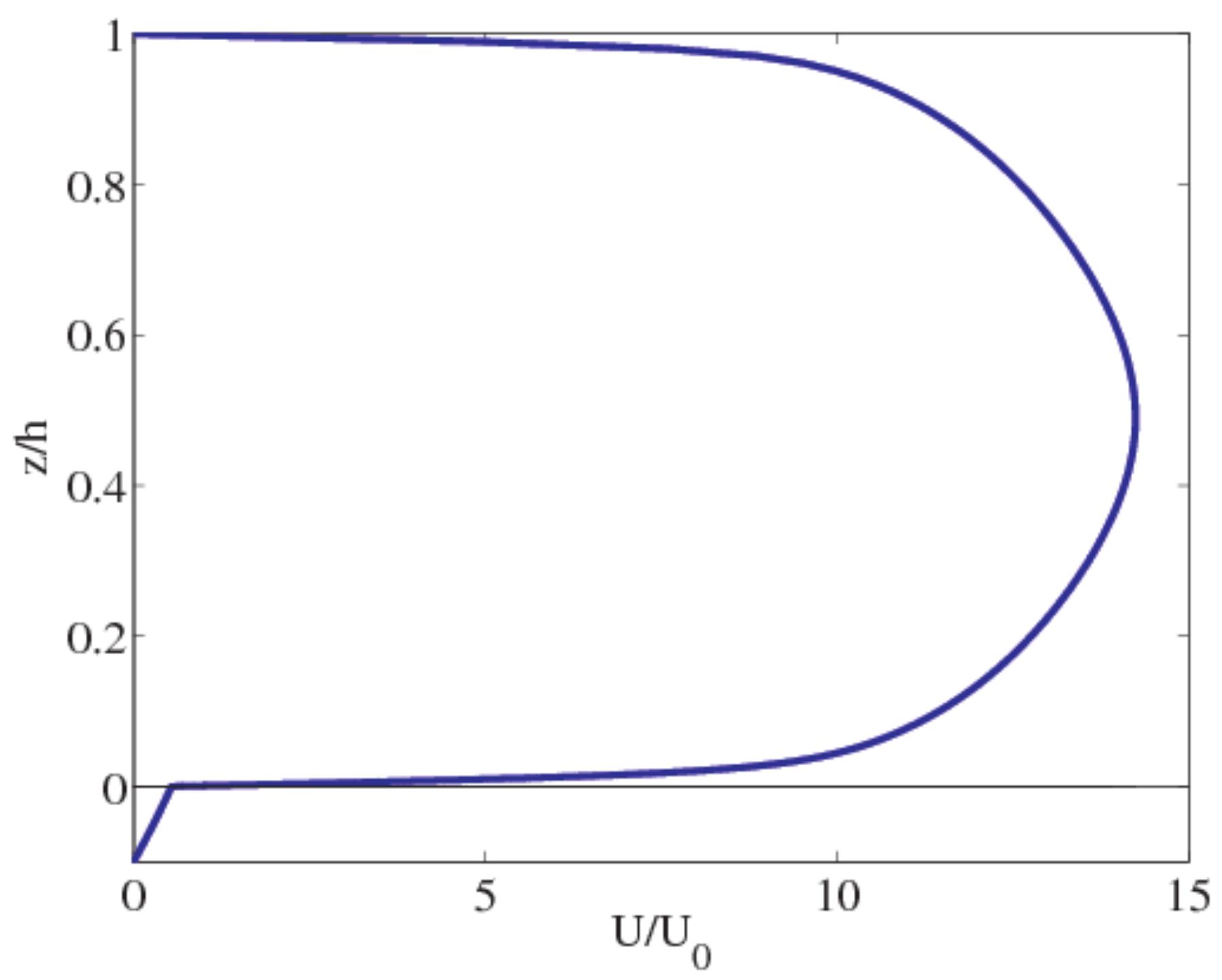}}
\subfigure[]{
\includegraphics[width=0.29\textwidth]{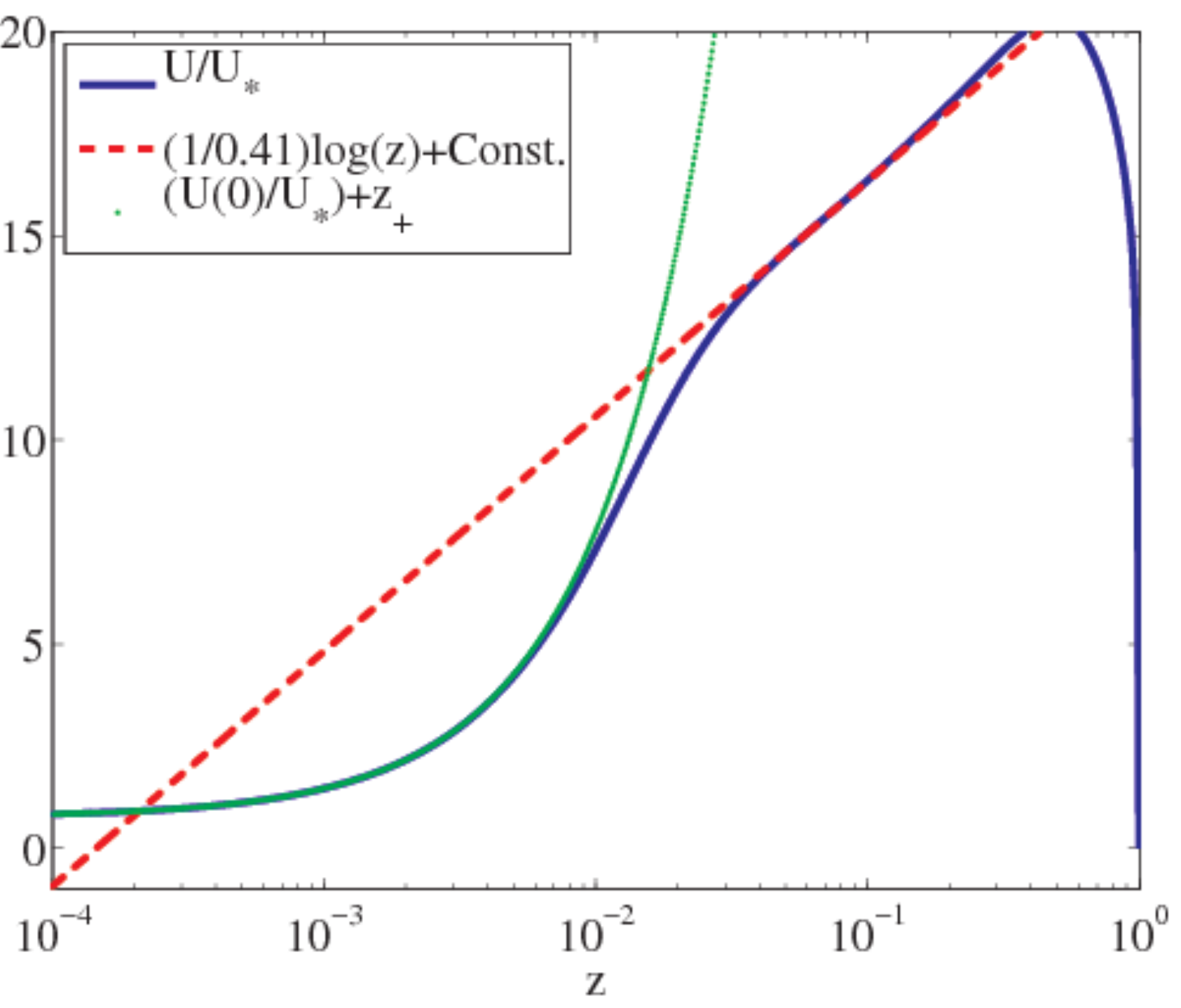}}
\subfigure[]{
\includegraphics[width=0.3\textwidth]{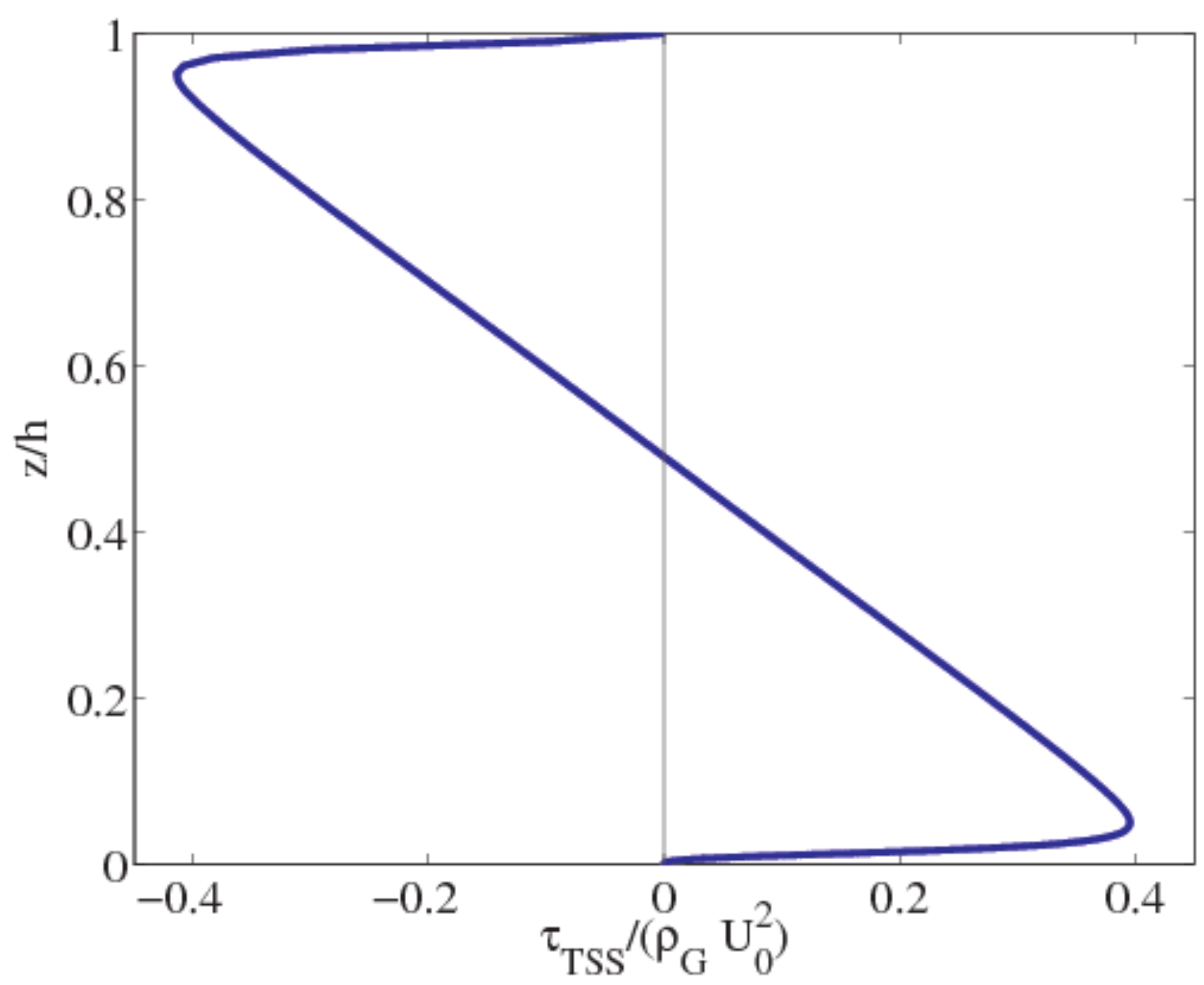}}
\end{center}
\caption{
Characteristics of the base profile for fixed parameter values $\left(\mu_L/\mu_G,\rho_L/\rho_G,d_L/h\right):=\left(m,r,\delta\right)=\left(100,1000,0.1\right)$,
and $Re_0=1000$.
(\textit{a}) The mean velocity profile; (\textit{b}) The mean velocity profile
in wall units, showing the logarithmic and viscous layers (the viscous layer
has a wall-unit thickness of approximately $5$); (\textit{c}) The
Reynolds stress profile corresponding to the basic velocity.}
\label{fig:base0}
\end{figure}

We now discuss in detail the choice of function in Eqs.~\eqref{eq:G}
and~\eqref{eq:psi}. The function $\psi_{\mathrm{i}}$
($\psi_{\mathrm{w}}$), as defined in Eq.~\eqref{eq:psi},
transitions rapidly from $\psi_{\mathrm{i}}\left(0\right)=0$
($\psi_{\mathrm{w}}\left(1\right)=0$)
 to unity, across a width
\begin{equation}
\frac{d_{\mathrm{v}}}{h}=\frac{\nu_G}{h U_{*\mathrm{i}}}=\frac{1}{Re_*},
\text{ or }
\frac{d_{\mathrm{v},\mathrm{w}}}{h}=\frac{\nu_G}{h U_{*\mathrm{w}}}=\frac{|R|}{Re_*},
\label{eq:dv}
\end{equation}
where the value of $A_{\mathrm{i}}$ ($A_{\mathrm{w}}$) is related to the
width of $\psi_{\mathrm{i}}$ ($\psi_{\mathrm{w}}$).
Note also the existence of the scale
\begin{equation}
\frac{h_{\mathrm{m}}}{h}=\frac{Re_*^2}{Re_0^2},
\label{eq:midpoint}
\end{equation}
which is the channel midpoint where ${\partial U}/{\partial z}=0$.
The choice of wall function $\psi_{\mathrm{i}}$ rapidly dampens the eddy viscosity to zero near the interface,
but has little effect elsewhere.  Using this choice, together with the
form
\begin{equation}
G\left(s\right)=s\left(1-s\right)\mathcal{V}\left(s\right),
\label{eq:G1}
\end{equation}
we obtain the correct viscous behaviour for the velocity profile near $\tilde{z}=0$:
\begin{eqnarray}
\tilde{U}&\sim&\mathrm{Const.}+\frac{Re_*^2}{Re_0}\int_0^{\tilde{z}}\left[1+O\left(s\right)\right]ds,\qquad\text{as }\tilde{z}\rightarrow0,\nonumber\\
&=&\mathrm{Const.}+\frac{Re_*^2}{Re_0}\tilde{z}+O\left(\tilde{z}\right)^2,\qquad
\text{as }\tilde{z}\rightarrow0.
\label{eq:ulog}%
\end{eqnarray}%
The form of $G$ given in Eqs.~\eqref{eq:G} and~\eqref{eq:G1} is chosen such
that the basic velocity profile possesses a log layer close to, but not at
the interface (wall).
 The $G$-function we use (Eq.~\eqref{eq:G}) was derived in the paper of \citet{Biberg2007}.
 Our model generalizes this work by taking account of the dynamically important
 viscous sublayers.  This extra detail has the added advantage that logarithmic
 singularities are no longer present in the velocity profile.
Thus, for those $z$-values in the part of the domain sandwiched between the
interface and the channel midpoint, that is, for
\[
\frac{1}{Re_*}\ll \tilde{z}\ll \frac{Re_*^2}{Re_0^2},
\]
\label{tag:ulog}%
the function $G\left(\tilde{z}\right)$ has the property that
\begin{eqnarray}
G\left(\tilde{z}\right)&=&\tilde{z}\left[\mathcal{V}\left(0\right)+\frac{\mathd\mathcal{V}}{\mathd \tilde{z}}\bigg|_0\tilde{z}+...\right],\nonumber\\
&=&\tilde{z}\sqrt{|R|}\left[1-\left(1+\frac{1}{R}\right)\tilde{z}\right]+...,\nonumber\\
&=&\tilde{z}\sqrt{|R|}\left[1-\frac{Re_0^2}{Re_*^2}\tilde{z}\right]+...,
\label{eq:dG}
\end{eqnarray}
and hence,
\begin{eqnarray*}
\tilde{U}&\sim&\mathrm{Const.}+\frac{Re_*^2}{Re}\int^{\tilde{z}}\frac{ds}{\kappa
Re_* s},\\
&=&\mathrm{Const.}+\frac{Re_*}{\kappa Re}\ln\left(\tilde{z}\right).
\end{eqnarray*}
A similar calculation near $\tilde{z}=1$ establishes the existence of a log
layer close to the upper wall.  Finally, we establish the values of $A_{\mathrm{i,w}}$ and
$n$ in Eq.~\eqref{eq:psi}.  Close to the interface $\tilde{z}=0$, the Reynolds stress has the form
\[
\frac{\tau_{TSS}}{\rho_G U_0^2}\sim\kappa\frac{Re_*^3}{Re_0^2}\frac{\tilde{z}^{n+1}}{A_{\mathrm{i}}},
\]
Now a good approximation to the interface in two-phase
turbulence with a large density contrast is in fact a solid wall~\citep{Banerjee2002}.
Thus, we set $n=2$, the value appropriate for wall-bounded turbulence~\citep{TurbulencePope}.
In a similar manner, we fix $A_{\mathrm{i}}$ and $A_{\mathrm{w}}$
with reference to single-phase theory,  wherein there is a
one-to-one correspondence between the values of $A_{\mathrm{i}}$ and
$A_{\mathrm{w}}$ and the additive constant $B$ in the single-phase
log law.  With $\kappa=0.4$, the specification
\begin{equation}
A_{\mathrm{i},\mathrm{w}}=e^{6.3} Re_{*\mathrm{i},\mathrm{w}}^{-2}
\label{eq:adef}
\end{equation}
corresponds to the known value $B=5.3$.  It is this relationship that we use throughout our study.
%
%
 We combine these modelling assumptions to obtain a velocity profile in Fig.~\ref{fig:base0}.
 This velocity profile is computed for $Re_0=1000$, for which the corresponding superficial Reynolds based on the gas flow rate is approximately $12,000$.
  The near-interfacial viscous and logarithmic layers are visible in Fig.~\ref{fig:base0}~(b).

\begin{figure}
  \begin{center}
\subfigure[]{
\includegraphics[width=0.4\textwidth]{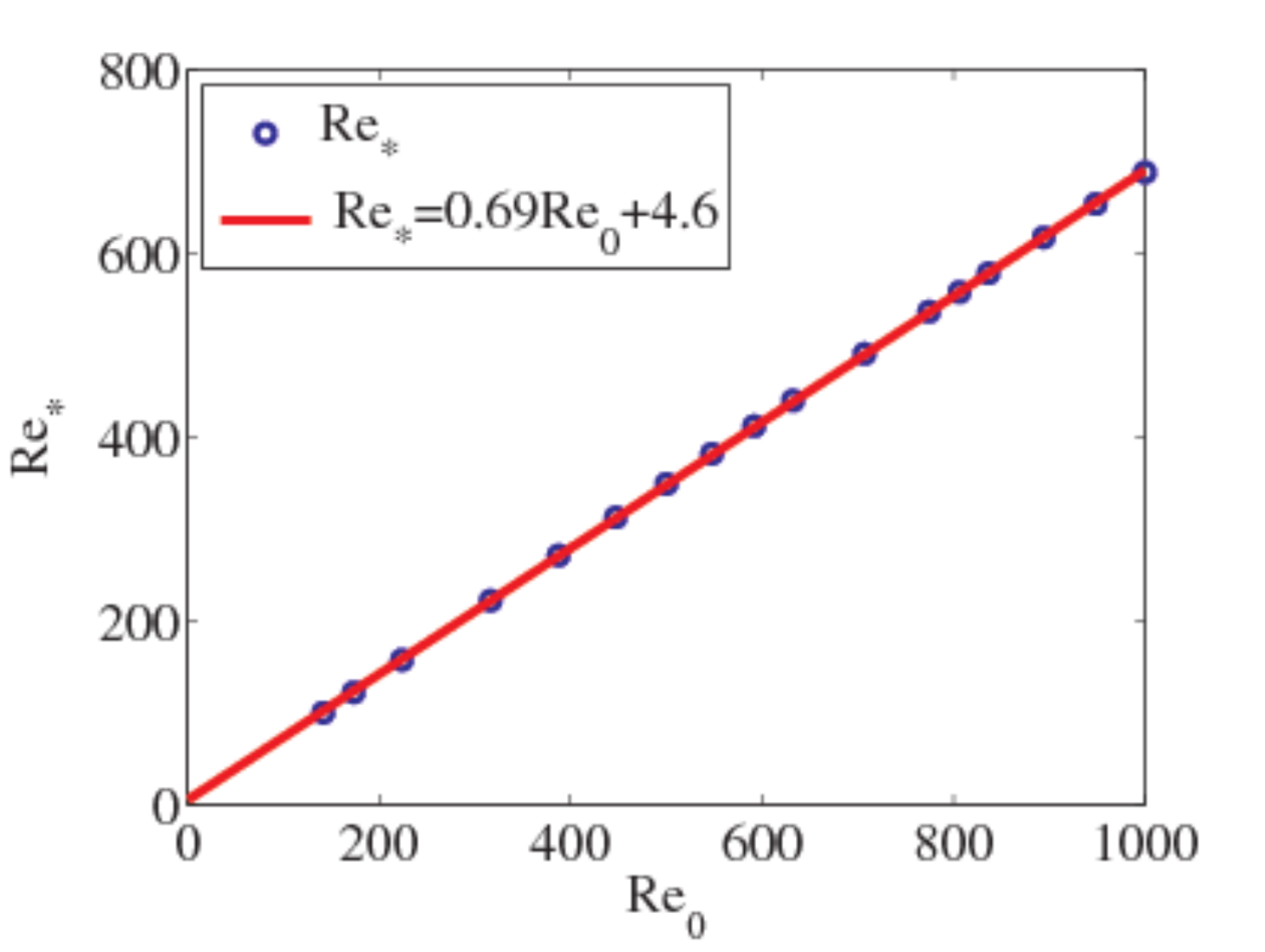}}
\subfigure[]{
\includegraphics[width=0.4\textwidth]{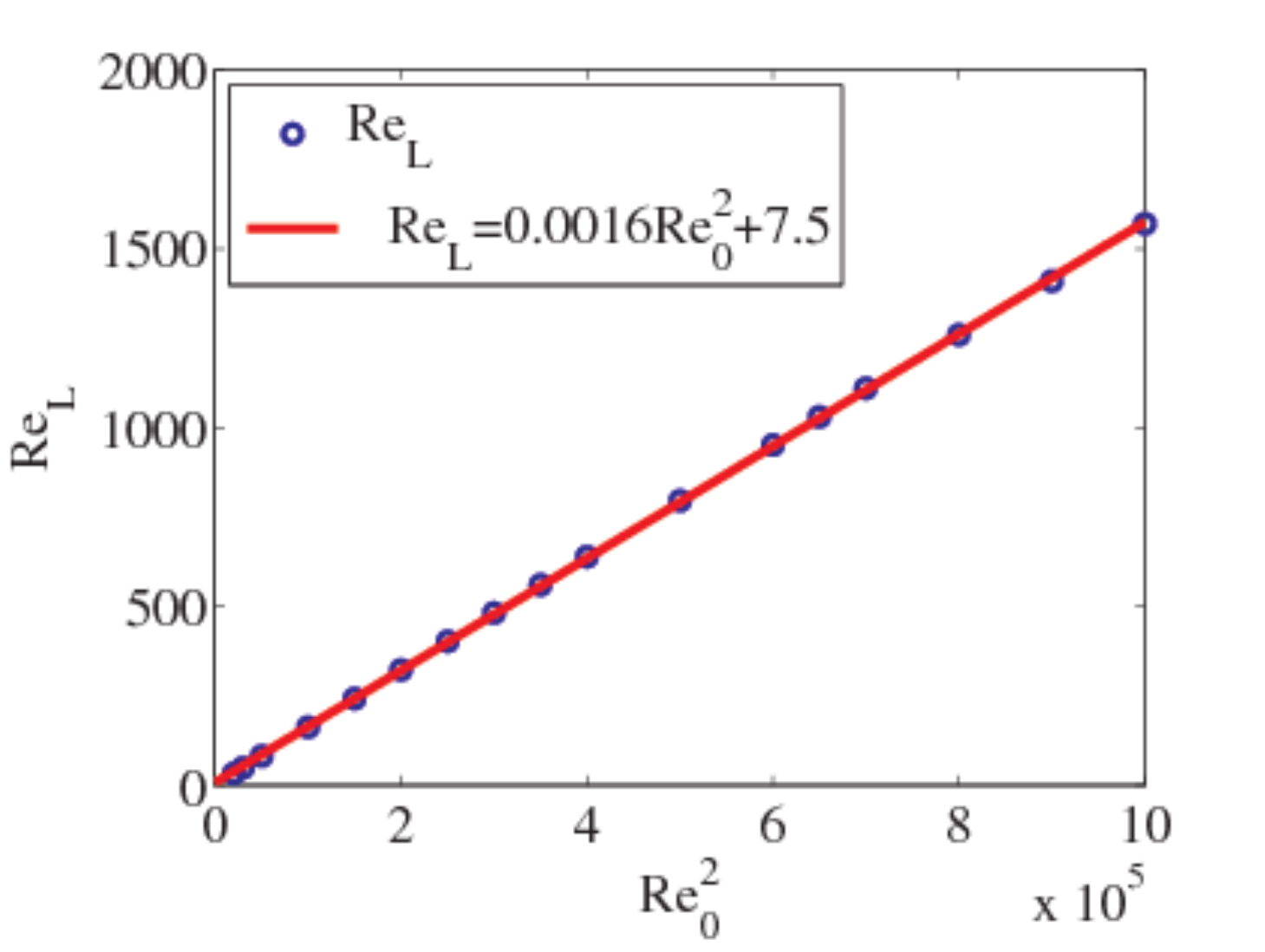}}
\end{center}
\caption{Properties of the base state as a function of the Reynolds number
$Re_0$.  We have set $\delta=0.1$ and $\left(m,r\right)=\left(55,1000\right)$.  (a)  Dependence of the friction Reynolds number $Re_*$ on $Re_0$;
(b) the Reynolds number $Re_L$.
}
\label{fig:base1}%
\end{figure}

The numerical solution for the base-state profile also enables us to determine
the friction Reynolds number and the liquid Reynolds number
as a function of the control parameter $Re_0$: this is done in Fig.~\ref{fig:base1}.
  This liquid Reynolds number is defined as
\begin{equation}
Re_L=\frac{\rho_L d_L U_{\mathrm{i}}}{\mu_L},\qquad U_{\mathrm{i}}=\frac{\tau_{\mathrm{i}}d_L}{\mu_L}.
\label{eq:ReL}
\end{equation}
Figure~\ref{fig:base1}~(a) shows the dependence of the friction Reynolds
number $Re_*$ on the control parameter $Re_0$.  The relationship is approximately
linear.  This is a consequence of the very small velocities in the liquid,
compared with the maximal gas velocity.  Thus, the gas layer closely resembles
single-phase channel flow, and the condition $U\left(0\right)\ll U_{\mathrm{max}}$
mimics the zero interfacial-velocity condition in the single-phase channel.  The channel midpoint
where $\partial U/\partial z=0$ is thus approximately equal to half the gas-layer
depth $h_{\mathrm{m}}\approx h/2$.  Using this guess in Eq.~\eqref{eq:midpoint}
gives $Re_*\approx Re_0/\sqrt{2}$, which is close to the slope calculated
in the figure.  Figure~\ref{fig:base1}~(a) shows the dependence of the liquid
Reynolds number $Re_L$ on the control parameter $Re_0$.  The approximately
quadratic relationship is a consequence of the definition of $Re_L$ (Eq.~\eqref{eq:ReL}),
and the linear relationship between $Re_0$ and $Re_*$.
%
%


\subsection{Slow and fast waves}
\label{subsec:slowandfast}

\begin{figure}
\begin{center}
\includegraphics[width=0.5\textwidth]{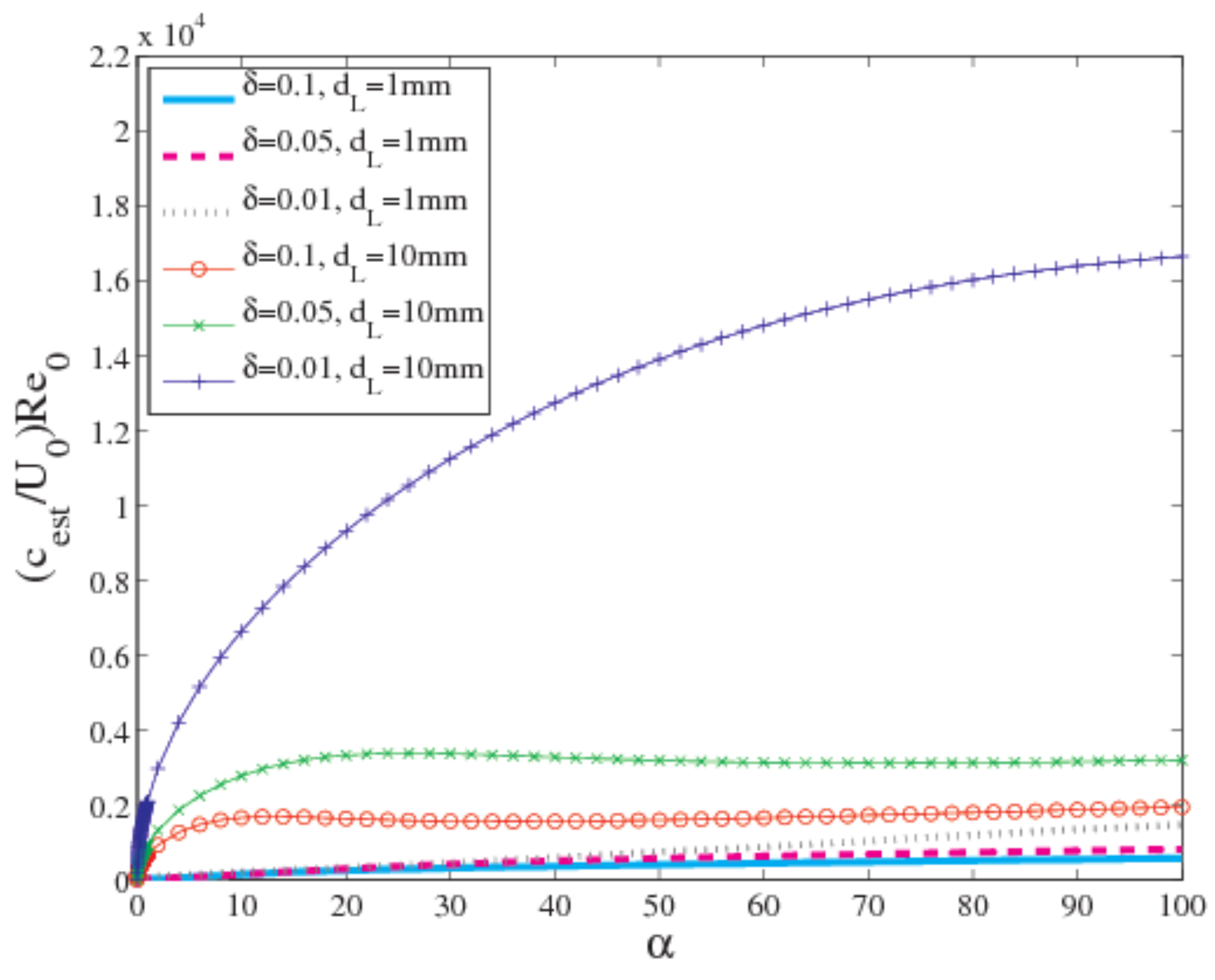}
\end{center}
\caption{An estimate of the boundary between slow and fast waves,
as a function of Reynolds number $Re_0$ and wavenumber $\alpha$.
For very thin films ($d_L=1\,\mathrm{mm}$), slow waves are
guaranteed at almost all Reynolds numbers, while for thicker films
($d_L=10\,\mathrm{mm}$) the waves are faster for all but the highest
Reynolds numbers. } \label{fig:c_est}
\end{figure}

Our model also gives a way of predicting the values of $\left(Re_0,d_L\right)$ for which the
\begin{table}
  \begin{center}
  \begin{tabular}{|l|c|c|}
  \hline
  Symbol&Numerical value, S.I. Units\\
  \hline
  \hline
  $\mu_G$&$1.8\times10^{-5}$\\
  $m=\mu_L/\mu_G$&$55$\\
  $\rho_G$&$1$\\
  $r=\rho_L/\rho_G$&$10^3$\\
  $d_L$&$10^{-3}$--$10^{-2}$\\
  $\delta=d_L/h$&0.1\\
  $g$&$9.8$\\
  $\sigma$&$0.074$\\
  \hline
  \end{tabular}
  \caption{Table of parameter values used to estimate the wave speed.}
  \label{tab:values0}%
  \end{center}%
\end{table}%
critical-layer instability
could be relevant.  This mechanism depends sensitively on the shape of the base
state, and causes a tiny wave-like perturbation at the interface to grow
in time when $\left(\mathd^2 U_G/\mathd z^2\right)_{z=z_{\mathrm{c}}}<0$,
where the critical height $z_{\mathrm{c}}$ is the root of the equation $U_G\left(z\right)=c$,
and where $c$ is the wave-propagation speed.  When the critical height lies inside
the viscous sublayer, the curvature of the mean profile is negligible, and
this mechanism is not relevant.  We obtain an estimate for the Reynolds numbers
$Re_0$
that produce this regime where the critical layer is unimportant
by solving the equation
\[
\frac{Re_*^2}{Re_0}z_{\mathrm{c}}=\frac{Re_*}{Re_0}z_{\mathrm{c}}^+=c,\qquad
z_{\mathrm{c}}^+\leq5,
\]
or,
\begin{subequations}
\begin{equation}
\frac{c}{U_0}\leq 5\frac{Re_*}{Re_0}.
\label{eq:slow_def}
\end{equation}
Equation~\eqref{eq:slow_def} gives a formal definition of a slow wave.
Since $Re_*\approx Re_0/\sqrt{2}$ for thin liquid layers, this definition reduces to
\begin{equation}
\frac{c}{U_0}\apprle\frac{5}{\sqrt{2}}=O\left(1\right).
\label{eq:c_order_1}%
\end{equation}%
\label{subeq:master}%
\end{subequations}%
We estimate the wave speed $c$, which we denote by $c_{\mathrm{est}}$, by using the formula for gravity-capillary
waves on a quiescent free surface (recall that the tilde is used to denote dimensionless variables):
\begin{equation}
\frac{c_{\mathrm{est}}}{U_0}=\frac{1}{Re_0}\sqrt{\frac{g h}{\left(\mu_G/\rho_G h\right)^2}\frac{r-1}{r+1}\frac{1}{\tilde{\alpha}}+\frac{1}{r+1}\frac{\sigma}{\mu_G^2/\rho_G
h}\tilde{\alpha}}\sqrt{\tanh\left(\tilde{\alpha}\delta\right)}.
\label{eq:c_grav}
\end{equation}
We test the accuracy of this formula in a number of cases (see Sec.~\ref{subsec:prelim_numerics}): it gives an order-of-magnitude prediction of the wave speed.  We use the values from Tab.~\ref{tab:values0}
and obtain a graphical description of the boundary between slow and
fast waves,  as a function of the parameters $\left(d_L,\delta,
Re_0, \alpha\right)$ (Fig.~\ref{fig:c_est}).  When
$c_{\mathrm{est}}/U_0\ll1$, we expect the critical-layer mechanism
to be unimportant.  This is precisely the regime of small
$d_L$-values and high Reynolds numbers, which is the subject of this
report.
 According to the classification of \citet{Boomkamp1996},
 the other two mechanisms of instability that exist for two-phase flow are
%
%
the viscosity-driven instability, and the liquid internal mode.  We must
therefore be on the lookout for these instabilities in the linear stability
analysis that follows.

One final note concerning Eq.~\eqref{eq:c_grav}.  It can be re-written as
\begin{eqnarray}
\frac{c_{\mathrm{est}}}{U_0}&=&\sqrt{\frac{Fr}{r+1}\frac{1}{\tilde{\alpha}}+\frac{S}{r+1}\tilde{\alpha}}\sqrt{\tanh\left(\tilde{\alpha}\delta\right)},\nonumber\\
&=&\sqrt{\frac{gh}{U_0^2}\frac{r-1}{r+1}\frac{1}{\tilde{\alpha}}+\frac{S}{r+1}\tilde{\alpha}}\sqrt{\tanh\left(\tilde{\alpha}\delta\right)},
\label{eq:c_est_froude}
\end{eqnarray}
where we have introduced the inverse Froude and inverse Weber numbers, respectively
\begin{equation}
Fr=\frac{g\left(\rho_L-\rho_G\right)h}{\rho_GU_0^2}=\frac{gh}{U_0^2}\left(r-1\right),\qquad S=\frac{\sigma}{\rho_GU_0^2h}.
\label{eq:Fr_def}
\end{equation}
By varying the inverse Froude number $Fr$, a transition between slow and fast waves is accomplished.  Equation~\eqref{eq:c_est_froude}
makes the role of $Fr$ manifest in this process.  Moreover, it suggests the
possibility of generating fast waves by fixing $Fr$ and reducing the density contrast, or increasing the strength of the surface tension.  We shall return to this
question in the parameter study in Sec.~\ref{sec:linear_stability}.
It is important, however, to treat this analysis as preliminary, since we
have no right to assume that Eq.~\eqref{eq:c_grav} is valid.  Indeed, a central
message of \S~\ref{sec:linear_stability} is
that the wave speed must be determined, along with the growth rate, by an
Orr--Sommerfeld type of analysis.

\subsection{Interfacial roughness}
\label{sec:flat_roughness}

So far we have been concerned with flow profiles where the interface is a
perfectly smooth surface separating the phases.  Now, we allow for surface
roughness by modifying the eddy-viscosity law~\eqref{eq:tss0} and~\eqref{eq:G}. %
The work of \citet{lin2008} gives one possible
explanation for the generation of such roughness.   This work
indicates that the so-called Phillips mechanism~\citep{Phillips1957}
may be important, whereby instantaneous turbulent pressure
fluctuations give rise to a regime of linear wave growth.  This is
later followed by an exponential growth regime, which is primarily
governed by the disturbances in the flow induced by the waves
themselves.  In the present context, we regard the surface roughness
as a consequence of the gas-phase turbulence, which then acts on the
interfacial waves, thereby modifying the growth in the wave
amplitude.

 Two approaches to the modelling of the surface roughness present themselves.  The first,
 and more rigorous approach, is to use an eddy-viscosity model like that of \citet{Biberg2007}.
 Such a model has the effect of shrinking the viscous sublayer near the interface.  In our formalism,
 this is achieved by altering the form of the mixing length near the interface:
 before it was
\[
\mathcal{L}\sim z\psi_{\mathrm{i}}\left(z/h\right),\qquad\text{as } z\rightarrow0,
\]
where $\psi_{\mathrm{i}}\left(z/h\right)$ is the damping function that operates
in the near-wall region $z\apprle 5U_{*\mathrm{i}}/\nu_G$; now, instead,
we propose the behaviour
\[
\mathcal{L}\sim \ell_\mathrm{i},\qquad\text{as } z\rightarrow0.
\]
Thus,
\begin{eqnarray*}
\frac{\tau_{TSS}}{\rho_G U_0^2}&=&\frac{\kappa}{\sqrt{|R|}}\left[s+K\left(1-s\right)\right]\left(1-s\right)\psi_{\mathrm{w}}\left(1-s\right)\mathcal{V}\left(s\right)\frac{\mathd
U}{\mathd z},\qquad s=z/h\\
&=&\frac{\kappa}{\sqrt{|R|}}G\left(s\right)\psi_{\mathrm{w}}\left(1-s\right)\frac{\mathd
U}{\mathd z},
\end{eqnarray*}
where $K=\ell_{\mathrm{i}}/\left(\kappa h\right)$ is the nondimensional interfacial
roughness parameter.
%
%
Now the second, and more ad hoc approach is simply to reduce
the interfacial viscous sublayer region in our flat-interface model.  This
is accomplished by reducing the parameter $A_{\mathrm{i}}$ in
the wall function $\psi_{\mathrm{i}}$.  \citet{Morland1993}
have used this approach, and have parametrized the effects of roughness
%
%
%
simply  by reducing the viscous-sublayer region
of the flat-interface model.  Such a reduction then gives rise to a reduced
growth rate.
This is in itself a rather trivial observation, although it does
have important implications for the linear stability analysis of
two-phase flow: when the interfacial growth rate is reduced, there
is the possibility of mode competition, and an internal mode can
come to dominate in the stability analysis.  We comment on this in
Sec.~\ref{subsec:roughness},  where we compare and contrast the
results of a linear stability analysis using both roughness models.

\subsection{Comparison with other studies}
\label{sec:compare}

To validate our model for the base state, we compare it with other
studies of both single- and two-phase flow.  A further comparison
with studies of single-phase flow over a wavy
wall~\citep{Zilker1976,Abrams1985}  is provided in Appendix~A.  We
first of all characterize the single-phase version of our model.
This is obtained by setting $U_{\mathrm{int}}=0$ and by ignoring the
liquid layer.
We compare with the experimental work of \citet{willmarth1987},  for single-phase pressure-driven channel
flow.  In the experiment, the Reynolds number based on the friction
velocity was $1.143\times 10^3$,  which corresponds to a model
Reynolds number $Re_0=\sqrt{2}\times 1.143\times10^{3}$.  The mean
Reynolds number in the experiment was $Re_\mathrm{m}=2.158\times
10^{4}$ -- a Reynolds number based on the mean velocity and channel
depth.
\begin{figure}
\begin{center}
\includegraphics[width=0.4\textwidth]{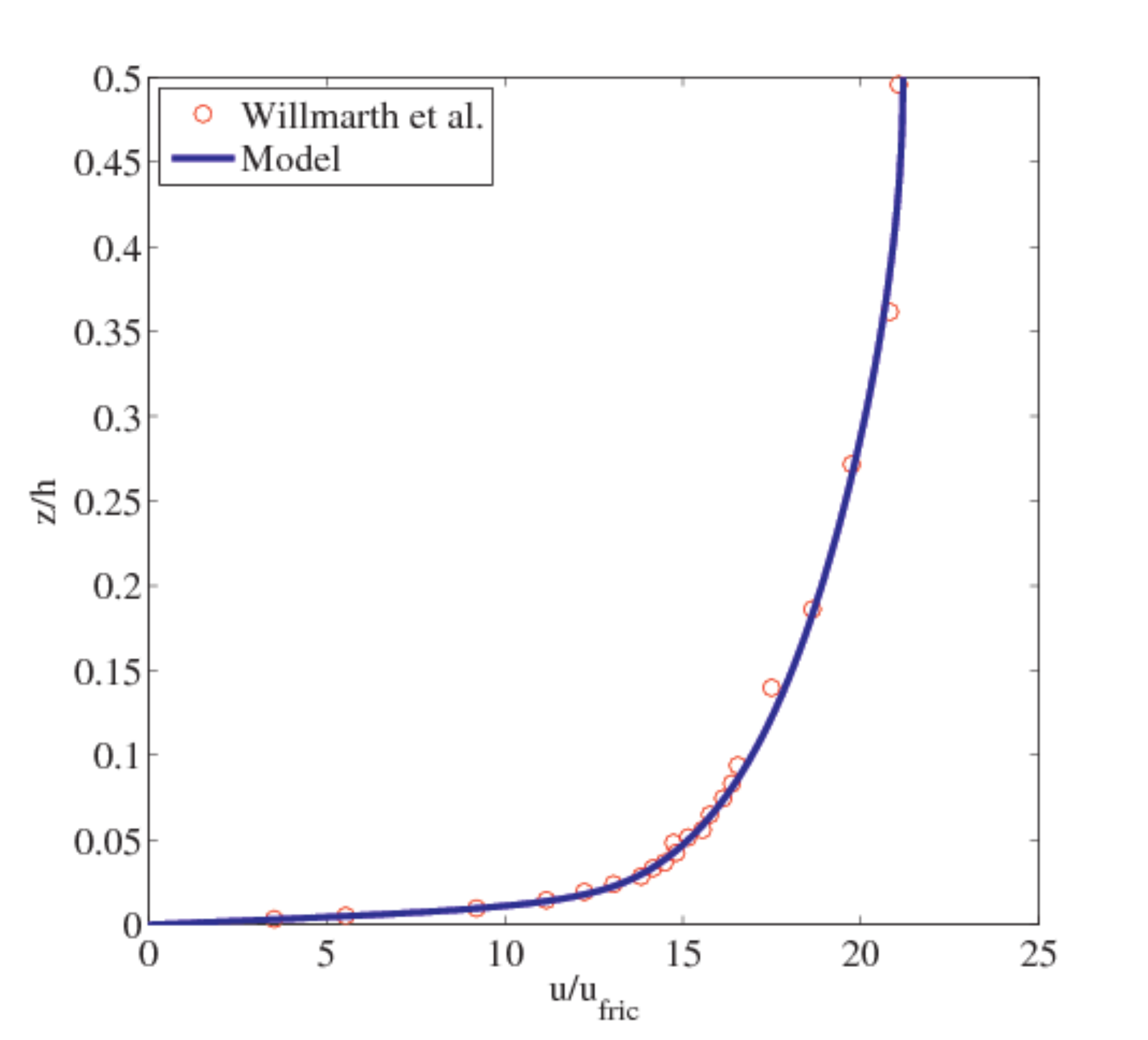}
\end{center}
\caption{A comparison with the work of \citet{willmarth1987} for single-phase channel flow for $Re_\mathrm{m}=2130\times 10^4$.  Excellent agreement between the model and the experiment is obtained, throughout the flow domain.}
\label{fig:compare_willmarth}
\end{figure}
We compute $Re_\mathrm{m}$ to be $2.13\times10^4$, close to the
value given in the experiment.   A plot of the profile is shown in
Fig.~\ref{fig:compare_willmarth}.  The model and the experimental
data are in excellent agreement.

To validate the two-phase version of our model, we first of all
compare  it with the work of \citet{akai1980,akai1981}  whostudied two-phase turbulence for
an air-mercury system, where
\[
m=77,\qquad r=1.120\times10^4,\qquad\text{at room temperature}.
\]
The liquid Reynolds number, based on the liquid-layer depth and the
mean liquid velocity is set to $Re_{\mathrm{m},L}=8040$ throughout
the experiments. Because the liquid is no longer laminar, we apply
the turbulence model Eq.~\eqref{eq:tss0} to both layers. The
gas-layer Reynolds number, based on the gas-layer height and mean
gas velocity, varies: we study the cases where
$Re_{\mathrm{m},G}=2340$ and $3690$,  for which the interface is
flat.   We have obtained the value of $\delta$ corresponding to the
flow rate $Re_{\mathrm{m},L}=8040$ through numerical
iteration\footnote{We use two parameters to specify the flow
configuration: $\delta$ and $Re_0$.  In many experiments, liquid and
gas growth rates are used instead.  The latter can be obtained from
the former within the framework of the model through numerical
iteration.}.  In a similar way, we have obtained the value of $Re_0$
corresponding to the flow rates $Re_{\mathrm{m},G}=2340$ and $3690$.
\begin{figure}
\begin{center}
\subfigure[]{\includegraphics[width=0.3\textwidth]{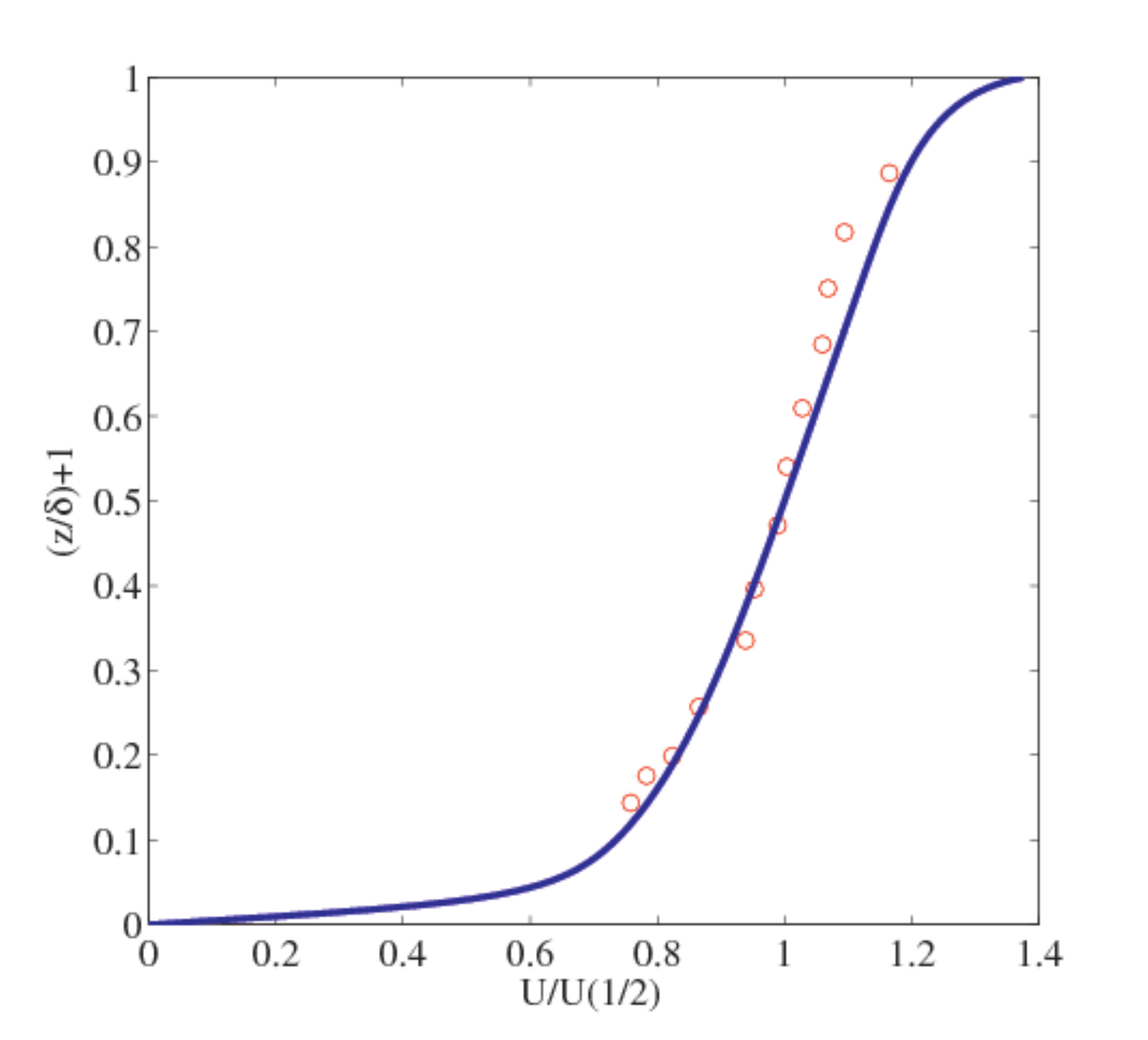}}
\subfigure[]{\includegraphics[width=0.3\textwidth]{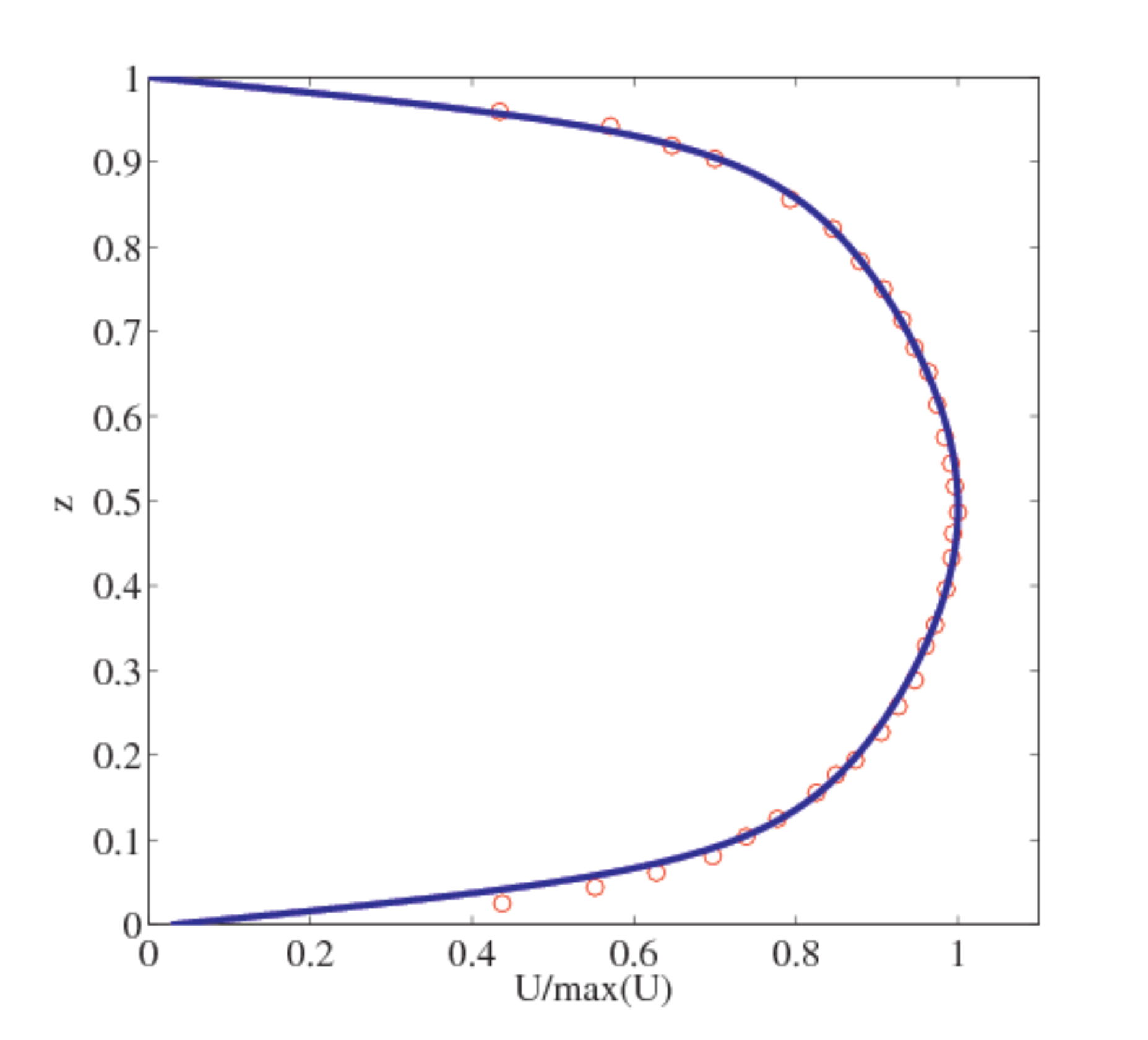}}
\subfigure[]{\includegraphics[width=0.3\textwidth]{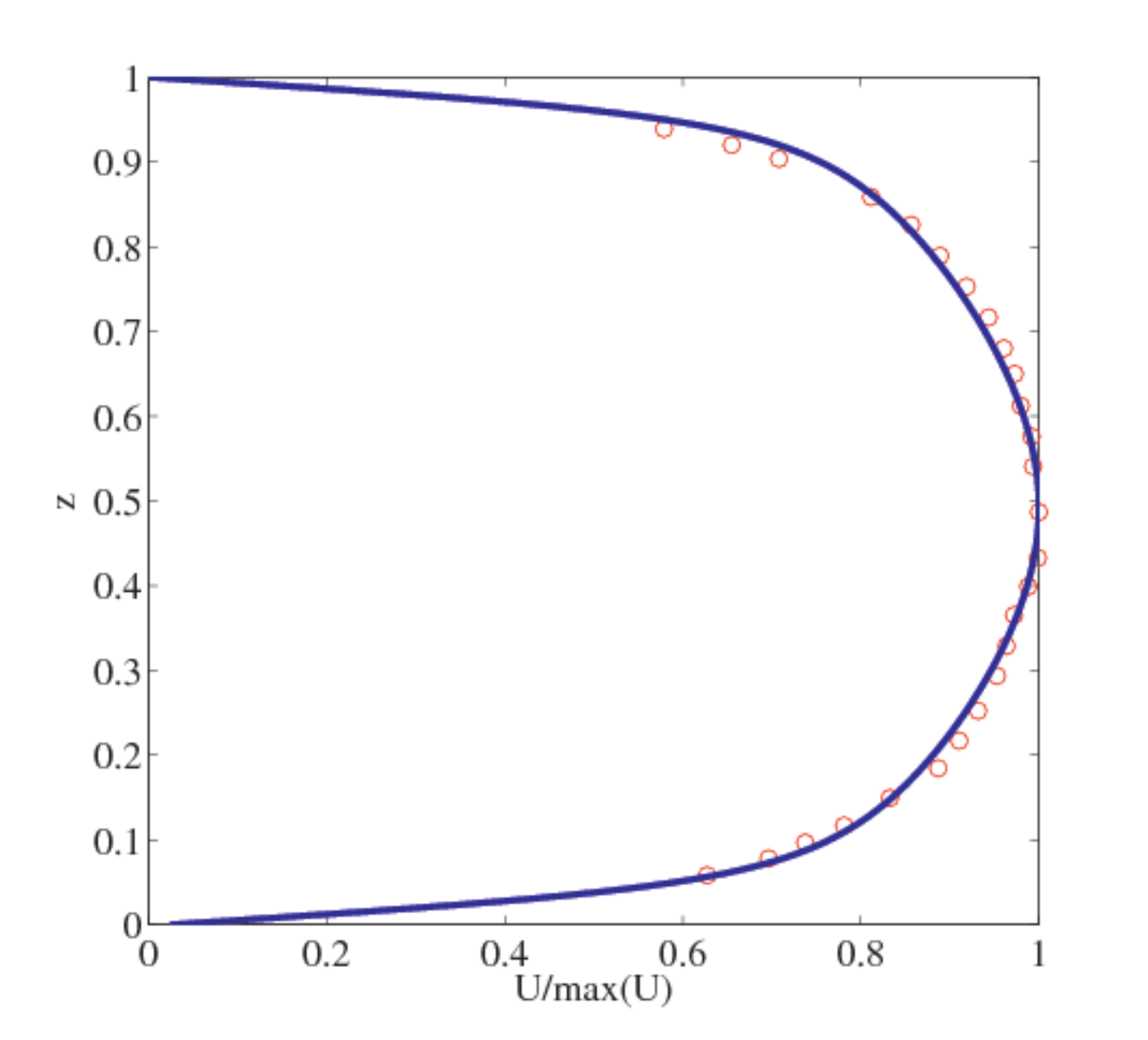}}
\end{center}
\caption{A comparison with the work of Akai \textit{et al.}.  Subfigures (a) and (b) show the comparison for $Re_{\mathrm{m},G}=2340$; (c) shows the result for $Re_{\mathrm{m},G}=3690$.  Here $U(1/2)$ denotes the liquid-phase velocity half-way between the bottom wall and the interface.  In both cases, $Re_{\mathrm{m},L}=8040$.  The agreement is excellent in the gas phase and reasonable in the liquid  phase, and is almost identical to the predictions given in the paper of \citet{Biberg2007}.}
\label{fig:compare_akai}
\end{figure}
The results of this comparison are shown in
Fig.~\ref{fig:compare_akai},  where excellent agreement is obtained,
particularly in the gas phase.  The agreement between the model and
the experiments is as good as in the paper of \citet{Biberg2007}.  This is not surprising, since our model
is designed to replicate his in the log-law regions of the flow, and
in the core regions.  Indeed, we conclude from the near-exact
agreement between our predictions and those of Biberg that our model
inherits all the results he obtained from experimental comparisons.
The added advantage of our model is that it can be continued down to
the wall and interfacial zones.

Finally, to validate the near-interface region of the model, we
compare it with the  DNS results of
\citet{Solbakken2004} for two-phase lubricated channel
\begin{figure}
\begin{center}
\subfigure[]{
\includegraphics[width=0.45\textwidth]{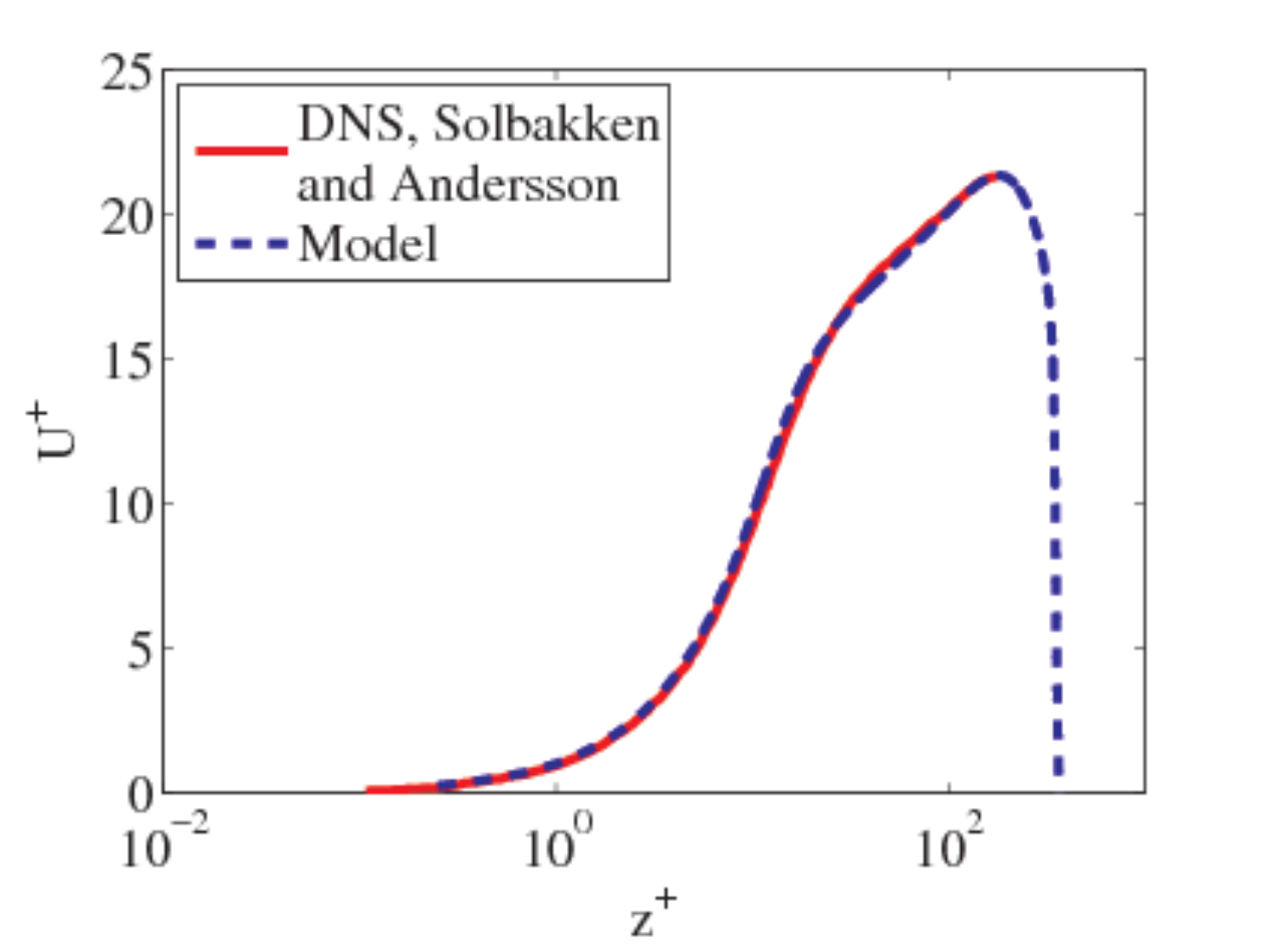}}
\subfigure[]{
\includegraphics[width=0.45\textwidth]{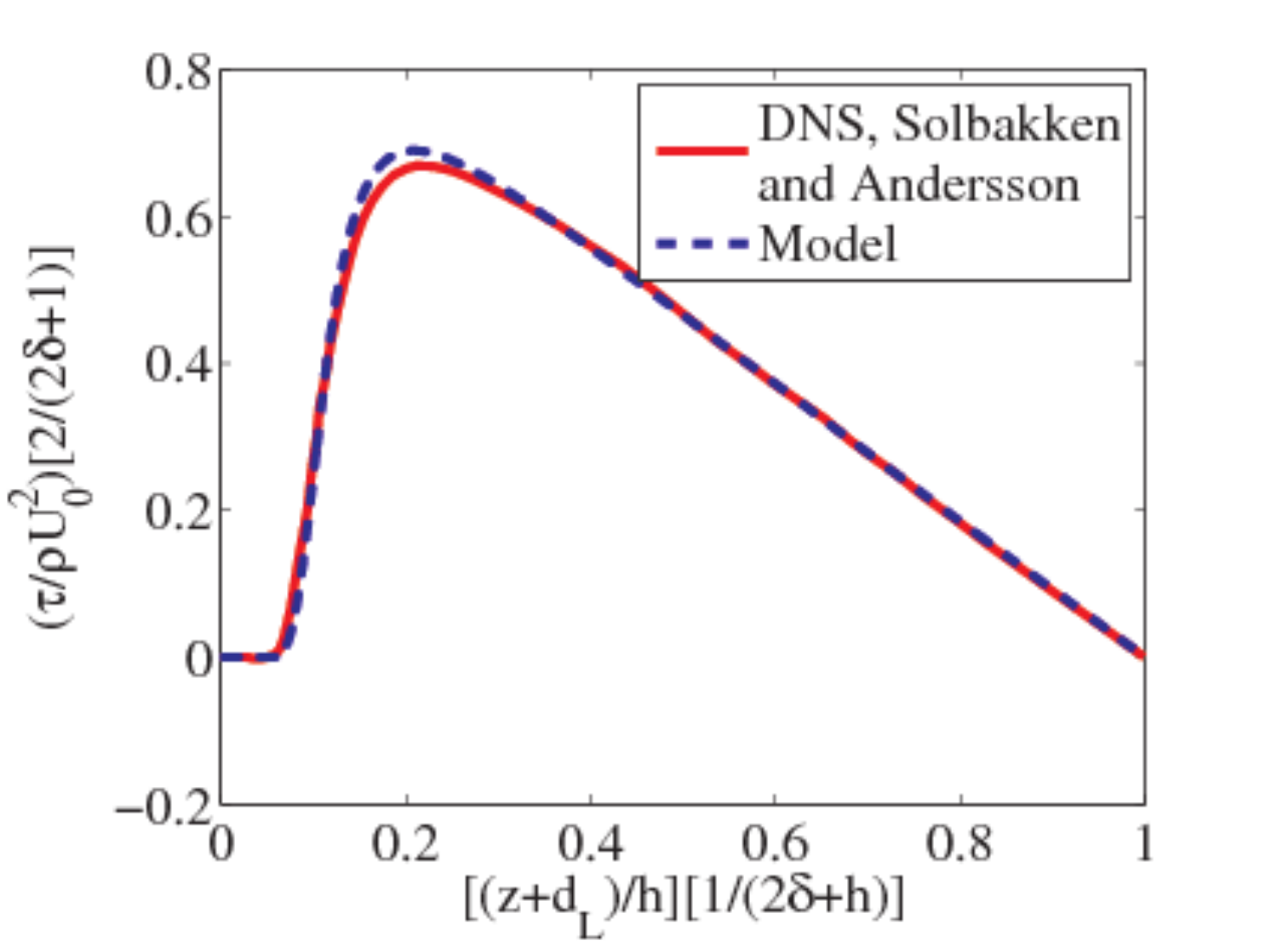}}
\end{center}
\caption{Comparison with the results of \citet{Solbakken2004}
for a lubricated channel. The broken-line curve gives the model profile across
the channel, while the solid-line curve describes the DNS results. The latter
results only extend to the channel midpoint.}
\label{fig:compare_dns}
\end{figure}
flow. To compare with their results, we take $\delta=1/34$, $m=2$, $r=1$ (hence $\rho_L=\rho_G=\rho$), and
\[
Re_{\tau}=\frac{\rho\left(h+2d_L\right)U_\tau}{2\mu_G}=\frac{\rho\left(h+2d_L\right)}{2\mu_G}\sqrt{\frac{h+2d_L}{2\rho}\left|\frac{\partial p}{\partial x}\right|}=180.
\]
Thus,
\[
\frac{U_0}{U_\tau}=\sqrt{\frac{2}{2\delta+1}},\qquad \frac{Re_0}{Re_\tau}=\left(\frac{2}{2\delta+1}\right)^{3/2}.
\]
Furthermore, we take
\[
y^+=\frac{z}{h}\frac{2Re_\tau}{2\delta+1},\qquad
U^+=\frac{U}{U_0}\sqrt{\frac{2}{2\delta+1}}.
\]
The results are shown in figure~\ref{fig:compare_dns}. There is
excellent agreement, in particular near the interface.  We have also
compared the model predictions with the  two-phase numerical
simulations of \citet{Magnaudet2009} for air and
water (not shown), and have found similar good agreement.  Thus, our
model is an adequate base state for the stability analysis we now
carry out.

\section{The model perturbation equations: derivation and numerical studies}
\label{sec:model_perturb}

\subsection{Derivation of the model perturbation equations}

We base
the dynamical equations for the interfacial motion on the Reynolds-averaged
Navier--Stokes (RANS) equations.  The turbulent velocity is decomposed into averaged
and fluctuating parts $\left(U,W\right)$ and $\left(u',w'\right)$ respectively.
 The averaged velocity depends on space and time through the RANS equations:
\begin{subequations}
\begin{equation}
\rho\left(\frac{\partial U_i}{\partial t}+\bm{U}\cdot\nabla U_i\right)=-\frac{\partial
P}{\partial x_i}+\mu\Delta
U_i-\rho\left(\frac{\partial}{\partial x}\langle u' u_i'\rangle+\frac{\partial}{\partial
z}\langle w' u_i'\rangle\right),
\label{eq:rans}
\end{equation}
\begin{equation}
\nabla\cdot\bm{U}=0,
\end{equation}%
\end{subequations}%
where $\langle\cdot\rangle$ denotes the averaging technique.
%
%
We use
these equations to model a flat-interface or base state of the two-phase
system shown in Fig.~\ref{fig:schematic}.  Next, we introduce a small
disturbance that shifts
the flat interface
at $z=0$ to $z=\eta$ (the dimensionless wave elevation), where $|\eta|\ll1$. This induces a change in the
average velocity and pressure fields, denoted as follows:
\[
\left(U,W,P\right)=\left(U_0\left(z\right)+\delta u\left(x,z,t\right),\delta
w\left(x,z,t\right),P_0\left(x,z\right)+\delta p\left(x,z,t\right)\right),
\]
where we denote base-state quantities by a subscript zero.
Since the flow is turbulent, and since the perturbations take the form of
a wave, they must satisfy the RANS equations for a linear wave, $\partial_t=-c\partial_x$:
\begin{eqnarray*}
\rho\left[\left(U_0-c\right)\frac{\partial}{\partial x} \delta u+\frac{\mathd
U_0}{\mathd{z}}\delta w\right]&=&-\frac{\partial}{\partial x}\left(\delta p-\rho\delta\sigma_z\right)+\mu\left(\frac{\partial^2}{\partial
x^2}+\frac{\partial^2}{\partial z^2}\right)\delta u+\rho\frac{\partial}{\partial
x}\delta\sigma+\rho\frac{\partial}{\partial
z}\delta\tau,\\
\rho\left(U_0-c\right)\frac{\partial}{\partial x}\delta w&=&-\frac{\partial}{\partial
z}\left(\delta p-\rho\delta\sigma_z\right)+\mu\left(\frac{\partial^2}{\partial x^2}+\frac{\partial^2}{\partial
z^2}\right)\delta w+\rho\frac{\partial}{\partial x}\delta\tau,\\
\frac{\partial}{\partial x}\delta u+\frac{\partial}{\partial z}\delta w&=&0.
\end{eqnarray*}
The quantities
\[
\delta\tau=-\langle u' w'\rangle-\tau^{(0)},\,\,
\delta\sigma_x=-\langle u'^2\rangle-\sigma_x^{(0)},\,\,
\delta\sigma_z=-\langle w'^2\rangle-\sigma_z^{(0)},\,\,\delta\sigma=\delta\sigma_x-\delta\sigma_z
\]
are the perturbation stresses due to the turbulence in the perturbed state,
while the quantities with the zero-superscript are base-state stresses.
 Using the streamfunction representation $\left(\delta u,\delta w\right)=\left(\phi_z,-\phi_x\right)$,
 and the normal-mode decomposition $\partial_x=i\alpha$, the perturbed
 RANS equations reduce to a single equation.   In non-dimensional form, the
 gas equation is
\begin{subequations}
\begin{equation}
i\alpha\left[\left(U_0-c\right)\left(\Diff^2-\alpha^2\right)\phi_G-\frac{\mathd^2U_0}{\mathd z^2}\phi_G\right]=\frac{1}{Re_0}\left(\Diff^2-\alpha^2\right)^2\phi_G+
i\alpha\Diff\delta\sigma+\left(\Diff^2+\alpha^2\right)\delta\tau,
\label{eq:os_turb}
\end{equation}
where $\Diff=\mathd/\mathd z$, while the liquid equation is simply
\begin{equation}
i\alpha r\left[\left(U_0-c\right)\left(\Diff^2-\alpha^2\right)\phi_L-\frac{\mathd^2U_0}{\mathd z^2}\phi_L\right]=\frac{m}{Re_0}\left(\Diff^2-\alpha^2\right)^2\phi_L.
\label{eq:os_liquid}%
\end{equation}%
\label{subeq:os}%
\end{subequations}%
Equations~\eqref{subeq:os} represent an Orr--Sommerfeld type of
system~\citep{orr_a,orr_b,orzag1971,Yiantsios1988}, with extra
turbulent stresses in the gas.  The problem of modelling the
additional stresses in Eq.~\eqref{eq:os_turb}  is considered
throughout the literature.  In this section, we use two stationary
turbulent models from this literature to describe these stress
terms.  Both models give the same result.  A stationary model is
appropriate for slow waves because the dynamically important region
for the instability located very close to the interface, on the gas
side.  This is precisely the region of the gas domain where the
turbulence stationarity condition is satisfied, namely that the eddy
turnover frequency $U_{*\mathrm{i}}/\left(\kappa z\right)$ should
greatly exceed the advection frequency
$\alpha|U_G\left(z\right)-c|$~\citep{Belcher1993,Belcher1994,OceanwavesBook2004}.
\paragraph*{The visco-elastic model:}  This is a stationary turbulence model wherein the perturbation
Reynolds stresses are assumed to be proportional to the
perturbation-induced turbulent kinetic energy (TKE). Such models
have been used by \citet{Townsend1972,Townsend1979}, and by \citet{Miles2001}.  The model described here fits
into the framework of the latter paper, with slight modifications:
the base-state quantites are computed according to the formalism in
Sec.~\ref{sec:flat},  and the dissipation rate is taken to be linear
in $\delta k$.  This last assumption is not necessary, but is
convenient from a mathematical point of view, since in this form,
the dissipation rate is well-behaved at the interface, unlike the
other models~\citep{Townsend1972,Townsend1979,Miles2001}.
The perturbation-induced TKE satisfies a balance law wherein the
advection of the kinetic energy is  balanced by production,
dissipation, and diffusive effects.  The production of TKE is
proportional to the stresses $\delta\rtensor_{12}=-\delta\tau$ and
$\delta\rtensor=-\delta\sigma$, the dissipation term is assumed to
be linear in $\delta k$, while in the dynamically-important
near-interfacial region, the molecular viscosity is expected to
dominate over the turbulent viscosity (whose effects are anyway
always negligible~\citep{Townsend1972,Townsend1979,Miles2001}).
Thus, we have the following balance law:
\begin{subequations}
\begin{multline}
\left[i\alpha\left(U_G-c\right)+\frac{Re_*^2}{Re_0}\right]\delta k
\\=\frac{1}{Re_0}\left(\Diff^2-\alpha^2\right)\delta
k-\delta\rtensor_{12}\frac{\mathd U_G}{\mathd z}-\rtensor_{12}\left(\Diff^2+\alpha^2\right)\phi-i\alpha\rtensor\Diff\phi_G+i\alpha\frac{\mathd
k_0}{\mathd z}\phi_G,
\end{multline}
To close the system, the visco-elastic hypothesis is invoked:
\begin{equation}
\delta\rtensor_{12}
-\frac{\rtensor_{12}^G}{k_G}\delta{k}=0,
\label{eq:rs_interp1}
\end{equation}
\begin{equation}
\delta\rtensor-\frac{\rtensor^G}{k_G}\delta{k}=0.
\label{eq:rs_interp2}%
\end{equation}%
The base-state stress $\tau=-\rtensor_{12}^G$ is modelled as in
Sec.~\ref{sec:flat}, and the stress $\rtensor$ is set equal to
$k_0$, the base-state kinetic energy, consistent with the  DNS
results of \citet{Spalart1988}.  Finally, the base-state
turbulent kinetic energy is modelled by the equation
\begin{equation}
\tilde{k}_0=\frac{k_0}{\rho_GU_0^2}=\frac{1}{C^2}\frac{Re_*^2}{Re_0^2}\psi\left(\tilde{z}\right)\psi\left(1-\tilde{z}\right),
\label{eq:k_model}%
\end{equation}%
\label{subeq:k_model}%
\end{subequations}%
where $C$ is another constant, here taken to be $0.55$, which is the
value appropriate for the logarithmic  region of the mean velocity
in a boundary layer.  This form is chosen because it accurately
models the viscous and logarithmic zones in single-phase flow.

\paragraph*{The zero-equation model:}  We shall compare the results of the
visco-elastic study with the results for an eddy-viscosity model.  In this
formalism, the normal stresses are set to zero, and the shear stress is modelled
as
\begin{subequations}
\begin{equation}
\delta\tau=\mu_T\left(\Diff^2+\alpha^2\right)\phi
\label{eq:evm_eqn}
\end{equation}
where
\begin{equation}
\mu_T=\frac{\kappa Re_{*}}{\sqrt{|R|}Re_0}G\left(\tilde{z}\right)\psi_{\mathrm{i}}\left(\tilde{z}\right)\psi_{\mathrm{w}}\left(1-\tilde{z}\right),
\end{equation}%
\label{subeq:evm}%
\end{subequations}%
as in Sec.~\ref{sec:flat}.  This is a rather basic model of the eddy
viscosity, which does not take account of perturbations in the eddy viscosity
function itself.  In particular, small changes in pressure will modifiy the
Van Driest coefficient $A_{\mathrm{i}}$, thus changing the viscous-sublayer
thickness~\citep{Zilker1976}.  This contribution is expected to be negligible in our small-amplitude
analysis, and
our finding that the details of the instability depend on conditions at the
interface, and that a small modification in the extent of the viscous sublayer
has little effect, strengthens this contention.
Moreover, Belcher and co-workers have used a similar model to Eqs.~\eqref{subeq:evm},
where it was thought to capture the physics of the equilibrium turbulence.

 To close Eqs.~\eqref{subeq:os}, they are matched
 across the interface $z=0$, where we have the following conditions:
\begin{subequations}
\begin{align}
\phi_{L}&=\phi_{G},\\
\Diff\phi_{L}&=\Diff\phi_{G}+\frac{\phi_1}{c-U_L}\left(\frac{\mathd U_{G}}{\mathd{z}}-\frac{\mathd
U_{L}}{\mathd{z}}\right),\\
%
%
%
%
%
m\left(\Diff^2+\alpha^2\right)\phi_L&=\left(\Diff^2+\alpha^2\right)\phi_{G},
\end{align}
\vskip -0.2in
\begin{multline}
m\left(\Diff^3\phi_L-3\alpha^2\Diff\phi_L\right)+\mathrm{i}\alpha rRe\left(c-U_L\right)\Diff\phi_L+i\alpha
rRe  \frac{\mathd U_L}{\mathd z}\phi_L-
\frac{\mathrm{i}\alpha r Re}{c-U_L}\left(Fr+\alpha^2 S\right)\phi_L
\\
=\left(\Diff^3\phi_{G}-3\alpha^2\Diff\phi_{G}\right)+\mathrm{i}\alpha  Re\left(c-U_L\right)\Diff\phi_{G}+i\alpha Re\frac{\mathd U_{G}}{\mathd z}\phi_{G}.
\label{eq:ic_normal}%
\end{multline}%
\label{eq:ic_interface}%
\end{subequations}%
The no-slip conditions are applied at $z=-d_L$ and $z=h$:
\[
\phi_L\left(-d_L\right)=\Diff\phi_L\left(-d_L\right)=\phi_G\left(h\right)=\Diff\phi_G\left(h\right)=0,
\]
and, where applicable, the following conditions are applied to the turbulent
kinetic energy:
\begin{equation}
\delta k=0\qquad z=0,\,\text{and for } z=h.
\end{equation}
Finally, at the top of the gas domain $z=h$, we have the no-slip conditions
\[
\phi_G\left(h\right)=\Diff\phi_G\left(h\right)=0.
\]
The OS equations~\eqref{subeq:os} and the turbulence model reduce to
an eigenvalue problem in the eigenvalue $c$.  This is solved
numerically according to a standard method, described and validated
elsewhere by the current authors in the context of  absolute and
convective instabilities in laminar two-phase
flows~\citep{Valluri2010}.

\subsection{Preliminary numerical studies}
\label{subsec:prelim_numerics}

To compare the turbulence models, we carry out a stability analysis based on the values in Tab.~\ref{tab:values0},
with $d_L=2.5\,\mathrm{mm}$ and $\delta=0.05$.
The inverse Froude and Weber numbers are computed as
\begin{eqnarray}
Fr&=&\frac{g h}{\left(\mu_G/\rho_G h\right)^2}\frac{r-1}{Re_0^2}=\left(3.7809\times10^6\right)\frac{r-1}{Re_0^2},\nonumber\\
S&=&\frac{\sigma}{\mu_G^2/\rho_G h}\frac{1}{Re_0^2}=\frac{1.1420\times10^7}{Re_0^2}.
\label{eq:fr_values}
\end{eqnarray}
%
%
%
%
%
We select a Reynolds number that produces substantial shear in the
liquid, but is such that the liquid remains laminar.   Thus, $Re_P=5\times10^5$,
$Re_0=1000$, and $Re_L=\tau_\mathrm{i}d_L/\mu_L\approx460$.  According to Fig.~\ref{fig:c_est}, these values
of $\left(Re_0,\delta,d_L\right)$ will produce the slow-wave instability.
We verify \textit{a posteriori} that the liquid remains laminar, in
the sense that the  TS mode associated with the liquid has a
negative growth rate.  This growth rate can become positive,
however, when mode competition takes place; this is the subject of
Sec.~\ref{subsec:internal_tau}.
Although our numerical method has been validated elsewhere~\citep{Valluri2010,ONaraigh2010}
\begin{figure}[htb]
\centering\noindent
\subfigure[]{\includegraphics[width=0.3\textwidth]{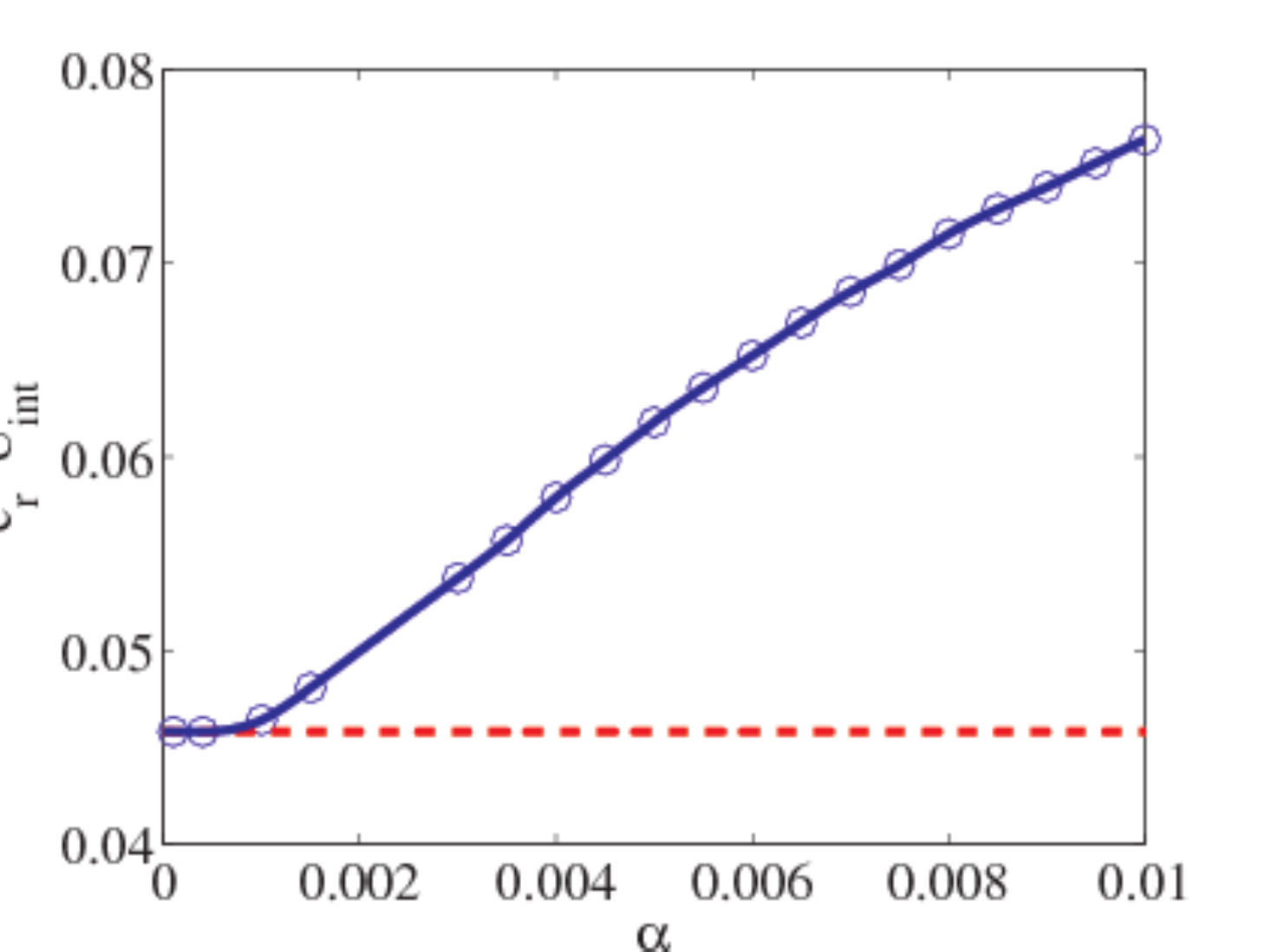}}
\subfigure[]{\includegraphics[width=0.305\textwidth]{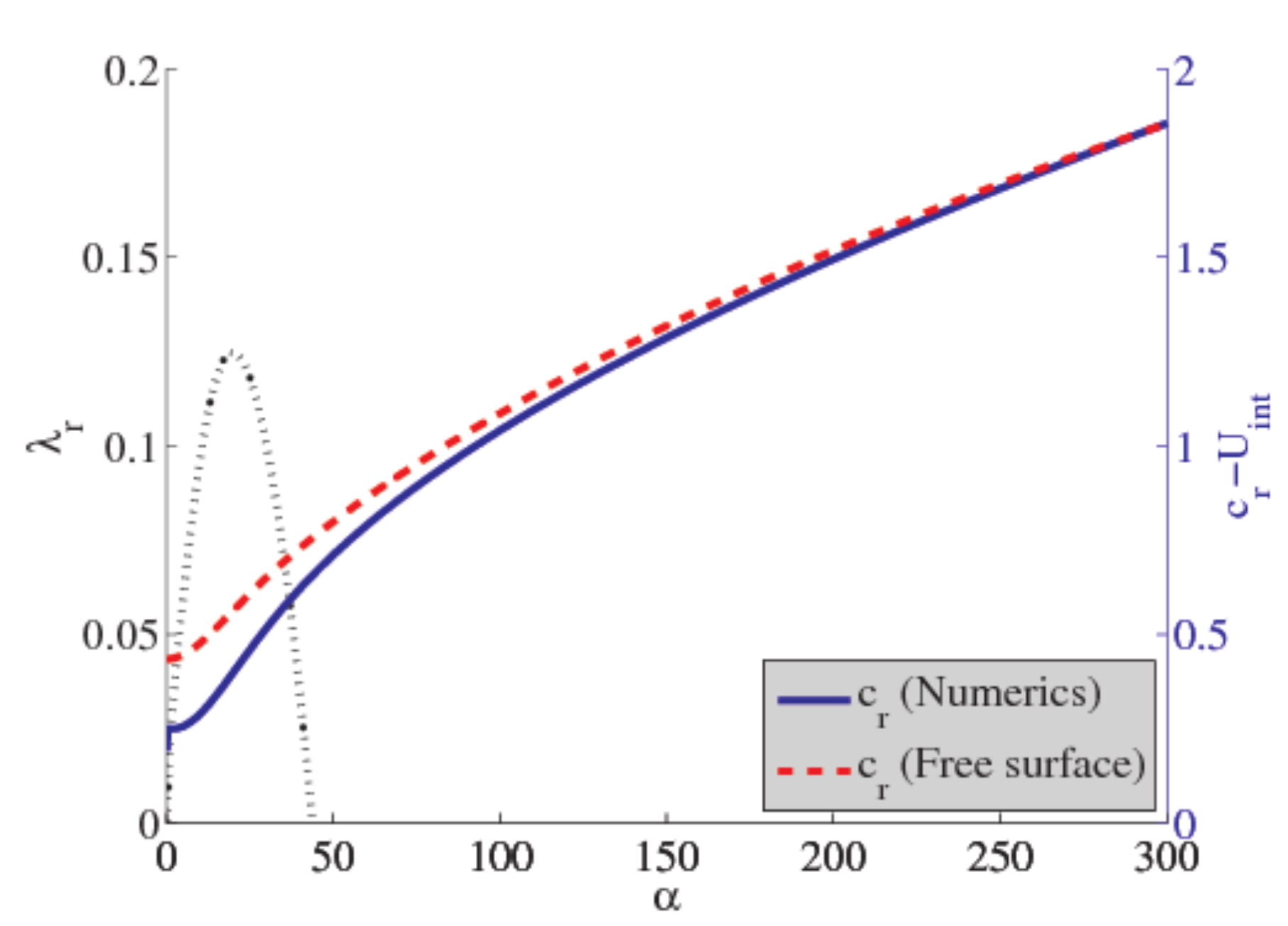}}
\subfigure[]{\includegraphics[width=0.27\textwidth]{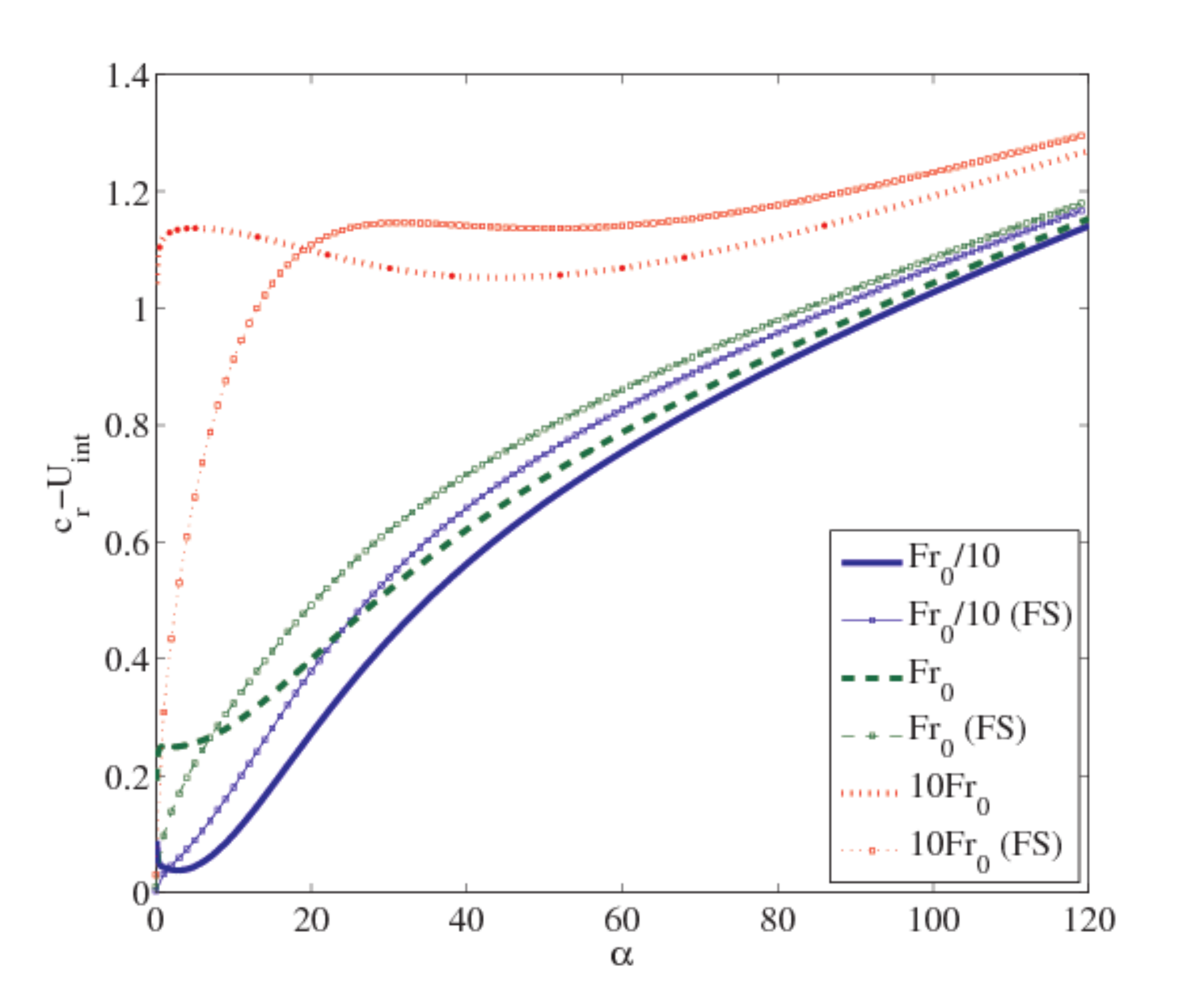}}
\caption{Test bed for the numerical solver.  The parameters are given at the start of the section~(\ref{subsec:prelim_numerics}).  (a) Long-wave analysis of the
OS model~\eqref{subeq:os}, without the PTS.  The numerical wave speed at
$\alpha=0$ agrees with the analytical formula, obtained by solving the OS
model analytically at $\alpha=0$.  At lowest order, the growth rate is zero.
(b)  Short-wave analysis of the OS model.  The numerical wave speed agrees
with Eq.~\eqref{eq:c_grav} for free-surface waves, as $\alpha\rightarrow\infty$.  The maximum discrepancy between the free-surface estimate and the true value of the wave speed is in the unstable part of the spectrum.
The results in (b) hold over a wide parameter range: in (c) we compare the formula  with the numerical calculations for a range of $Fr$-values.
The agreement between the calculations confirms the correctness of our numerical
technique.}
\label{fig:test_numerics}
\end{figure}
we provide a further quick by comparing
the zeroth-order long-wave analytical solution to the OS equation with our
numerical method.  At lowest order in $\alpha$, the OS equation without the
PTS reduces to the following matrix problem:
\begin{equation*}
\left(
\begin{array}{cccc}
\delta^2&\delta^3&-1&1\\
\delta^2\frac{Re_*^2}{Re_0}\left(1-\frac{1}{m}\right)&\delta^3\frac{Re_*^2}{Re_0}\left(1-\frac{1}{m}\right)&0&0\\
2m&6m\delta&-2&6\\
0&6m&0&-6\end{array}\right)\bm{x}=
c_0\left(\begin{array}{cccc}
0&0&0&0\\
2\delta&3\delta^2&2&-3\\
0&0&0&0\\
0&0&0&0\end{array}\right)\bm{x},
\end{equation*}
with solution
\begin{equation}
c_0=U_{\mathrm{int}}+\frac{2\delta^2Re_*^2\left(m-\delta-1+m\delta\right)}{Re_0\left(\delta^4+4\delta^3m+6\delta^2m+4m\delta+m^2\right)}.
\label{eq:c0_anal}
\end{equation}
Unlike in studies of laminar flow~\citep{Yiantsios1988,Sahu2007}, the first-order term is not
available explicitly, since the complicated (i.e. non-polynomial) form for the base state precludes an explicit solution to the first-order streamfunction.
 Nevertheless, the formula~\eqref{eq:c0_anal} serves as an adequate test
 for our numerical scheme, as demonstrated in Fig.~\ref{fig:test_numerics}~(a).
 One further test is to check that the wave speed $c_{\mathrm{r}}$ agrees
 with the free-surface formula~\eqref{eq:c_grav} in the limit of large $\alpha$.
  This is shown in Fig.~\ref{fig:test_numerics}~(b).  This figure also vindicates
  the use of the free-surface wave-speed formula in our estimate for the
  parameter range where the critical-layer mechanism is important.

\begin{figure}[htb]
\centering\noindent
\subfigure[]{
\includegraphics[width=0.35\textwidth]{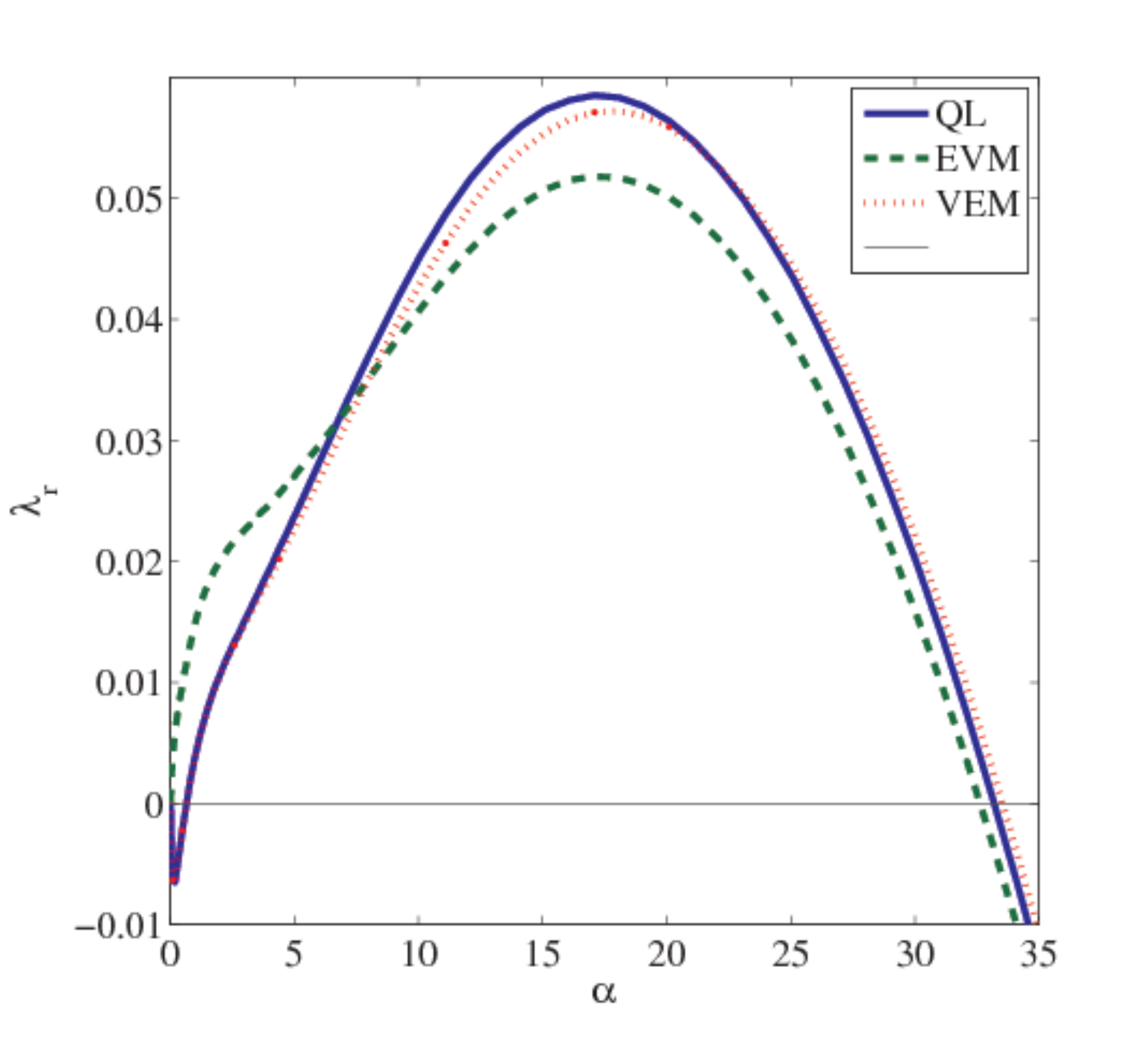}}
\subfigure[]{
\includegraphics[width=0.35\textwidth]{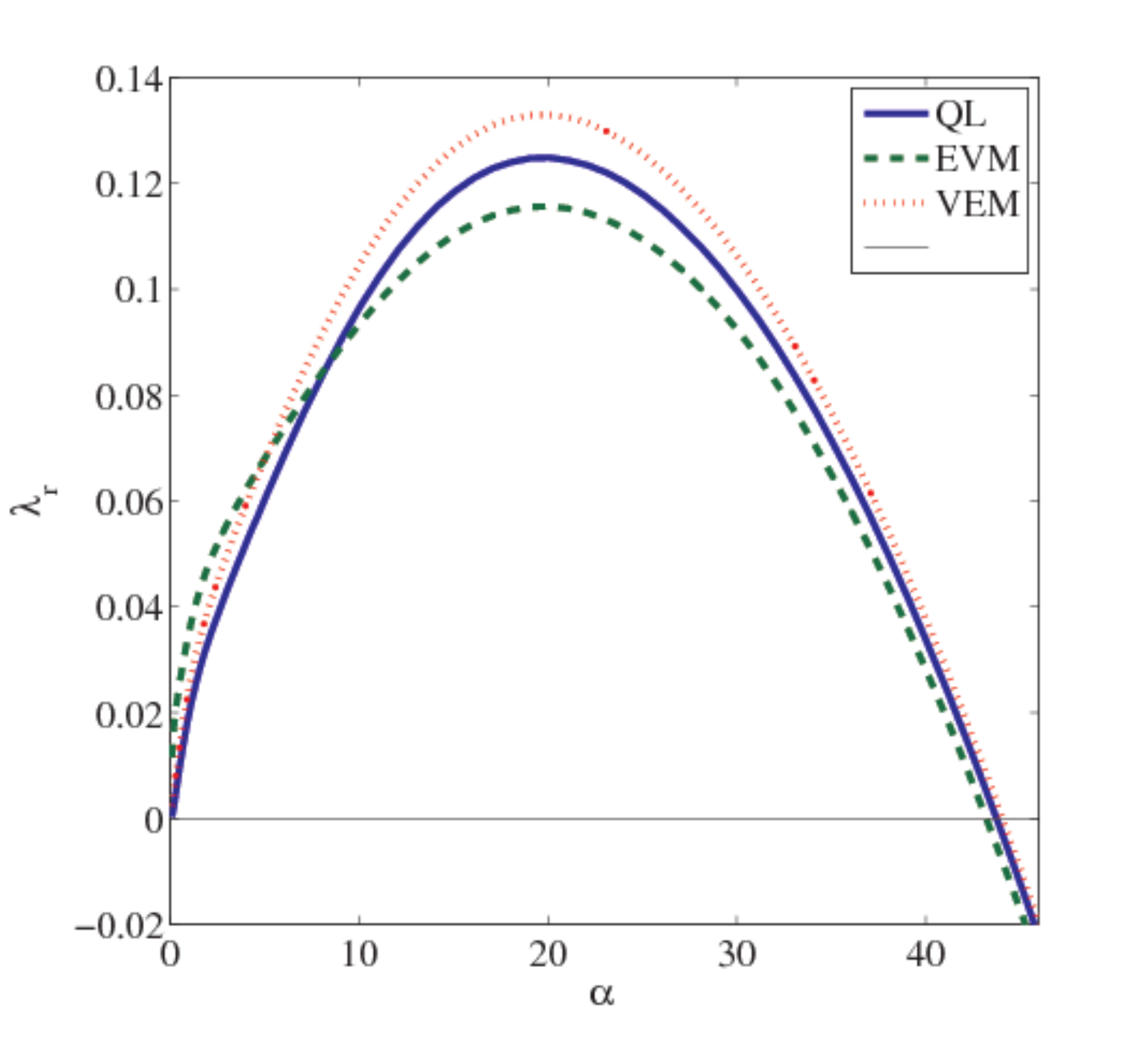}}
\subfigure[]{
\includegraphics[width=0.35\textwidth]{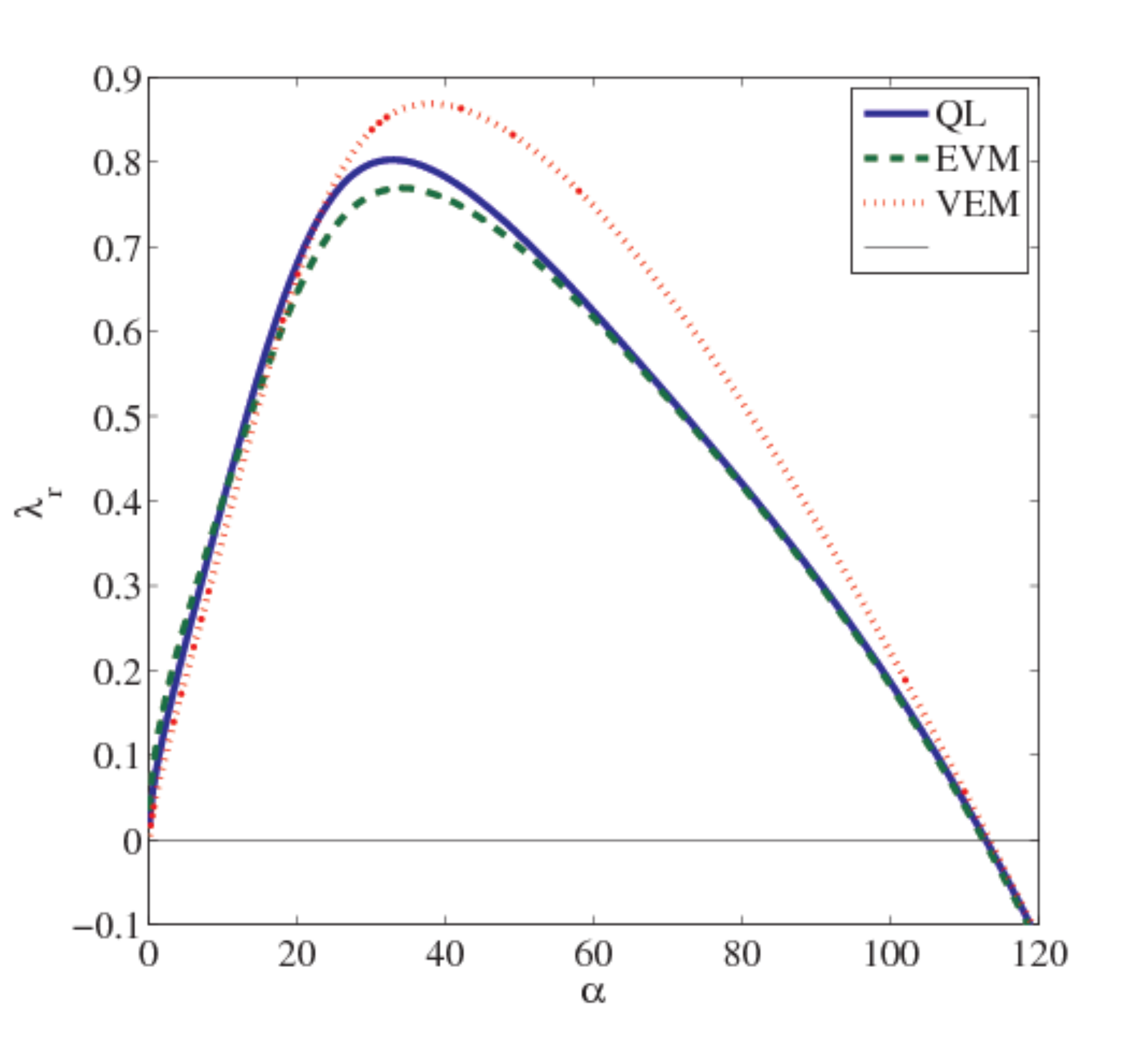}}
\subfigure[]{
\includegraphics[width=0.35\textwidth]{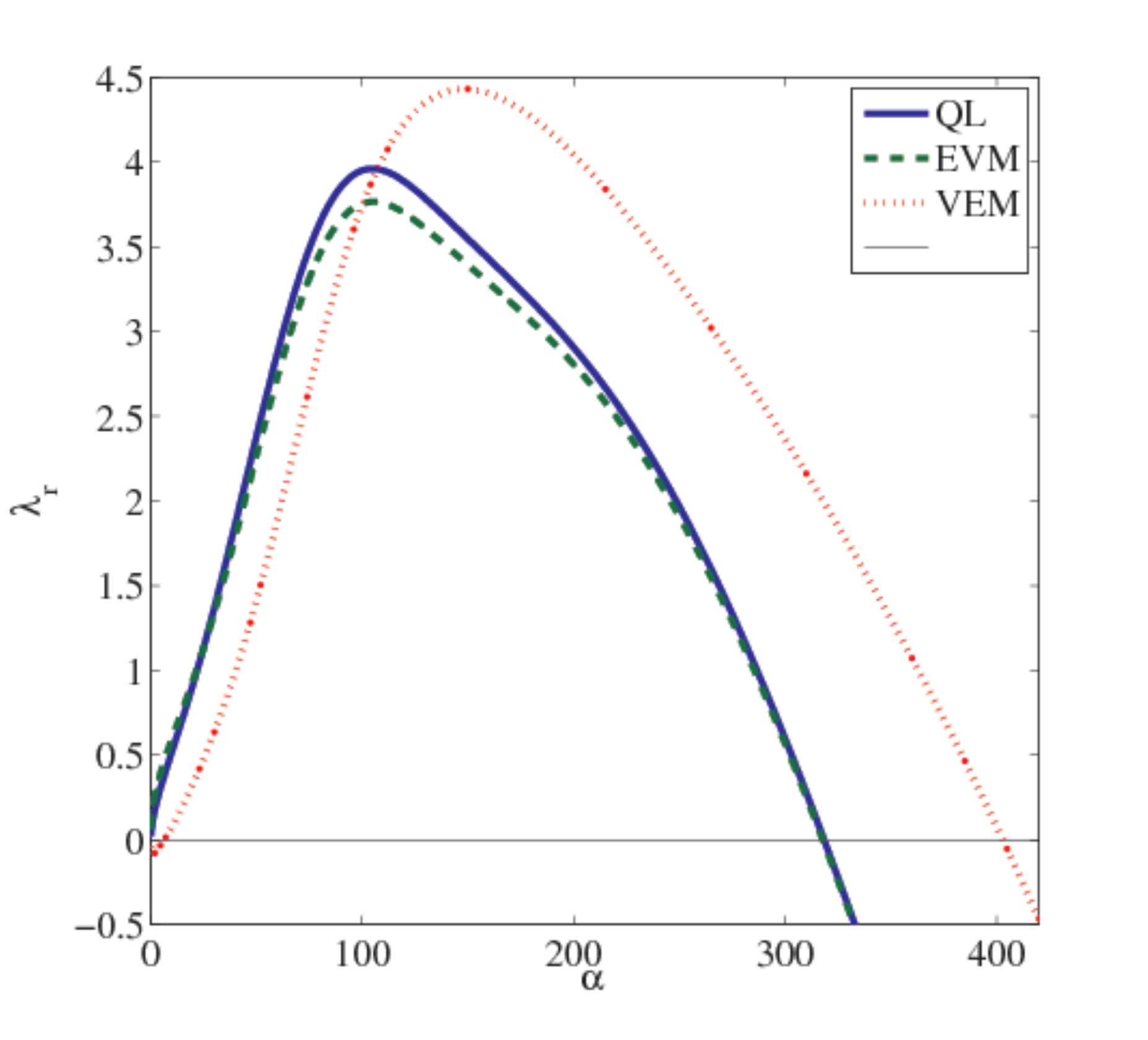}}
\caption{Comparison between the models for the PTS.  Solid line: quasi-laminar model; dashes: eddy-viscosity model; dots: `visco-elastic' model.  Shown is the parametric dependence of the growth rate on the Reynolds number for (a) $Re_0=875$; (b) $Re_0=1000$; (c) $Re_0=2000$; (d) $Re_0= 5000$.   We have set
$m=55$, $r=1000$, and $\delta=0.05$.  In each case, the difference of the maximum growth rate between the models is less than, or equal to 10\%, which justifies the choice of the quasi-laminar model throughout the rest of this work.}
\label{fig:comparemodels}
\end{figure}

Having validated our numerical technique, we turn to a full-spectrum calculation.
 In Fig.~\ref{fig:comparemodels}, we obtain the growth rate
using three models: the basic OS
 equation without the PTS (the so-called \textit{quasi-laminar} approach), the visco-elastic
model~\eqref{subeq:k_model}, and the eddy-viscosity
model~\eqref{subeq:evm}.  Over a large range of Reynolds numbers
($Re_0=500$--$5000$, $Re_{U_{\mathrm{max}}}=\rho_G
U_{\mathrm{max}}h/\mu_G=10^3$--$10^5$), the growth rates for the
different models differ only quantitatively.  In particular, the
differences between the quasi-laminar calculation  and the
eddy-viscosity calculation are small: the shift in the maximum
growth rate upon including the PTS is less than 10\% in the cases
considered here, while the cutoff wavenumbers are virtually
unchanged.  The differences between the quasi-laminar calculation
and the visco-elastic calculation are slightly larger.  In
particular, the cutoff wavenumber is shifted to a higher value in
the $Re_0=5000$ case (Fig.~\ref{fig:comparemodels}~(d)).
Nevertheless, the shift in the maximum growth rate upon including the visco-elastic terms is not more than 10\% in the cases considered in Fig.~\ref{fig:comparemodels}. %
The minor discrepancy in behaviour between the visco-elastic model
and the other two models  is due to the lack of understanding in
modelling the kinetic-energy dissipation function, here assigned the
simple linear form $\left(Re_*^2/Re_0\right)\delta k$.  Accurate
modelling of this term will be the subject of future work.
%
%
%
%
%

Our conclusion from the small differences evinced by these
comparisons is that we are justified  in considering the
quasi-laminar approximation for the rest of this work.  Furthermore,
we can provide a physical justification for the smallness of the
contribution made by the PTS in the eddy-viscosity model.
%
%
We use the analogy
between the Reynolds-averaged Navier--Stokes equations and the equations
for a laminar non-Newtonian fluid~\citep{Zou1997}.  In the latter case, a linear stability
analysis has been performed on a two-phase stratified flow, where the bottom
layer is a Bingham fluid~\citep{Sahu2007}.  In that case, the authors found a small difference
($\sim10\%$)
between the results of the Orr--Sommerfeld analysis, depending on whether
the perturbation non-Newtonian stresses were included.  The difference in
the stability results was driven by the presence of extra terms in the perturbation
equations
for the bulk flow, and by the existence of extra terms in the interfacial
conditions, which enhance the viscosity contrast.  In our case, the additional
terms in the bulk equations scale as $\kappa Re_*/Re_0$, which for thin layers is approximately $\kappa/\sqrt{2}$, and
thus has a small effect.  Moreover, in our case, the additional interfacial
terms are turbulent in nature, and are thus damped to zero in the viscous
sublayer, and vanish at $z=0$.  Hence, this second contribution to the modified
growth rate is also small.  Thus, in the case of equilibrium turbulence
considered here, the effects of turbulence are felt almost entirely through
the choice of base state.  The only possibility for the effects of turbulence
to enter through the PTS is when the critical-layer instability is present,
since then rapid distortion effects are possible, and an interaction
occurs between the turbulence and the critical layer.   This will be the subject of future work.
In the present paper however, we consider slow waves, and thus the effects of turbulence need be considered only in the base state.

\begin{figure}[htb]
\centering\noindent
\subfigure[]{
\includegraphics[width=0.45\textwidth]{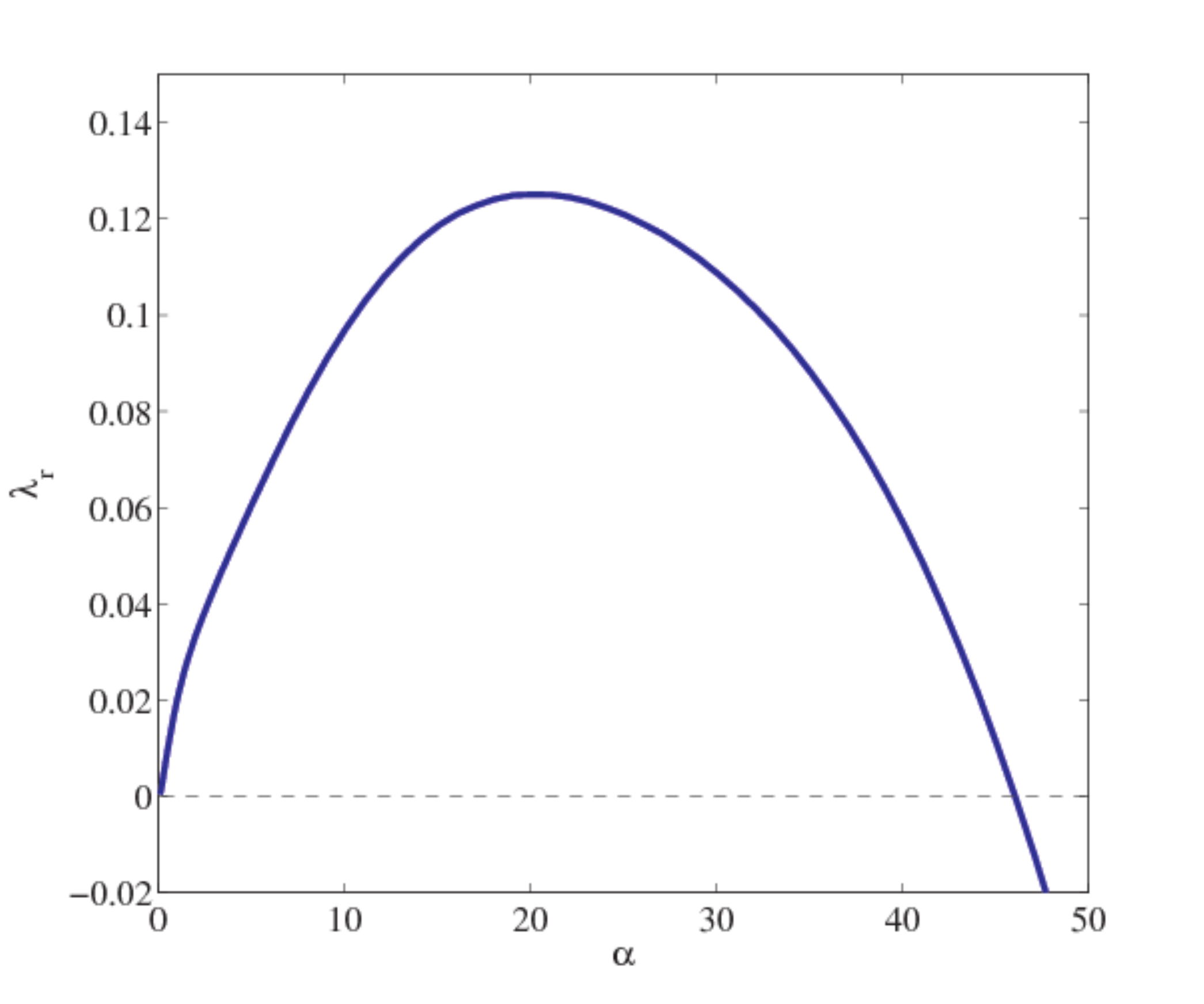}}
\subfigure[]{
\includegraphics[width=0.45\textwidth]{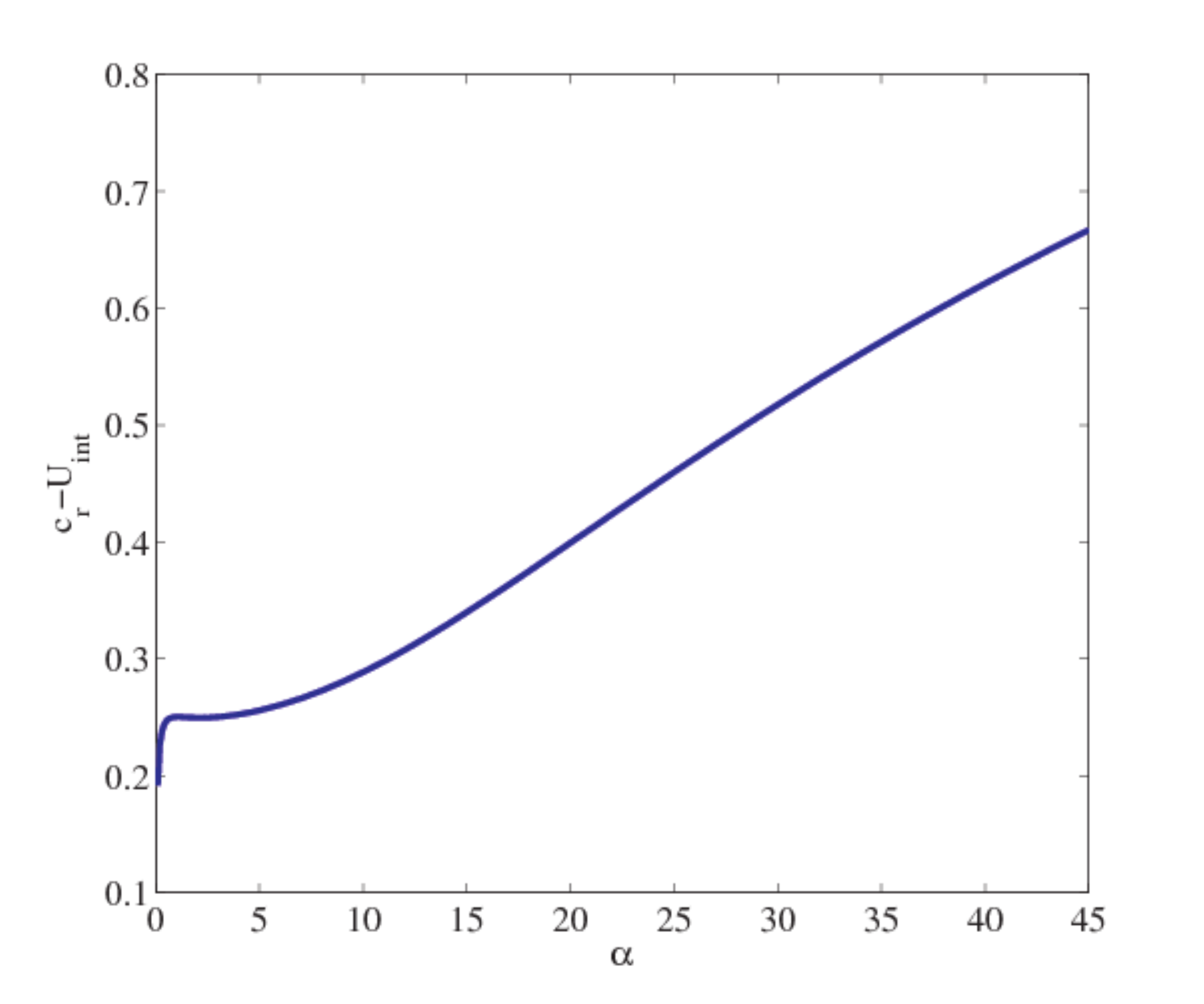}}
\caption{Growth rate and wave speed for pressure-driven channel flow, with
turbulent base state.  We have set $Re_0=1000$ (hence $Re_L\approx460$),
$m=55$, $r=1000$, and $\delta=0.05$.  The most dangerous mode is at $\alpha\approx20$,
which equates to a wavelength $\ell\approx1.0 d_L$.}
\label{fig:growth_rate1}
\end{figure}

Next, we present the growth rate only for the quasi-laminar model in Fig.~\eqref{fig:growth_rate1},
 where $m=55$, $r=1000$, $\delta=0.05$, and $Re_0=1000$ ($Re_L\approx460$).
Maximum growth
occurs at a wavenumber $\alpha\approx20$, that is, for a wavelength
$\ell/d_L=2\pi/\left(20\delta\right)\approx6$.  The wave speed $c_r/U_0$
is less than unity
for the dynamically important unstable waves.
Thus, according to the analysis of Sec.~\ref{sec:flat}, neither the critical-layer
mechanism, nor rapid-distortion effects, will be relevant.  Note finally
the convexity of the small-$\alpha$ part of the growth rate in Fig.~\ref{fig:growth_rate1}.
The corresponding part of the growth rate is concave for shear-driven flow
(see \citet{Miesen1995}).
Having validated and compared our models, we identify
 the source of the instability and investigate its dependence on the various
 parameters in the system.

\section{Linear stability analysis}
\label{sec:linear_stability}

\begin{figure}
  \begin{center}
\subfigure[]{
\includegraphics[width=0.45\textwidth]{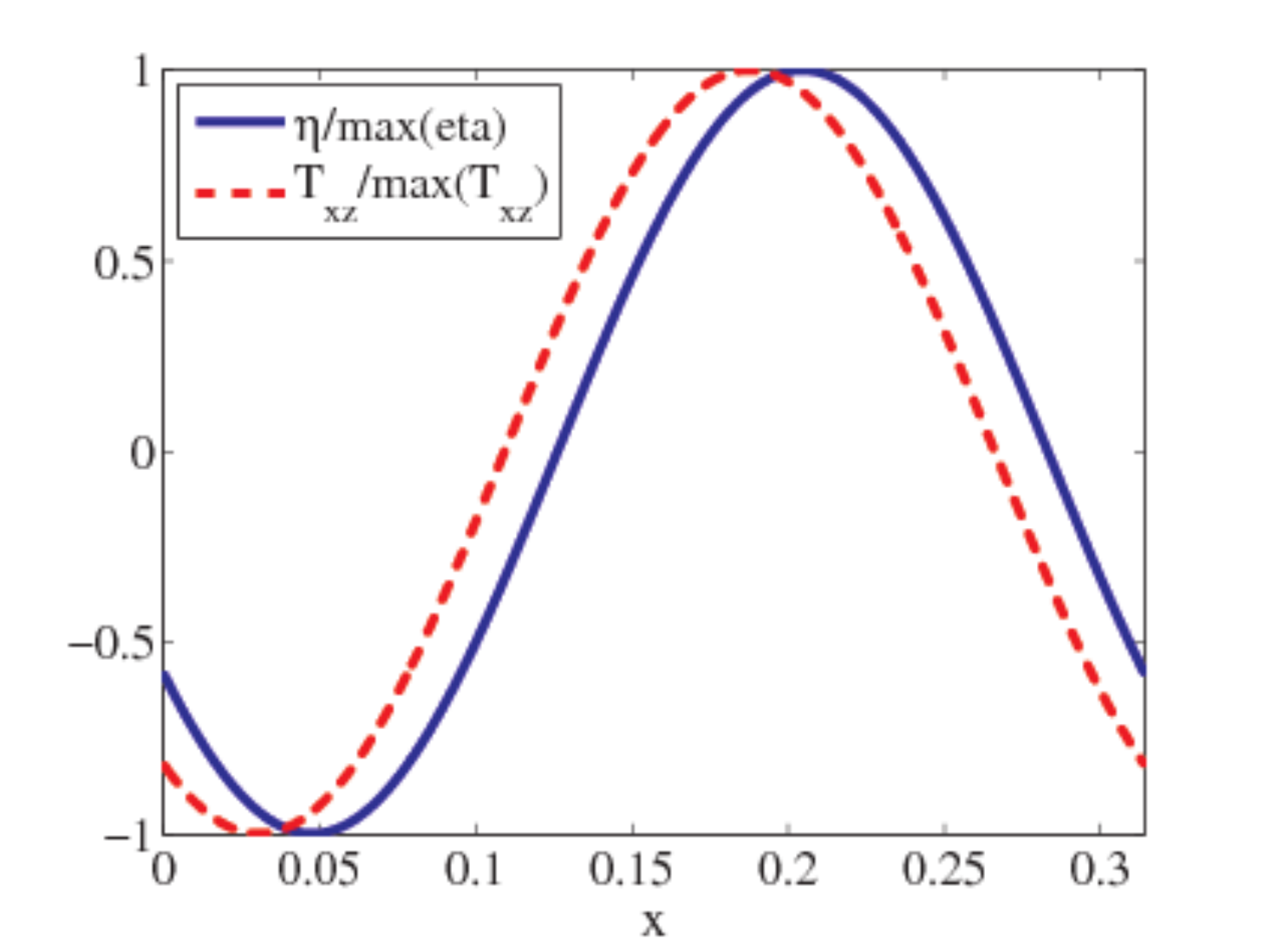}}
\subfigure[]{
\includegraphics[width=0.45\textwidth]{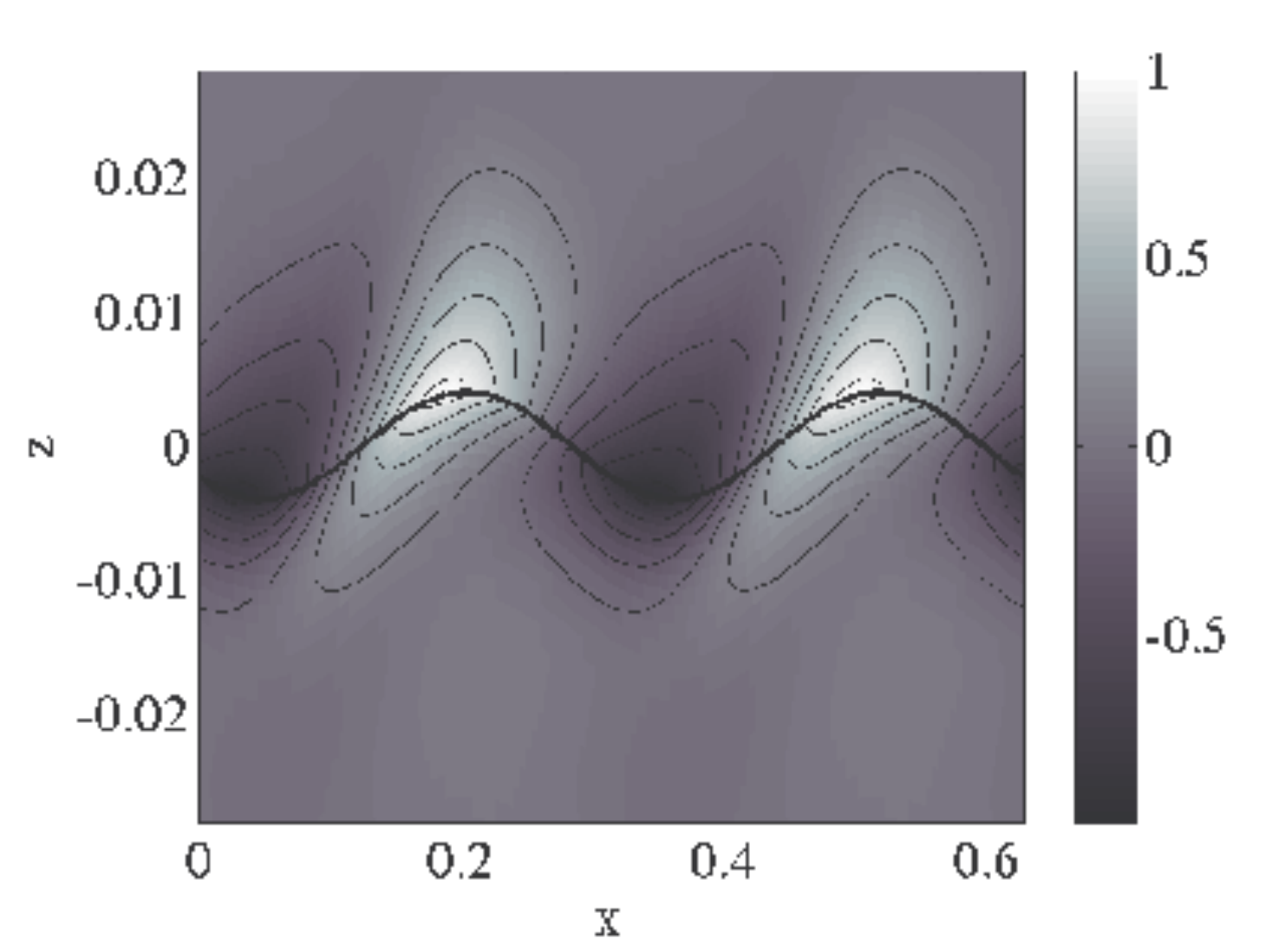}}
\end{center}
\caption{(\textit{a})  The phase shift between the viscous shear stress at
the interface, $\Ttensor_{xz}\left(x,z=0\right)$ and the interface shape
$\eta\left(x\right)$ for the most dangerous mode $\alpha=20$.
The shift
is small $\delta\phi=0.22\times\left(\pi/2\right)$.
Thus the tangential term in the energy budget is positive (destabilizing),
as required by~\eqref{eq:tan_term}; (\textit{b}) the behaviour of the viscous
shear stress across the interface.
}
\label{fig:phases}
\end{figure}

In this section we present detailed results of the Orr--Sommerfeld (OS) analysis, based on the reference values of the inverse Froude and Weber numbers already described in Eq.~\eqref{eq:fr_values}.
 First of all, by performing an energy-decomposition, we confirm that the instability at work is the viscosity-contrast
 mechanism.
 This decomposition or budget is obtained from the RANS equations, and was
 introduced by \citet{Boomkamp1996}:
\begin{subequations}
\begin{equation}
r_j\left(\frac{\partial}{\partial t}\delta\bm{u}_j+\bm{U}_j^{(0)}\cdot\nabla\delta\bm{u}_j+\delta\bm{u}_j\cdot\nabla\bm{U}_j^{(0)}\right)=\nabla\cdot\delta\Ttensor^{(j)}-r_j\nabla\cdot\delta\rtensor^{(j)},
\end{equation}
\begin{equation}
\delta\Ttensor=\left(\begin{array}{cc}-\delta p + \mu\partial_x\delta u&\mu\left(\partial_x\delta
w+\partial_z\delta u\right)\\\mu\left(\partial_x\delta
w+\partial_z\delta u\right)&-\delta p + \mu\partial_z\delta w
\end{array}\right),\qquad
%
%
\delta\rtensor=\left(\begin{array}{cc}-\delta\rtensor_{11}+\delta\rtensor_{22}&-\delta\rtensor_{12}\\-\delta\rtensor_{12}&0
\end{array}\right),
\end{equation}
\begin{equation}
\nabla\cdot\delta\bm{u}_j=0,
\end{equation}%
which we multiply by the velocity $\delta\bm{u}_j$ and integrate over space.  We obtain the following
balance equation:
\begin{equation}
\sum_{j=L,G}{KIN}_j=\sum_{j=L,G}{REY}_j+\sum_{j=L,G}{DISS}_j+\sum_{j=L,G}{TURB}_j+{INT},
\end{equation}
where
\begin{eqnarray}
{KIN}_j&=&\tfrac{1}{2}\frac{\mathd}{\mathd t}\int dx \int dz\, r_j\delta\bm{u}_j^2,\\
{REY}_j&=&-r_j\int dx \int dz\, \delta u_j\delta w_j\frac{\mathd U_j}{\mathd z},\\
{DISS}_j&=&-\frac{m_j}{Re}\int dx \int dz\,\left[2\left(\frac{\partial{}}{\partial{x}}\delta
u_j\right)^2
+\left(\frac{\partial}{\partial z}\delta u_j+\frac{\partial }{\partial
x}\delta w_j\right)^2+2\left(\frac{\partial }{\partial z}\delta w_j\right)^2\right],\\
{TURB}_j&=&
\delta_{j,G}\bigg\{r\int
dx \int dz\,\left[\delta\rtensor\frac{\partial}{\partial x}\delta u+\delta\rtensor_{12}\left(\frac{\partial}{\partial
z}\delta u+\frac{\partial}{\partial x}\delta w\right)\right]\bigg\}.
\end{eqnarray}%
\label{eq:eb}%
\end{subequations}%
Additionally,
\[
{INT}=\int dx \,\left[\delta u_L \delta\Ttensor_{L,zx}+\delta w_L\Ttensor_{L,zz}\right]_{z=0}
-\int dx \,\left[\delta u_G \delta\Ttensor_{G,zx}+w_G\delta\Ttensor_{G,zz}\right]_{z=0},
\]
which is decomposed into normal and tangential contributions,
\[
{INT}={NOR}+{TAN},
\]
where
\[
{NOR}=\int dx\,\left[\delta w_L\delta\Ttensor_{L,zz}-\delta w_G\delta\Ttensor_{G,zz}\right]_{z=0},
\]
and
\[
{TAN}=\int dx \,\left[\delta u_L\delta\Ttensor_{L,zx}-\delta u_G \delta\Ttensor_{G,zx}\right]_{z=0}.
\]
For the quasi-laminar model under consideration throughout this paper, the term $TURB$ is set to zero.
 We perform the energy decomposition for the most dangerous
 mode in Fig.~\ref{fig:growth_rate1} ($\alpha=20$),
 and demonstrate the result in Tab.~\ref{tab:energy_budget0}.
\begin{table}[htb]
\centering\noindent
\begin{tabular}{|c|c|c|c|c|c|c|c|}
\hline
$KIN_G$&$KIN_L$&$REY_L$&$REY_G$&$DISS_L$&$DISS_G$&$NOR$&$TAN$\\
\hline
0.18&0.82&2.34&-11.90&-4.28&-57.42&-2.73&74.99\\
\hline
\end{tabular}
\caption{Energy budget for the most dangerous mode $\alpha=20$ at $Re_0=1000$
(hence $Re_L\approx460$), $m=55$, $r=1000$, and $\delta=0.05$.  It is the
$TAN$ term that gives rise to a net positive energy, and thus destabilizes
the interface.}
\label{tab:energy_budget0}
\end{table}
The results in this table indicate that it is the $TAN$ term that is the main source of the instability.
 This is the work done by tangential stresses on the interface, and can be
 written as
\[
{TAN}=\int_0^{\ell}dx\left[\left(\delta u_L-\delta u_{G}\right)\delta\Ttensor_{xz}\right]_{z=0},\qquad
\ell=2\pi/\alpha.
\]
Since the kinematic condition implies the following jump condition on $\delta
u_L-\delta u_{G}$,
\[
\delta u_{L}-\delta u_{G}=\eta\left(U_{G,1}'-U_{L}\right)=\eta U_L'\left(m-1\right),\qquad\text{on
}z=0,
\]
the tangential term is non-zero because $m\neq1$:
\begin{equation}
{TAN}=\left(m-1\right)\frac{Re_*^2}{Re_0}\int_0^{\ell}dx\,
\eta\left(x\right)\delta\Ttensor_{xz}\left(x,z=0\right).
\label{eq:tan_term}
\end{equation}
Thus, the viscosity contrast $m>1$ induces instability, provided
the interfacial shape $\eta\left(x\right)$ and the disturbance stress $\Ttensor_{xz}\left(x,z=0\right)$
possess a phase shift in the range $\left[-\tfrac{\pi}{2},\tfrac{\pi}{2}\right]$ (see Fig.~\ref{fig:phases}).

So far we have chosen a set of parameters that are comparable
in magnitude to those describing an air-water system under particular conditions.
 However, we wish to quantify
the stability properties of the system in full generality (and in particular,
to delineate the boundary between slow and fast waves), and we therefore
investigate the implications of varying the pressure gradient, the density
contrast, and the Froude and Weber numbers.

\subsection{The interfacial mode}
\label{subsec:results_TSS}

\begin{figure}[htb]
\centering\noindent
\subfigure[]{\includegraphics[width=0.3\textwidth]{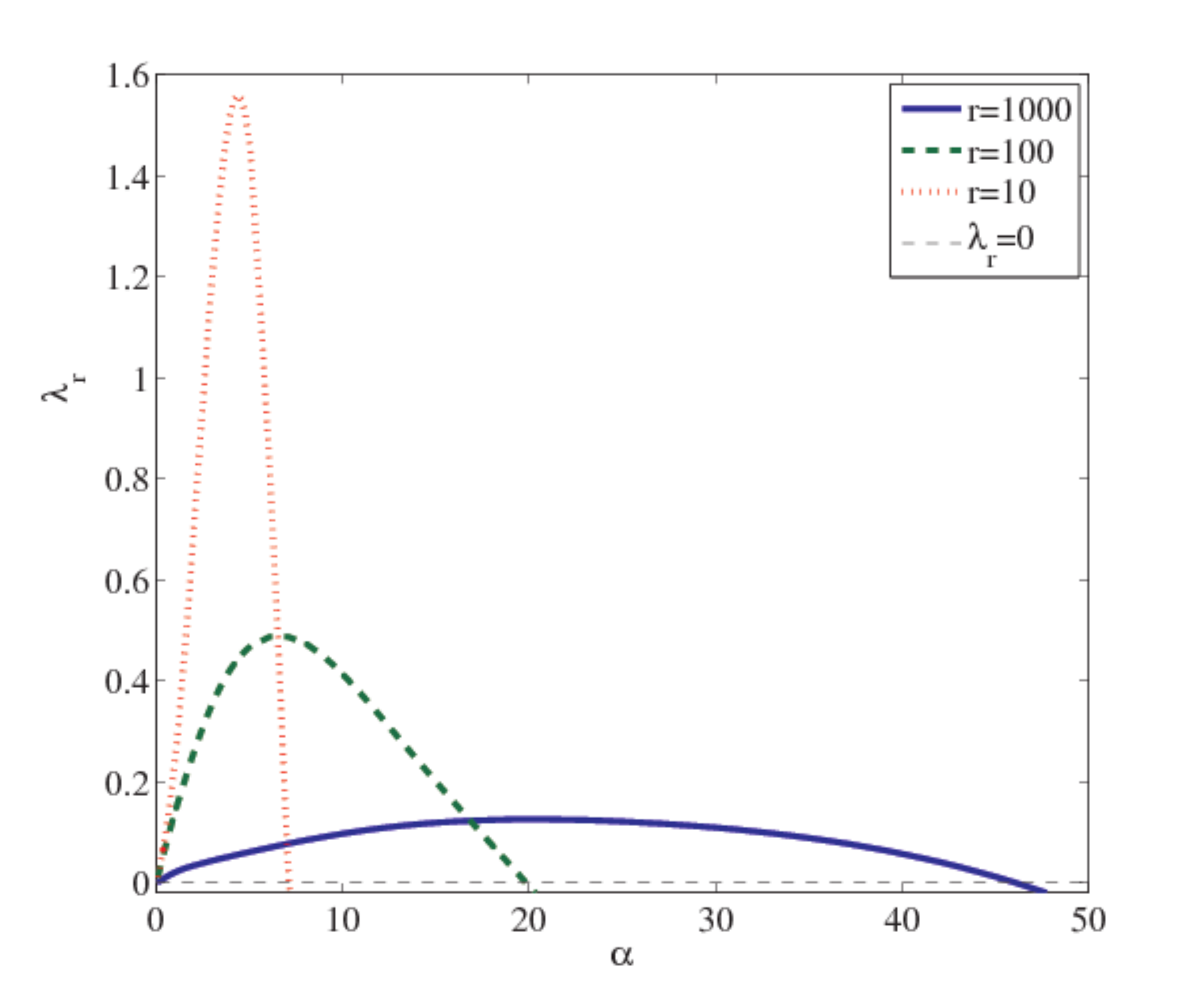}}
\subfigure[]{\includegraphics[width=0.3\textwidth]{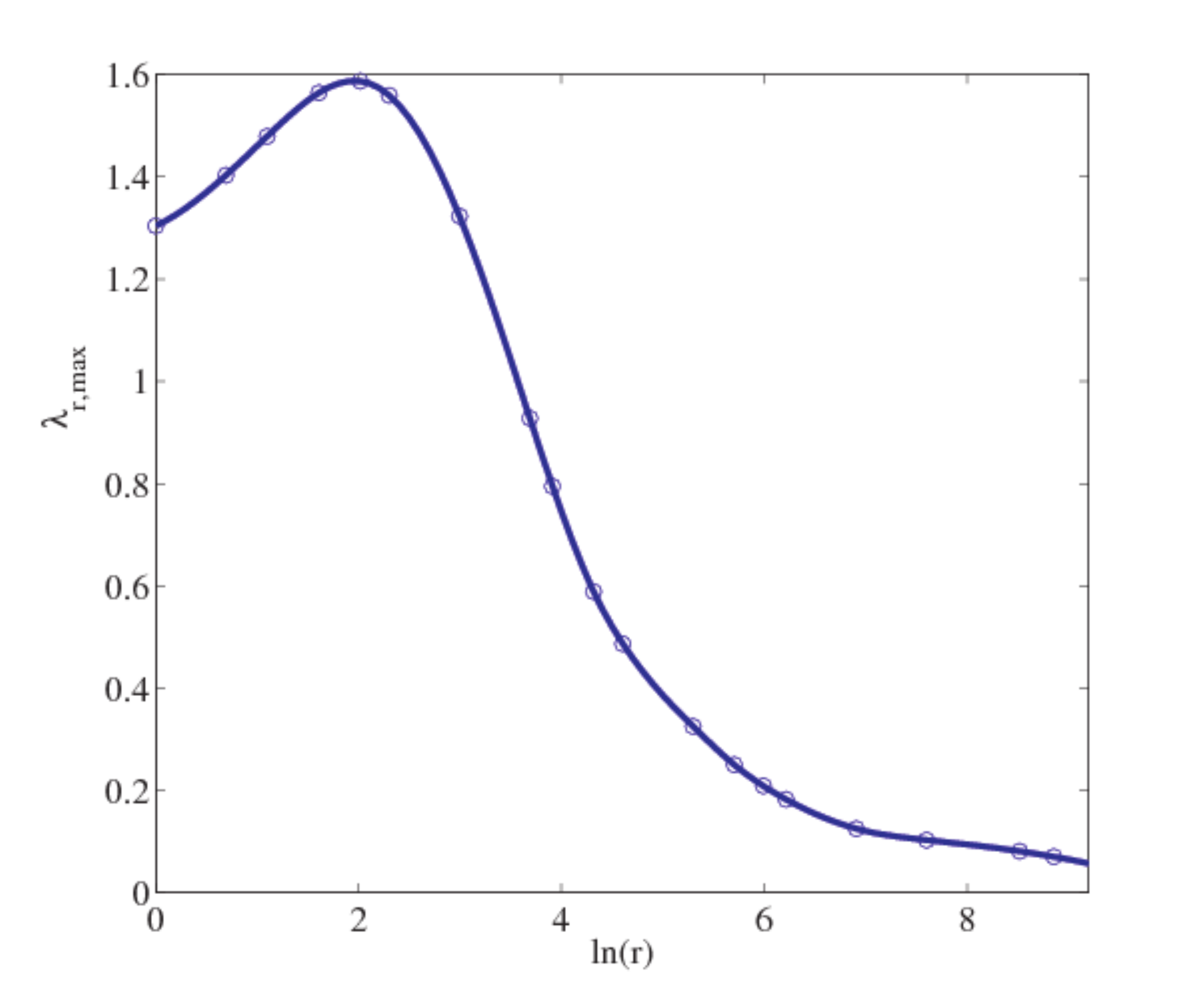}}
\subfigure[]{\includegraphics[width=0.3\textwidth]{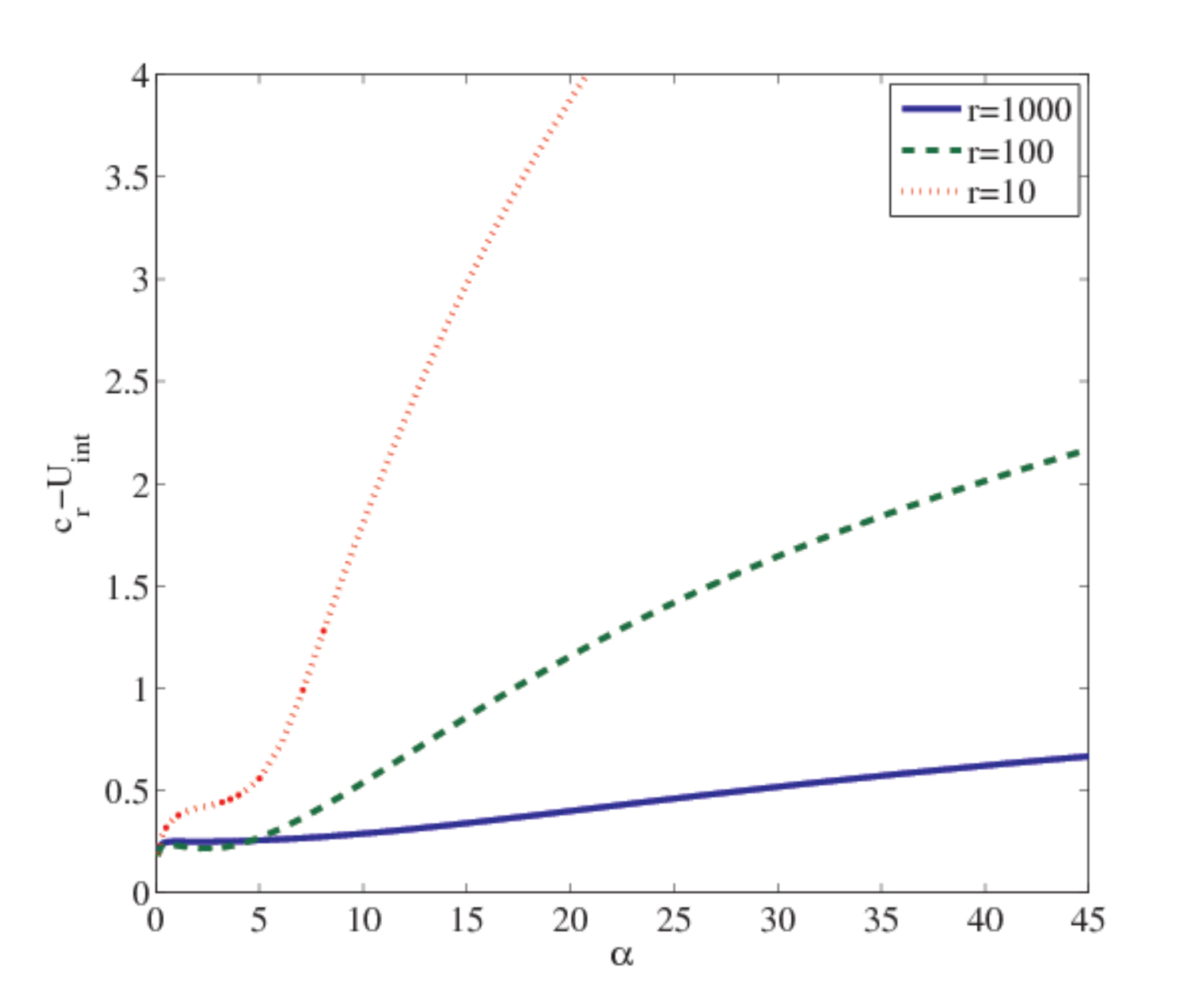}}
\caption{The effects of varying the density ratio $r$.  We have set $Re_0=1000$, $\left(m,\delta\right)=\left(55,0.05\right)$, and we have taken
$Fr=3.7809\times10^6\left(r-1\right)/Re_0^2$
and $S=1.1420\times10^7/Re_0^2$.
(a)  Decreasing the density contrast is destabilizing;
(b) the dependence of $\lambda_{\mathrm{max}}$ on $r$; (c)  the dependence of the wave
speed on $r$.  Decreasing $r$ leads to faster waves, although the unstable
waves are still slow.
}
\label{fig:vary_r}
\end{figure}

In this section we set $Re_P=5\times10^5$ ($Re_0=1000$).  Fig.~\ref{fig:vary_r}
shows the result of varying the parameter $r$ through a range $r=10$--$10,000$.
For large $r$-values, the maximum growth rate decreases upon decreasing $r$.
This is not surprising: a decreasing value of $r$ implies that the density
of the liquid approaches that of the gas, and thus the liquid has less inertia.
 The interface is then expected to be less stable.  The plot in Fig.~\ref{fig:vary_r}~(b)
neatly sums up this dependence.  However, for smaller $r$-values $r\apprle10$, this dependence
is reversed.
This is explained by the energy budget in Tab.~\ref{tab:energy_budget_r},
where the energy decomposition at the maximum growth
rate is shown, as a function of $r$.  The principal source of instability
is the $TAN$ term, consistent
\begin{table}[htb]
\centering\noindent
\begin{tabular}{|c|c||c|c|c|c|c|c|c|c|}
\hline
$r$&$\alpha$&$KIN_G$&$KIN_L$&$REY_L$&$REY_G$&$DISS_L$&$DISS_G$&$NOR$&$TAN$\\
\hline
\hline
1000&20&0.18&0.82&2.34&-11.90&-4.28&-57.42&-2.73&74.99\\
\hline
100&7.1&0.78&0.22&0.20&-2.60&-0.60&-14.21&-0.54&18.75\\
\hline
5&4.2&0.96&0.04&0.00&0.38&-0.35&-3.55&-0.13&4.65\\
\hline
1&3.9&0.99&0.01&0.00&0.32&-0.52&-4.26&-0.11&5.56\\
\hline
\end{tabular}
\caption{Energy budget for the most dangerous mode as a function of $r$,
for $Re_0=10^3$, $\delta=0.05$, and $m=55$.  In general, there are
two terms contributing to the instability: one interfacial, and one due to
effects in the liquid layer.  As $r$ is reduced, the latter term diminishes
importance, thus removing one of the sources of instability.}
\label{tab:energy_budget_r}
\end{table}
with a viscosity-contrast instability.  Note that the term $REY_L$ is positive
too, although this contribution diminishes with decreasing $r$, so removing a source
of instability and thus reversing the monotone dependence of the growth rate
(at large $r$, the growth rate decreases with increasing $r$).
As $r$ decreases further (in particular, for $r=1$,~$5$), there is a destabilizing net input of energy into the perturbations
from the $REY_G$ term, which implies that the critical-layer mechanism plays
a secondary role.

These findings raise two questions.  Is the turbulence model valid at these fast wave speeds? Is our base-state model valid at these low values of $r$?
%
%
%
%
Now when the critical-layer mechanism is relevant, the dynamically
important region moves into the bulk of the gas  flow (away from the
interface).  It is possible that this region will coincide with that
part of the gas domain where the rapidity of advection dominates
over the eddy turnover time (see Sec.~\ref{sec:model_perturb}).
Thus, in this case, rapid-distortion effects may be important.
These effects may alter both the growth rate and the structure of
the wave-induced velocity field, and will be considered in future
work, in a parametric study similar to this one.
%
%
%
%
The
  second question concerns the wall-interface equivalence in the base-state
  model.  This assumption was used in choosing the exponent in the wall function.
   Such an equivalence is only valid for large density contrasts.  However,
   this is
   a relatively unimportant ingredient in the model, since it determines
   the second-order term (but not the first-order term) in the Taylor expansion
   of the base-state velocity near the interface $z=0$.  Thus, this equivalence
   assumption is unlikely to affect the development of fast waves as $r\downarrow1$.
    Having verified the effect of density stratification on the character
    of the instability, consistent with our analysis in Sec.~\ref{sec:flat},
    we now perform a more systematic analysis on the effects of the Froude
    and Weber numbers on the instability.

\begin{figure}[htb]
\centering\noindent
\subfigure[]{\includegraphics[width=0.45\textwidth]{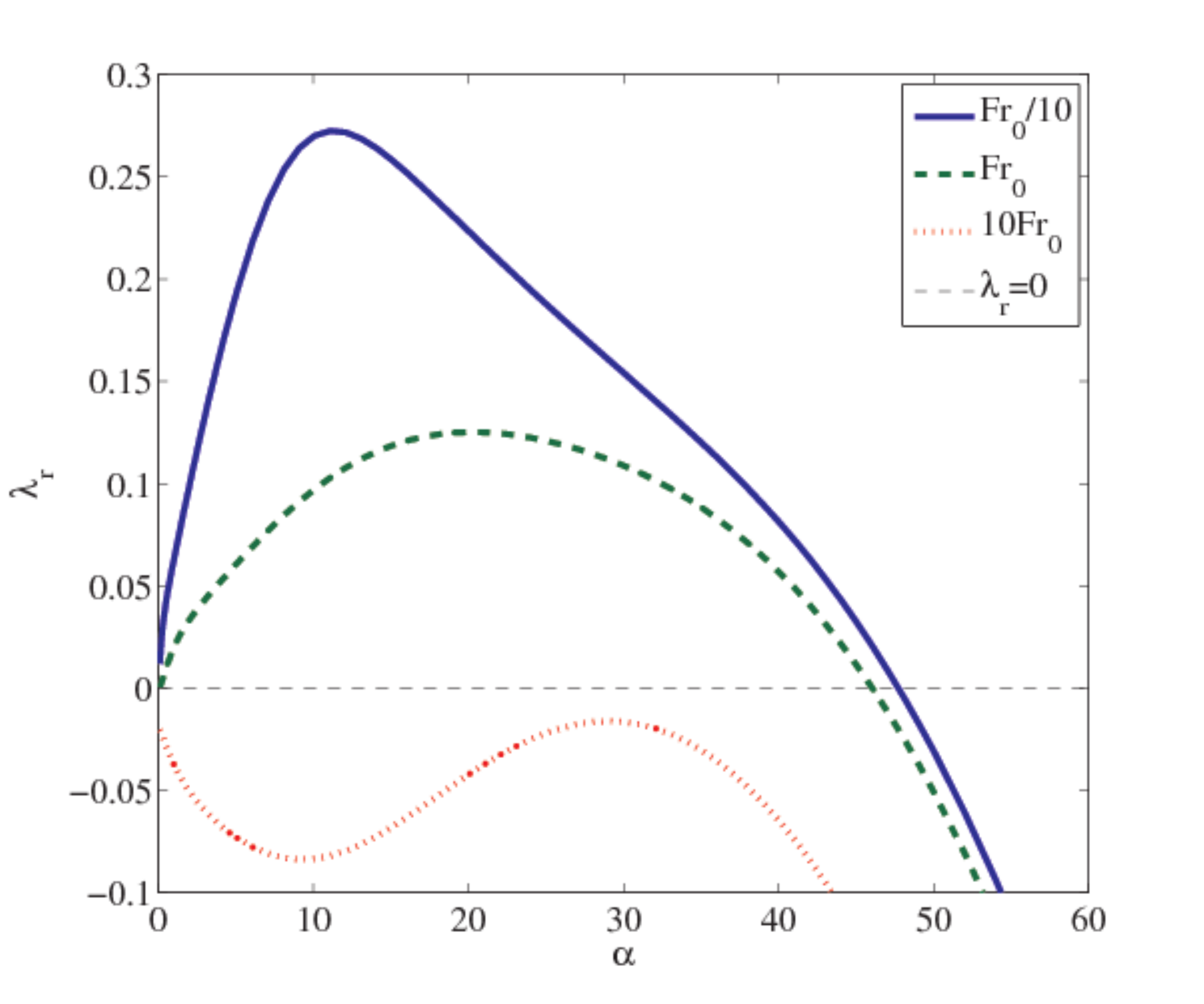}}
\subfigure[]{\includegraphics[width=0.45\textwidth]{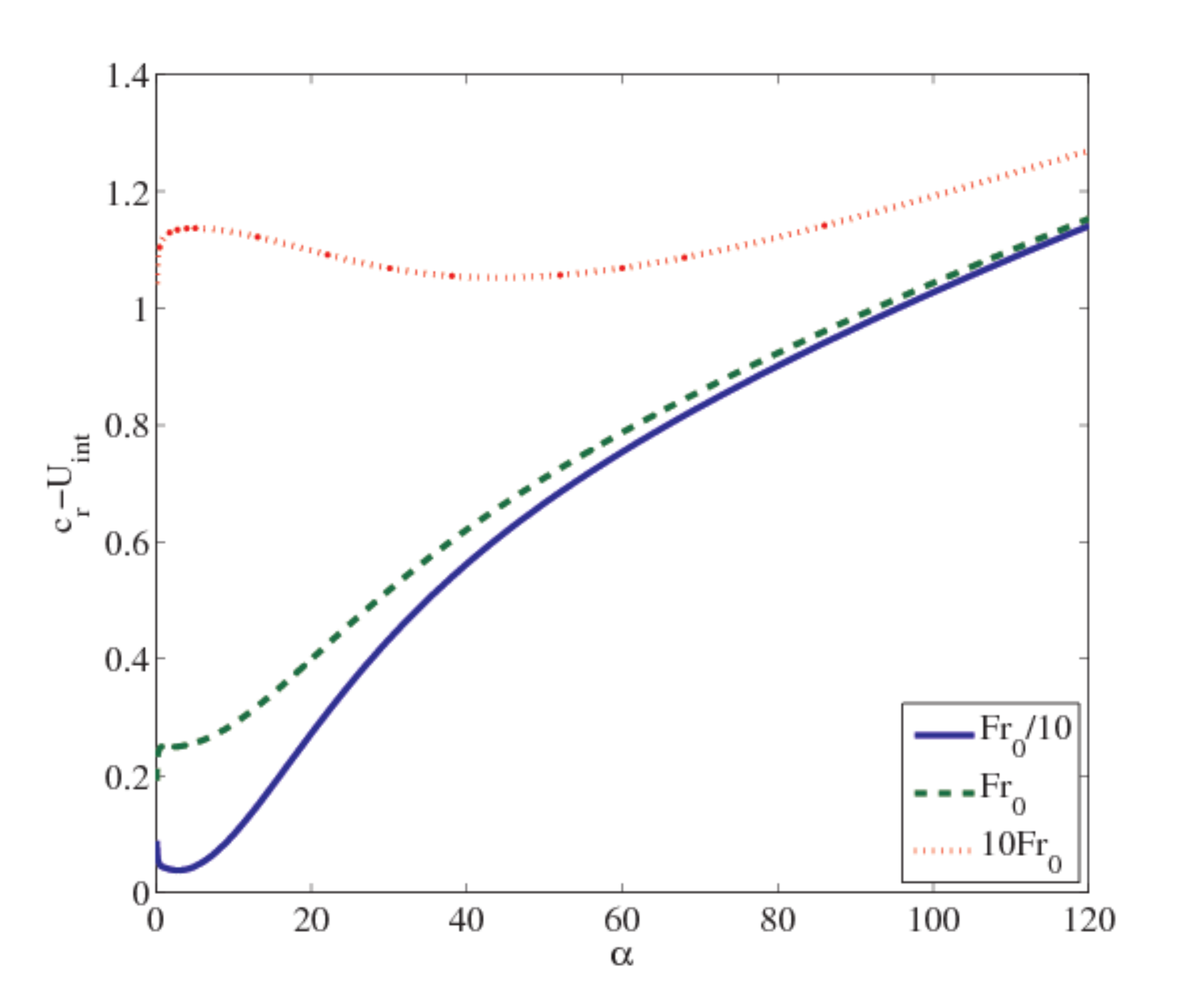}}
\caption{The effects of varying the parameter $Fr$ at fixed density ratio.   We have set $Re_0=1000$, $\left(m,r,\delta\right)=\left(55,1000,0.05\right)$, and have taken the reference values
$Fr_0=3.7809\times10^6\left(r-1\right)/Re_0^2$
and $S_0=1.1420\times10^7/Re_0^2$.  This corresponds to a liquid-film depth $0.0025\,\mathrm{m}$ and a gas-layer depth $0.05\,\mathrm{m}$.  (a)  Increasing $Fr$ is stabilizing; for
 a given parameter set $\left(m,r,\delta,S,Re_0\right)$ there is a critical Froude number
 for which the interface is stable at all wavenumbers; (b) increasing $Fr$
 increases the wave speed, in particular, the wave speed is increased in
 the wavenumber range for which instability is observed.
}
\label{fig:vary_Fr}
\end{figure}

\begin{figure}[htb]
\centering\noindent
\subfigure[]{\includegraphics[width=0.45\textwidth]{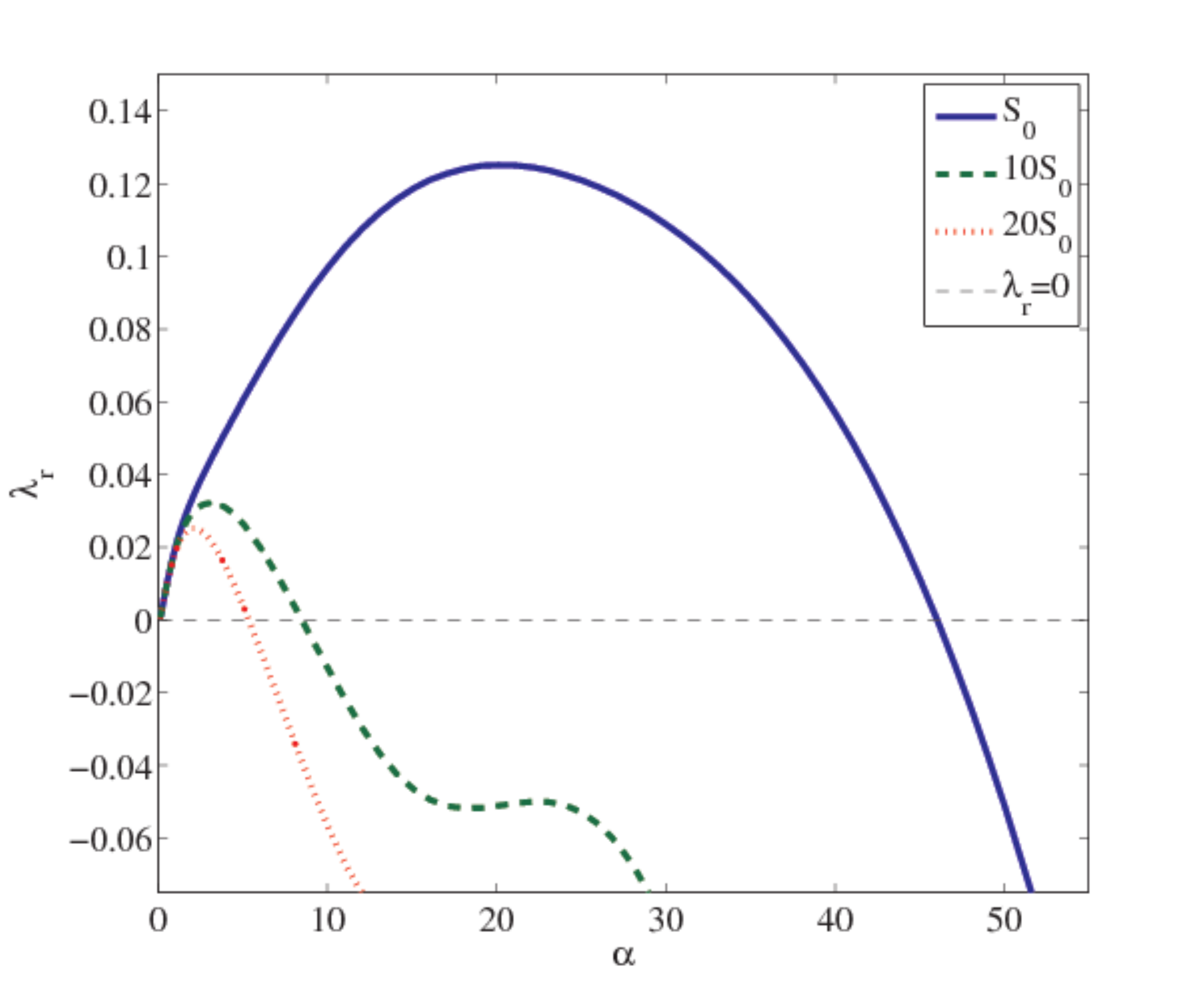}}
\subfigure[]{\includegraphics[width=0.45\textwidth]{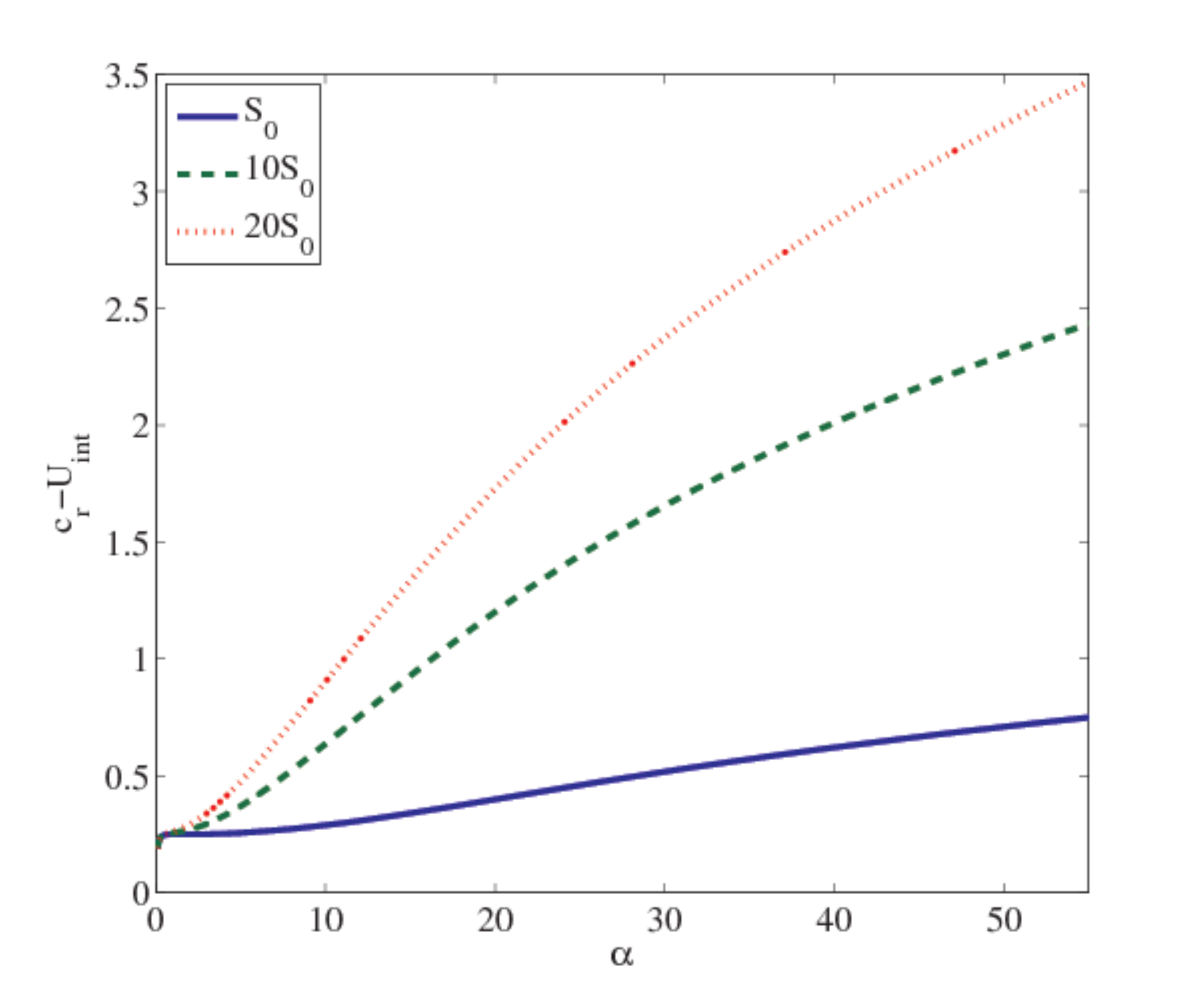}}
\caption{The effects of varying the inverse Weber number $S$.   We have set $Re_0=1000$, $\left(m,r,\delta\right)=\left(55,1000,0.05\right)$, and have taken the reference values
$Fr_0=3.7809\times10^6\left(r-1\right)/Re_0^2$
and $S_0=1.1420\times10^7/Re_0^2$.   (a)  Increasing $S$ is stabilizing; (b)
increasing $S$
 increases the wave speed, although this effect generates fast waves only
 in the wavenumber region of stable interfacial waves.
}
\label{fig:vary_S}
\end{figure}

We perform a parameter study based on the inverse Froude number $Fr$
in Fig.~\ref{fig:vary_Fr}, wherein $Fr$ is varied around  the
reference value $Fr_0$ given by Eq.~\eqref{eq:fr_values}, at fixed
density ratio and Reynolds number.
As expected, increasing $Fr$ is stabilizing; such an increase also leads to faster waves, and $c_r/U_0\apprge1$ in a
range of wavenumbers where the interface is unstable.
%
%
%
%
%
%
In Fig.~\ref{fig:vary_S} we demonstrate the effects varying the inverse Weber
number $S$ relative to the reference value $S_0$ (Eq.~\eqref{eq:fr_values}), at fixed Reynolds number.
As expected, increasing $S$ is stabilizing, and leads to faster waves.  However,
the fast waves are stable, while the slower waves are unstable.  This is
in contrast to Fig.~\ref{fig:vary_S}, where increasing $Fr$
led to fast, unstable waves.  This difference can be readily understood by
recourse to the free-surface formula~\eqref{eq:c_est_froude}, here recalled
to be
\[
\frac{c_{\mathrm{r}}}{U_0}\approx\sqrt{\frac{Fr}{r+1}\frac{1}{\tilde{\alpha}}+\frac{S}{r+1}\tilde{\alpha}}\sqrt{\tanh\left(\tilde{\alpha}\alpha\right)}.
\]
The effects of varying $S$ in this equation are felt at large wavenumbers,
while the effects of varying $Fr$ are more significant at smaller wavenumbers.
 Since the instability attains maximum growth at $\left(\ell/d_L\right)\approx
 1$--$10$, it is the variation in $Fr$ that is  prominent in
\begin{table}[htb]
\centering\noindent
\begin{tabular}{|c|c|c||c|c|c|c|c|c|c|c|}
\hline
$Fr$&$S$&$\alpha$&$KIN_G$&$KIN_L$&$REY_L$&$REY_G$&$DISS_L$&$DISS_G$&$NOR$&$TAN$\\
\hline
\hline
$Fr_0$&$S_0$&20&0.18&0.82&2.34&-11.90&-4.28&-57.42&-2.73&74.99\\
\hline
$Fr_0$&10$S_0$&3&1.00&1.00&0.00&-15.24&-2.72&-65.40&-6.78&91.14\\
\hline
$Fr_0$&$20S_0$&2&1.00&0.00&0.00&-33.29&-6.00&-155.23&-13.22&208.74\\
\hline
$0.1Fr_0$&$S_0$&10&0.00&1.00&0.00&-10.22&-0.53&-39.06&-2.22&53.04\\
\hline
$10Fr_0$&$S_0$&30&-3.16&2.16&0.00&0.04&-0.89&-1.53&-0.63&2.00\\
\hline
\end{tabular}
\caption{Energy budgets for the parameter study in which $S$ and $Fr$ are
varied, at fixed density ratio $r$.   We have set $Re_0=1000$, $\left(m,r,\delta\right)=\left(55,1000,0.05\right)$, and we have taken the reference values
$Fr_0=3.7809\times10^6\left(r-1\right)/Re_0^2$
and $S_0=1.1420\times10^7/Re_0^2$. In all cases considered, and for both
slow and fast waves, the instability is due to the viscosity-contrast mechanism.}
\label{tab:energy_budget_FS}
\end{table}
the wavenumber range of instability.  Thus, it is likely to be a shift in
 $Fr$, rather than $S$, that precipitates
 a change in the character of the unstable interfacial waves.  In Tab.~\ref{tab:energy_budget_FS}
 we verify that it is the viscosity-contrast mechanism, and not the critical-layer
 mechanism, that is at work in each case of instability studied, in spite
 of the change in the wave speed.  Careful parameter tuning is thus required
 to observe the critical-layer instability, which was highlighted in Fig.~\ref{fig:vary_r}
 and Tab.~\ref{tab:energy_budget_r}.

\begin{figure}[htb]
\centering\noindent
\subfigure[]{\includegraphics[width=0.45\textwidth]{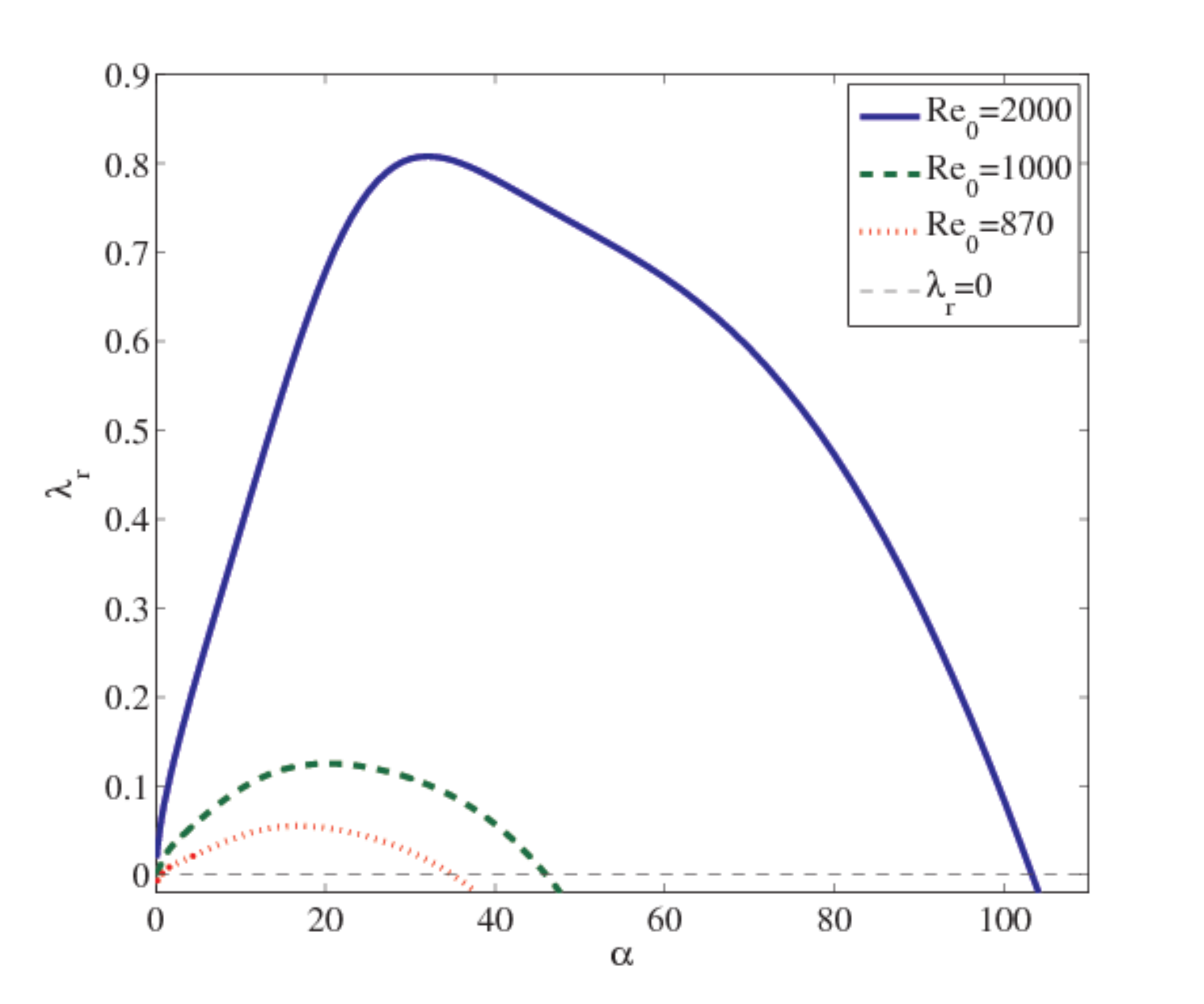}}
\subfigure[]{\includegraphics[width=0.45\textwidth]{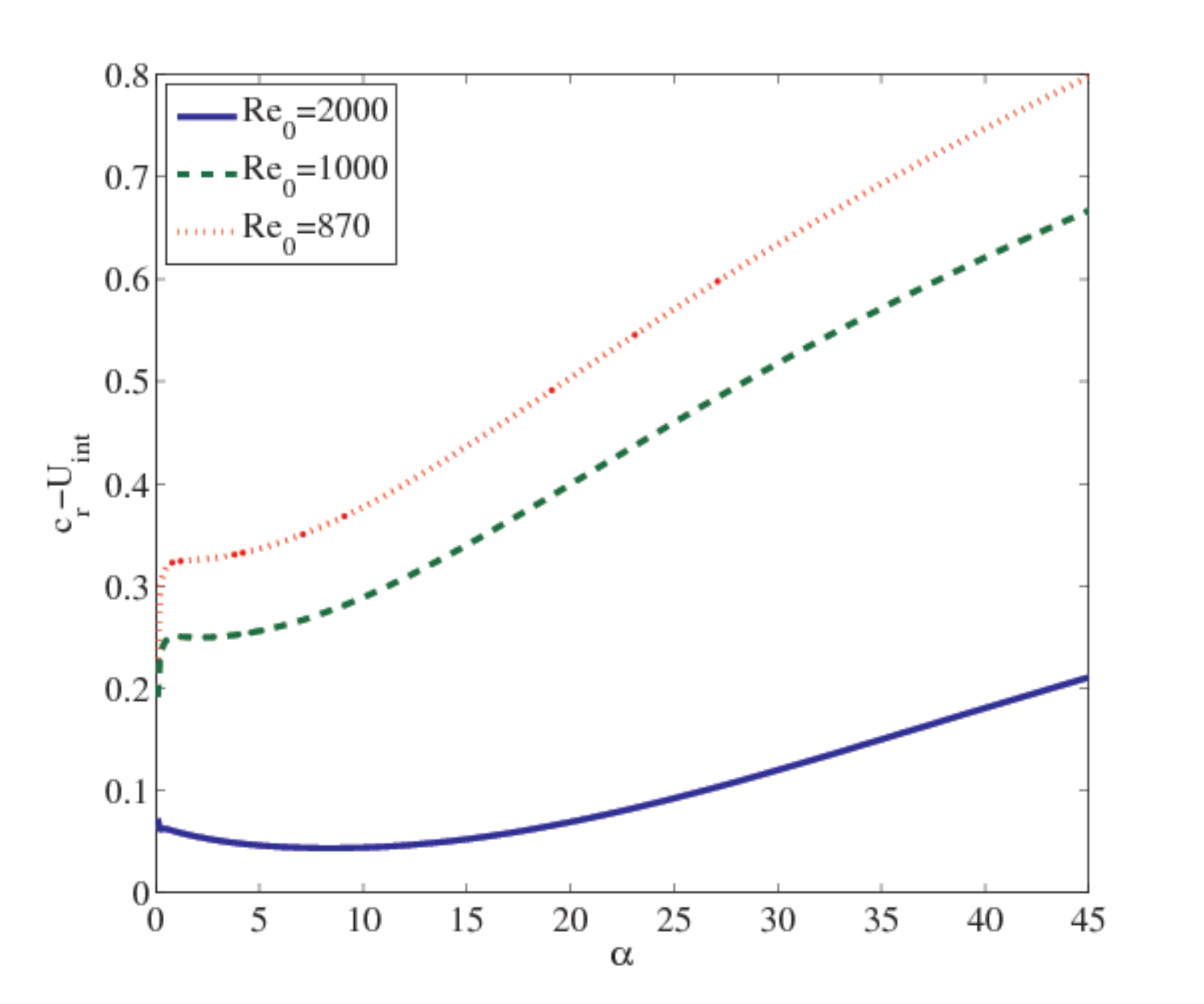}}
\caption{The effects of varying the Reynolds number $Re_0$ on (a) the growth rate; (b) the wave speed.  We have set $\left(m,r,\delta\right)=\left(55,1000,0.05\right)$, and have taken
$Fr=3.7809\times10^6\left(r-1\right)/Re_0^2$
and $S=1.1420\times10^7/Re_0^2$.
The wave speed $c_r/U_0$ is
less than unity for the unstable waves, confirming that these waves are in
fact slow.
}
\label{fig:vary_Re}
\end{figure}

Finally, we study the effect of varying $Re_0$ in Fig.~\ref{fig:vary_Re}.  The
maximum growth rate and the cutoff wavenumber are both shifted to higher
values with increasing
$Re_0$.  Typical values of the wave speed are higher for smaller values of
$Re_0$, as predicted by the formula for free-surface waves~\eqref{eq:c_grav}.
 However, for unstable waves, that is, for $\alpha$ less than the cutoff
 wavenumber, the wave speed $c_{\mathrm{r}}/U_0$ is less than unity, confirming
 that these are in fact slow waves.  Two further issues arise when studying
 the $Re_0$-dependence of the stability.  First, upon decreasing $Re_0$,
 the lower critical wavenumber shifts from $\alpha_{\mathrm{cl}}=0$ to some finite value
 $\alpha_{\mathrm{cl}}>0$.  This suggests that for a given parameter set $\left(m,r,\delta,Re_0^2S,Re_0^2Fr\right)$,
 there is a critical Reynolds number for stability.
\begin{figure}[htb]
\centering\noindent
\subfigure[]{\includegraphics[width=0.43\textwidth]{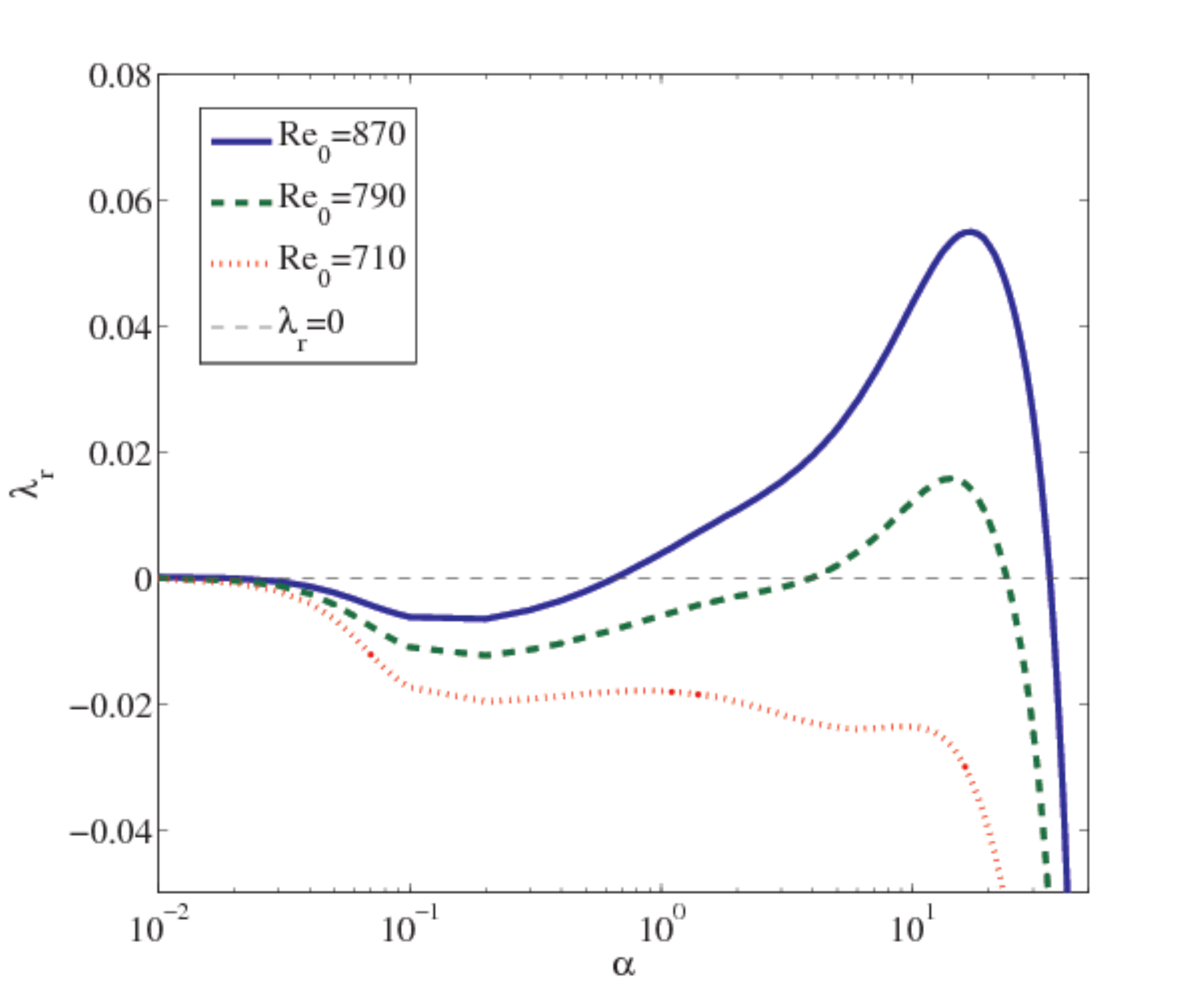}}
\subfigure[]{\includegraphics[width=0.45\textwidth]{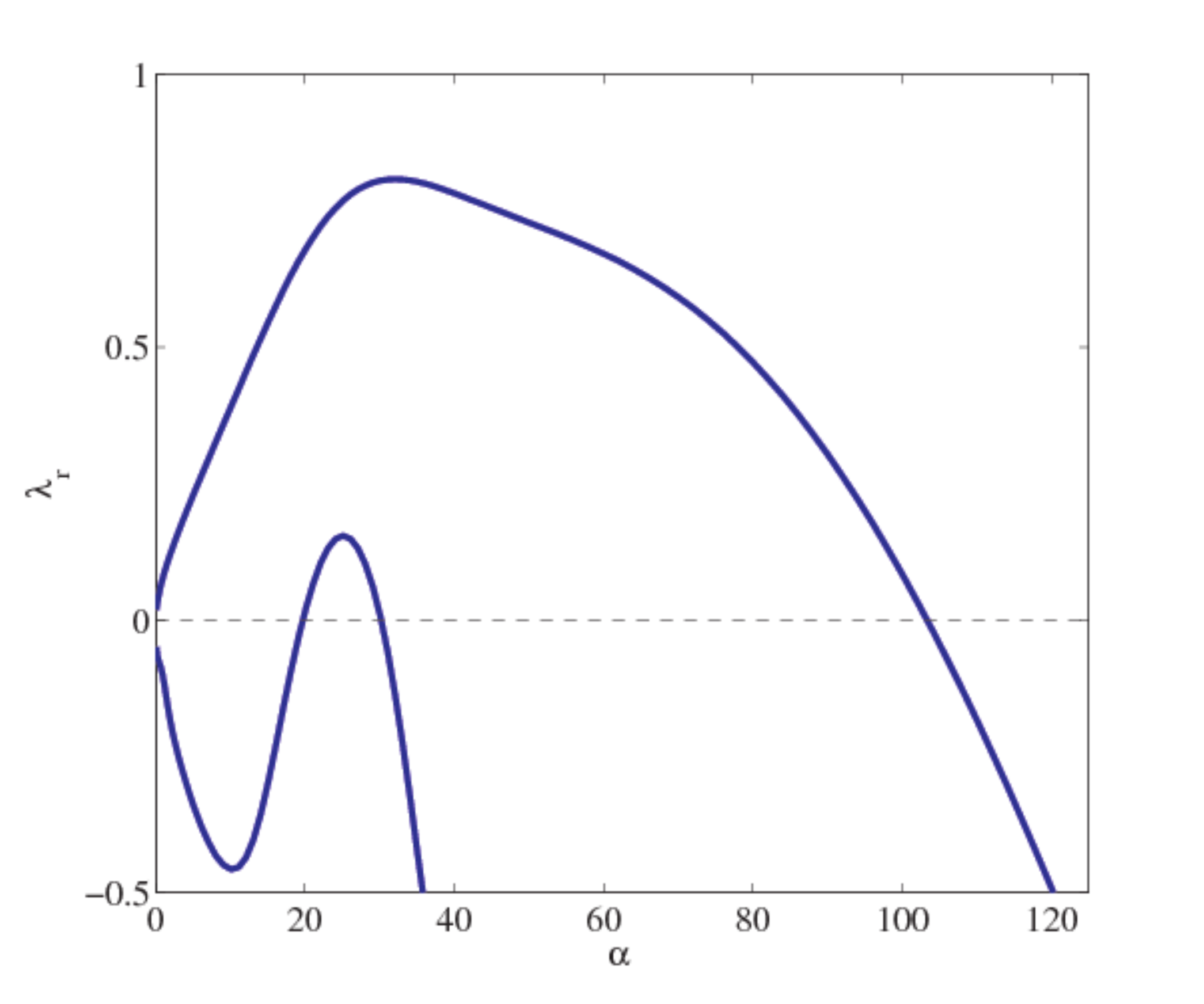}}
\caption{The effects of varying the Reynolds number $Re_0$.  We have set $\left(m,r,\delta\right)=\left(55,1000,0.05\right)$, and have taken
$Fr=3.7809\times10^6\left(r-1\right)/Re_0^2$
and $S=1.1420\times10^7/Re_0^2$.
 Subfigure~(a)
shows the existence of a critical Reynolds number below which the interface
is stable; (b) demonstrates the development of a second mode of instability
at higher Reynolds numbers ($Re_0=2000$).
}
\label{fig:vary_Re1}
\end{figure}
This is demonstrated in Fig.~\ref{fig:vary_Re1}, where the critical
Reynolds number is $Re_{0\mathrm{c}}\approx750$.  In a later
section, we use this result as a means of verifying our model against experiments,
since the critical Reynolds number for the onset of instability is readily
measured.  The second issue concerns the development of a second unstable
mode at higher Reynolds numbers, as demonstrated in Fig.~\ref{fig:vary_Re1}~(b).
 This is the so-called internal mode, which we now investigate in detail.

\subsection{The internal mode}
\label{subsec:internal_tau}

We examine the properties of the second unstable mode observed in Fig.~\ref{fig:vary_Re1}~(b).
 We first of all examine the energy  budget at $\alpha=25$, for both unstable
 modes.  This is shown in Tab.~\ref{tab:two_modes}.
\begin{table}[htb]
\centering\noindent
\begin{tabular}{|c|c|c|c|c|c|c|c|c|c|}
\hline
$\alpha$&$\lambda_{\mathrm{max}}$&$KIN_G$&$KIN_L$&$REY_L$&$REY_G$&$DISS_L$&$DISS_G$&$NOR$&$TAN$\\
\hline
\hline
25&0.77&0.85&0.15&0.45&-9.57&-0.41&-36.96&-1.17&48.67\\
\hline
25&0.15&0.15&0.85&3.80&-9.98&-0.80&-26.46&-0.18&34.62\\
\hline
\end{tabular}
\caption{Energy budget of the interfacial and internal modes at $\alpha=25$.
 Here we have set $\left(m,r,\delta,Re_0\right)=\left(55,1000,0.05,2000\right)$.
The values of $Fr$ and $S$ by Eq.~\eqref{eq:fr_values}.  Both modes
 enjoy destabilizing contributions from $TAN$ and $REY_L$, although this
 latter contribution is much larger for the internal mode.}
\label{tab:two_modes}
\end{table}
As usual, the first mode, associated with the eigenvalue branch that has
interested us until now, derives all but a small fraction of its destabilizing
energy from the $TAN$ term, which we have identified as a work done by the
tangential stress on the interface.  This term is positive when $m>1$, and
we designate this mode the `interfacial' mode.  The second mode derives the
majority of its destabilizing energy from this source too, although the term
\begin{figure}[htb]
\centering\noindent
\subfigure[]{\includegraphics[width=0.45\textwidth]{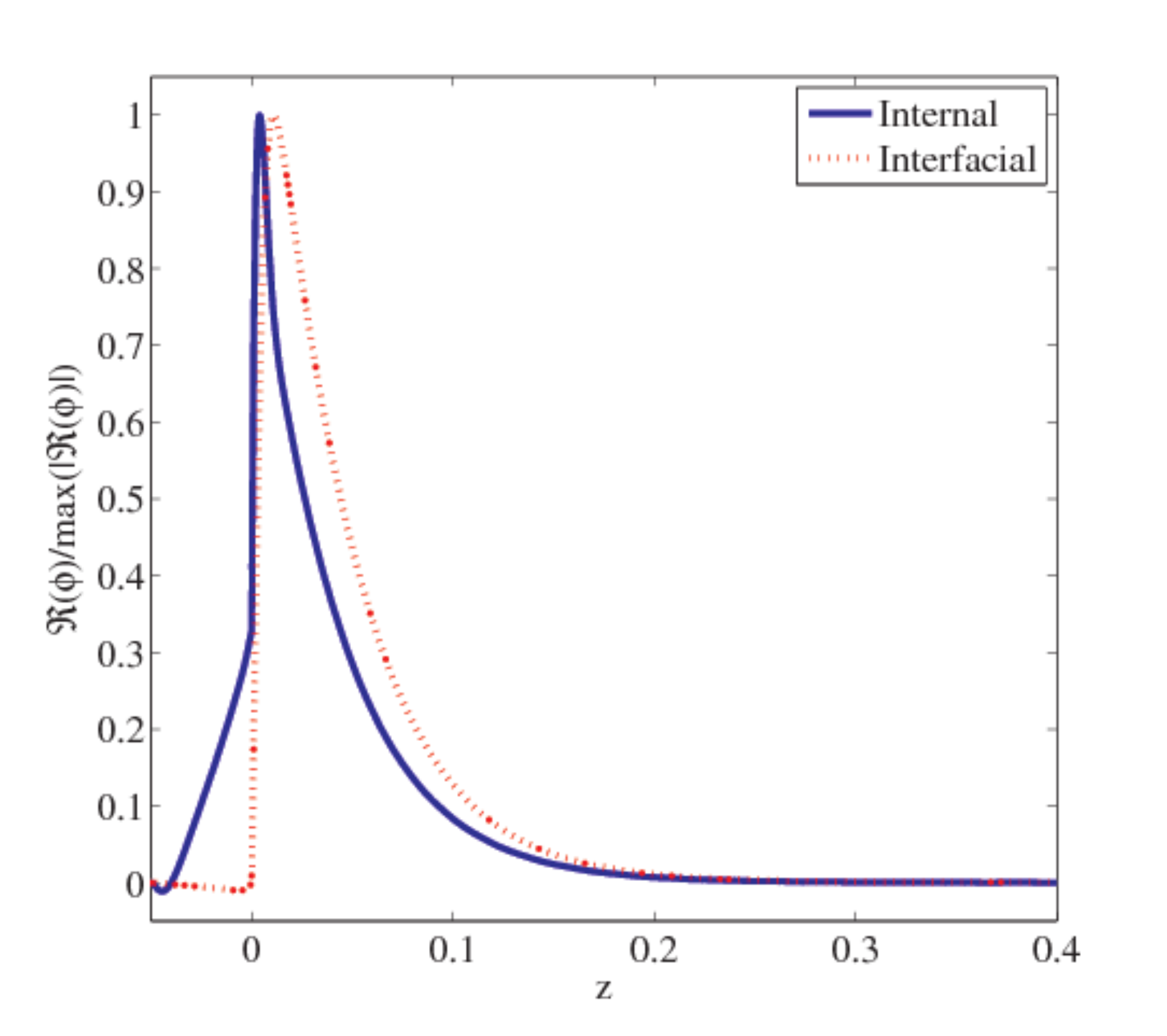}}
\subfigure[]{\includegraphics[width=0.45\textwidth]{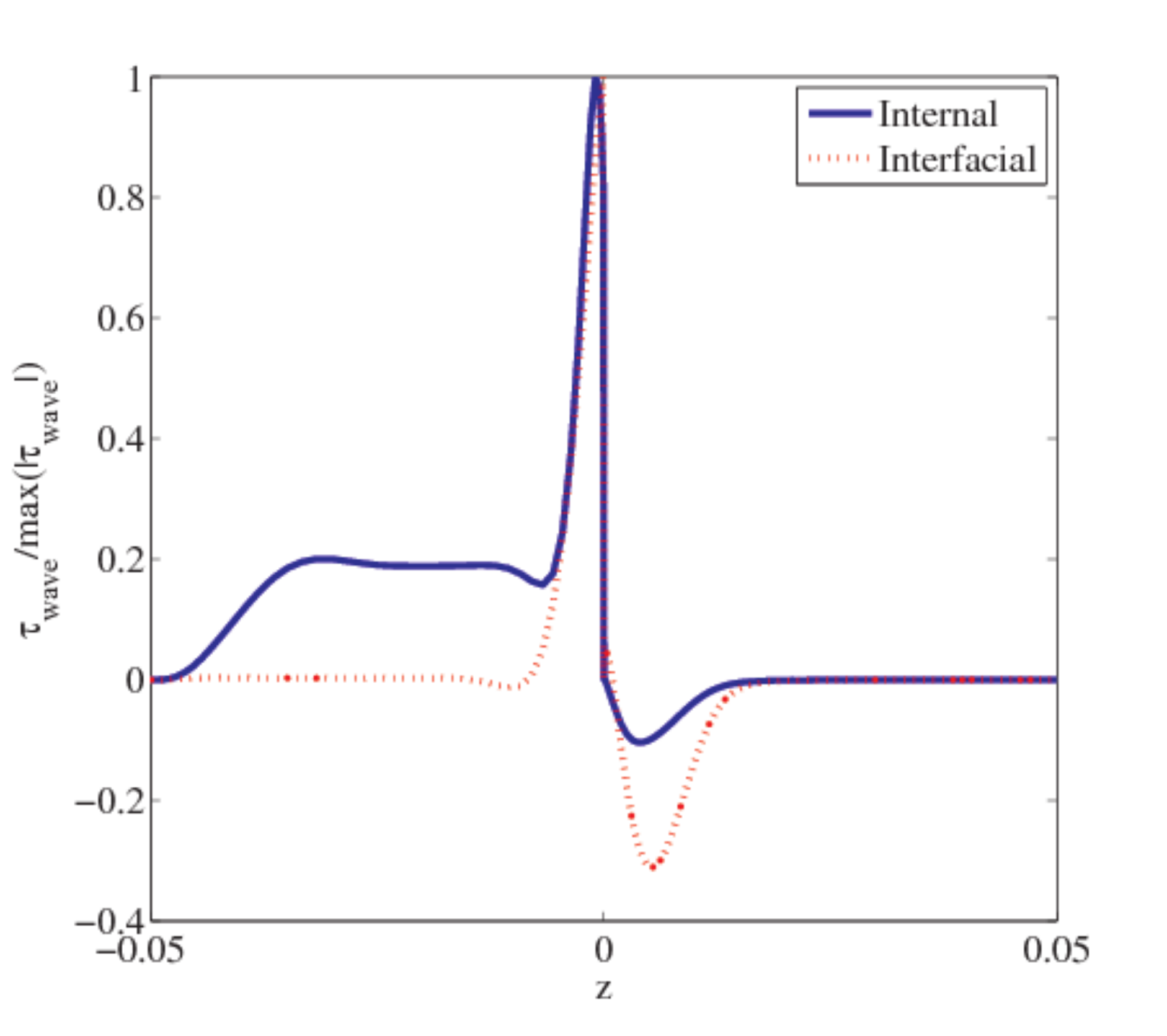}}
\caption{The streamfunction for the internal and interfacial modes, normalized
such that $\max\,|\Re\left(\phi\right)|=1$.  The wavenumber is $\alpha=25$, and the
other parameters are the same as those in the energy-budget table~\ref{tab:two_modes};
(b) the wave Reynolds stress function for the internal and interfacial modes,
normalized such that $\max\,\tau_{\mathrm{wave}}=1$.  The internal
mode exhibits a stronger flow in the liquid layer, and thus gives rise to
a larger wave Reynolds stress there.}
\label{fig:compare_interfacial_internal}
\end{figure}
$REY_L$ is now more important.  Thus, the transfer of energy from the mean
flow in the liquid, to the perturbation flow, is important.  We therefore
designate this mode the `internal' mode.  This justification is strengthend
further by examination of the streamfunction and wave Reynolds stress function
associated with these modes, shown in Fig.~\ref{fig:compare_interfacial_internal}.
 Figure~\ref{fig:compare_interfacial_internal}~(a) shows the streamfunction
 for these two modes.  The streamfunction of the internal mode possesses
 a large non-zero component in the liquid, in contrast to that of the interfacial
 mode.  This gives rise to significant flow  in the liquid, and hence gives
 an important contribution to the transfer term $REY_L$.  The development
 of a larger transfer term is shown in Fig.~\ref{fig:compare_interfacial_internal},
 where we examine the wave Reynolds stress.  This is the function
\begin{eqnarray*}
\tau_{\mathrm{wave}}^{(i)}\left(z\right)&=&-r_i\int_0^\ell \delta u_i\left(x,z\right)\delta w_i\left(x,z\right) \frac{\mathd
U_i^{(0)}}{\mathd z}\mathd x,\qquad i=L,G\\
REY_L&=&\int_{-d_L}^0\tau_{\mathrm{wave}}^{(L)}\left(z\right)\mathd z.
\end{eqnarray*}
Clearly, $REY_L$ is much larger for the internal mode, thus confirming the
importance of the dynamics of the liquid layer for the development of this
secondary instability.
Moreover, the critical layer of the internal mode is in the liquid,
a fact that has been used in the past to justify its designation  as
`internal'~\citep{Miesen1995} ($U\left(0\right)=0.92$, and
$c_\mathrm{r}=0.25$ at $\alpha=25$ for the internal mode).
%
%
%
%
%
%

\begin{figure}
\begin{center}
\subfigure[]{
\includegraphics[width=0.3\textwidth]{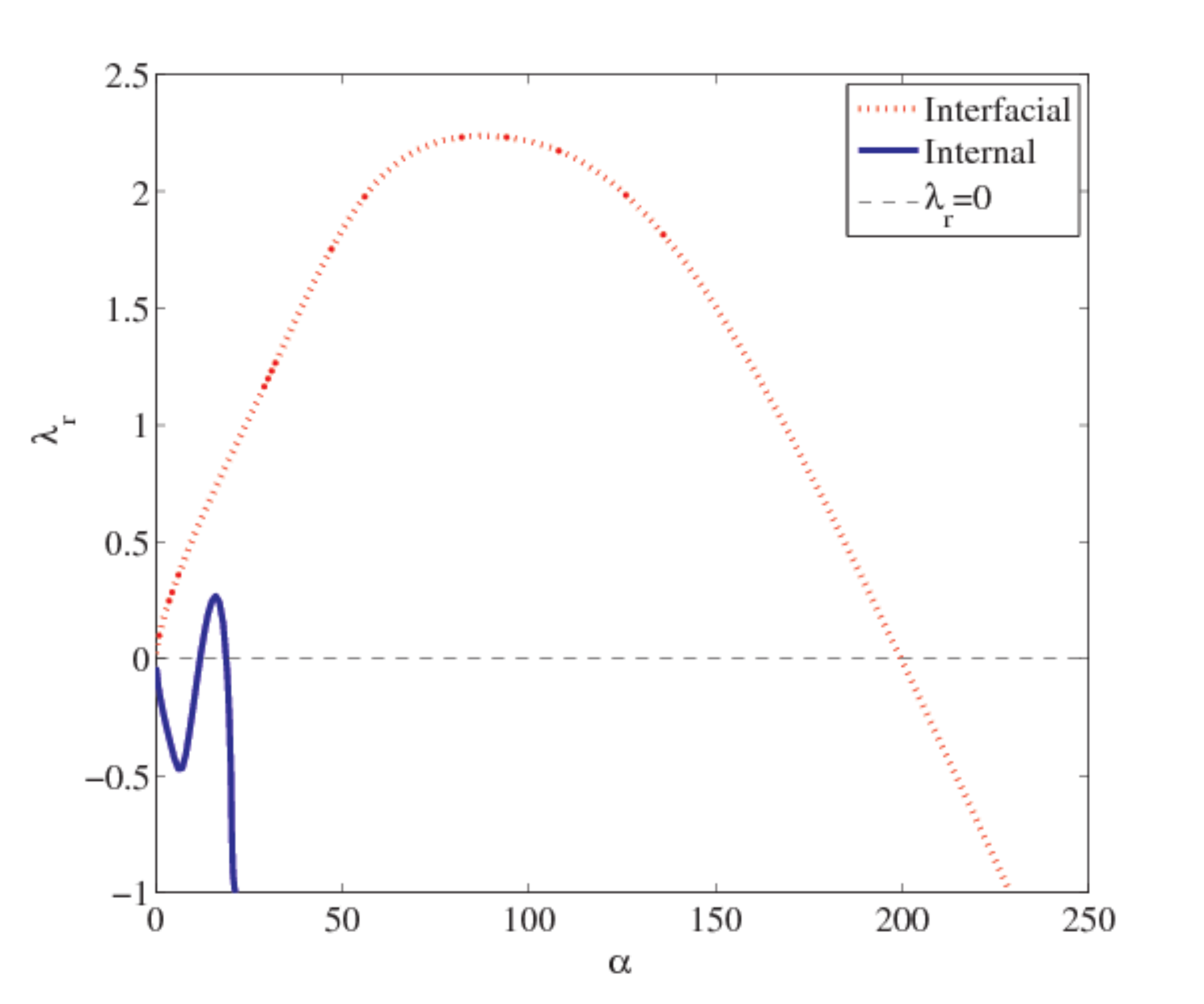}
}
\subfigure[]{
\includegraphics[width=0.3\textwidth]{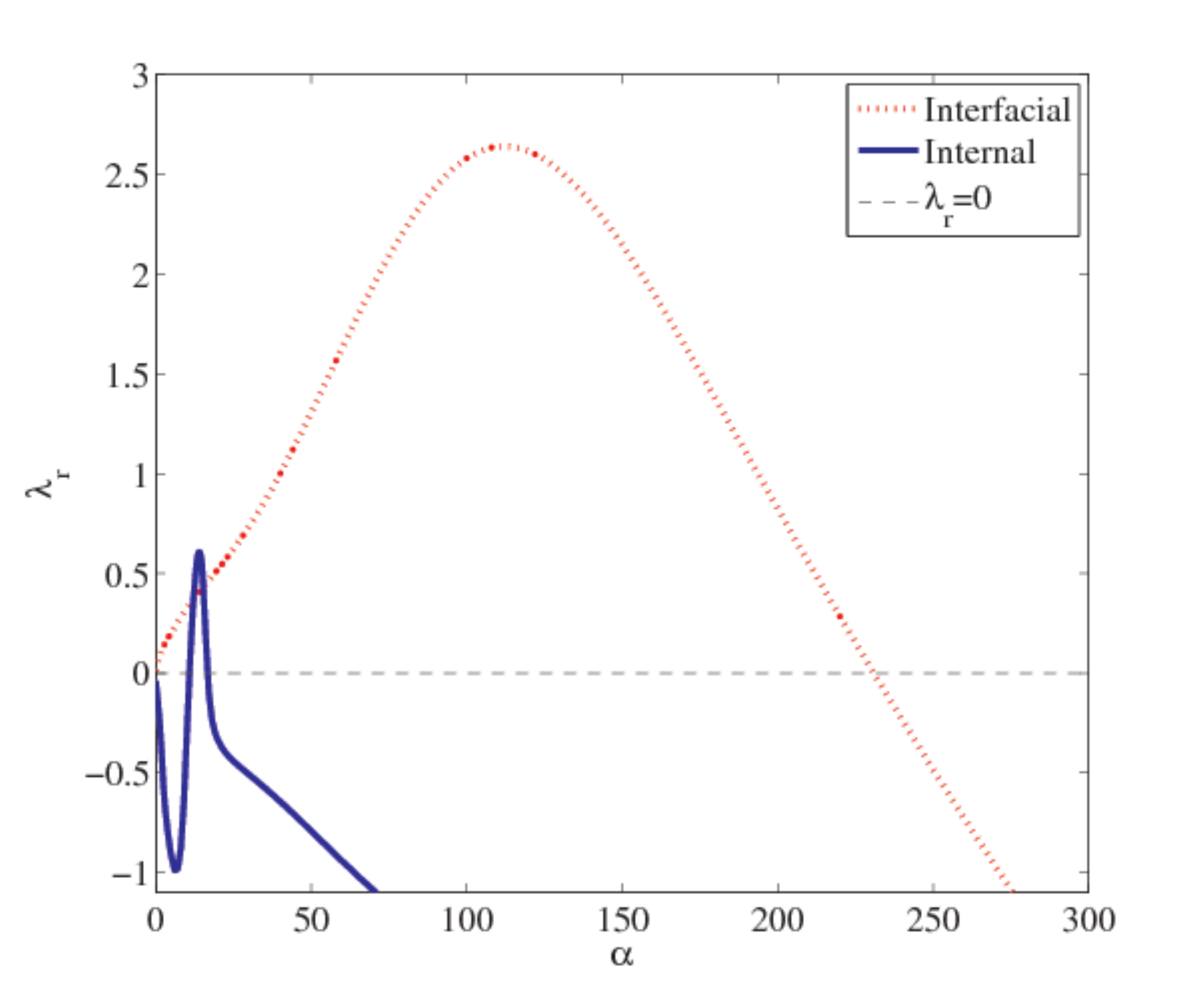}
}
\subfigure[]{
\includegraphics[width=0.3\textwidth]{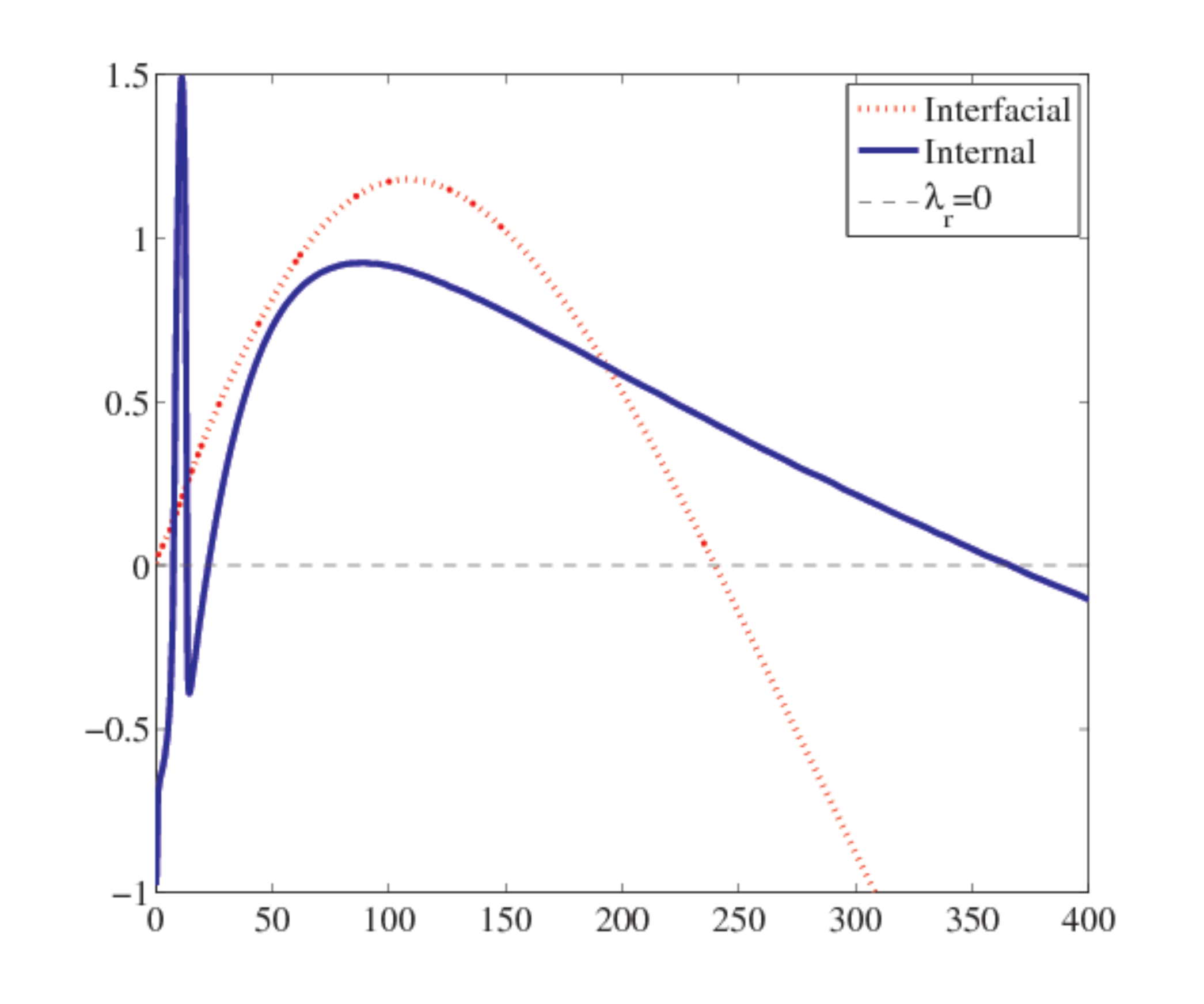}
}
\end{center}
\caption{Mode competition between the interfacial and internal modes. Here $\left(r,\delta,Re_0\right)=\left(1000,0.05,4000\right)$
The parameter $m$ takes the values $55$, $20$, and $5$ in subfigures
(\textit{a}), (\textit{b}), and (\textit{c}), respectively.  The wave speed corresponding to the internal mode is shown in Fig.~\ref{fig:competition_speed}.}
\label{fig:competition}
\end{figure}

\begin{figure}[tbp]
\begin{center}
\includegraphics[width=0.5\textwidth]{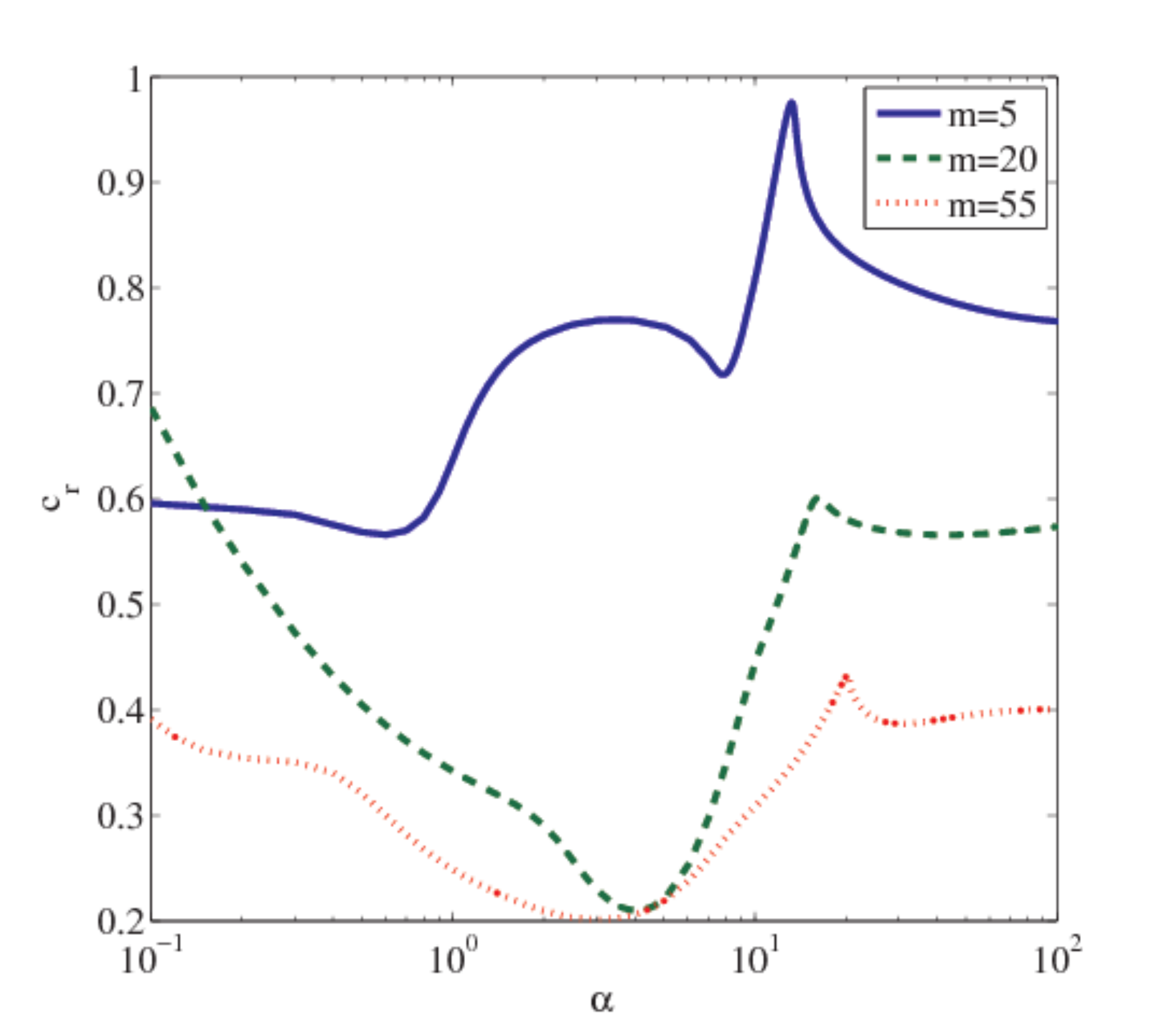}
\end{center}
\caption{Companion to Fig.~\ref{fig:competition}: internal-mode wave speed as a function of wavenumber for various $m$-values, at $Re_0=4000$.  The continuous nature of these curves confirms that the sharp changes in the growth rate of the internal mode are genuine.  Although the wave speeds plotted here are positive (to facilitate easy comparison between the different $m$-values), the wave speed $c_r-U_{\mathrm{int}}$ is negative, confirming that the critical layer for these waves is in the liquid, thus justifying the designation of this instability as \textit{internal}.  The sharp ``kinks' correspond to the turing points or inflection points in the associated growth-rate curves.  Such kinks often occur during modal coalescence, for example, in the Kelvin--Helmholtz instability~\citep{ChandraStability}, and in other two-phase flow scenarios~\citep{Shapiro2005}.}
\label{fig:competition_speed}
\end{figure}

The existence of a second unstable mode implies the possibility of mode competition,
in which the most dangerous mode changes type, from being interfacial to
internal.
In Figs.~\ref{fig:competition}--\ref{fig:competition_speed} we demonstrate how this competition can be
achieved by decreasing $m$.  This is expected to reduce the importance of
the interfacial mode relative to the internal mode, since $TAN\propto m-1$.
The figure does indeed confirm a change in the character of the most dangerous
mode as $m$ is reduced: when $m$ is reduced from $20$ to $5$, the most dangerous
mode becomes internal.  Note that this crossover depends not only on $m$,
but also on $Re_0$: we need first of all to identify a value of $Re_0$ for
which the internal mode is positive, and then carefully select $m$ to observe
mode competition.
%
%
%
%
 Now a similar modal competition has been observed in two-phase mixing layers by \citet{YeckoZaleski2002},
where again, the mode competition is a function of the viscosity contrast.
 What these examples share is the control of the modal competition by
 a parameter that requires a change in the properties of the two fluids under
 investigation.  There does however, exist a situation in which the mode
 competition can be engendered by a change in the flow properties (more precisely,
 a change in the properties of the turbulence), rather than in the fluid
 properties.  It is to this example that we now turn.

\subsection{Modelling surface roughness}
\label{subsec:roughness}

In this section we examine the effect of surface roughness on the internal
and interfacial modes.  Surface roughness is modelled by two distinct approaches.
 Now it is not inconsistent to
examine the effects of surface roughness on wave growth: the origin of the
surface roughness is not found in the waves we study, but rather in instantaneous
pressure fluctuations at the interface that give rise to a roughened interface,
where the vertical extent of the roughness elements is proportional to the strength of these pressure fluctuations.
Such fluctuations appear in the work of \citet{Phillips1957}, and direct numerical simulations by \citet{lin2008} indicate that these pressure
fluctuations are a precursor to the
 exponential wave growth we have described here.  As mentioned in Sec.~\ref{sec:flat},
 we have two distinct models for the surface roughness.  In the first
\begin{figure}
\begin{center}
\subfigure[]{
\includegraphics[width=0.3\textwidth]{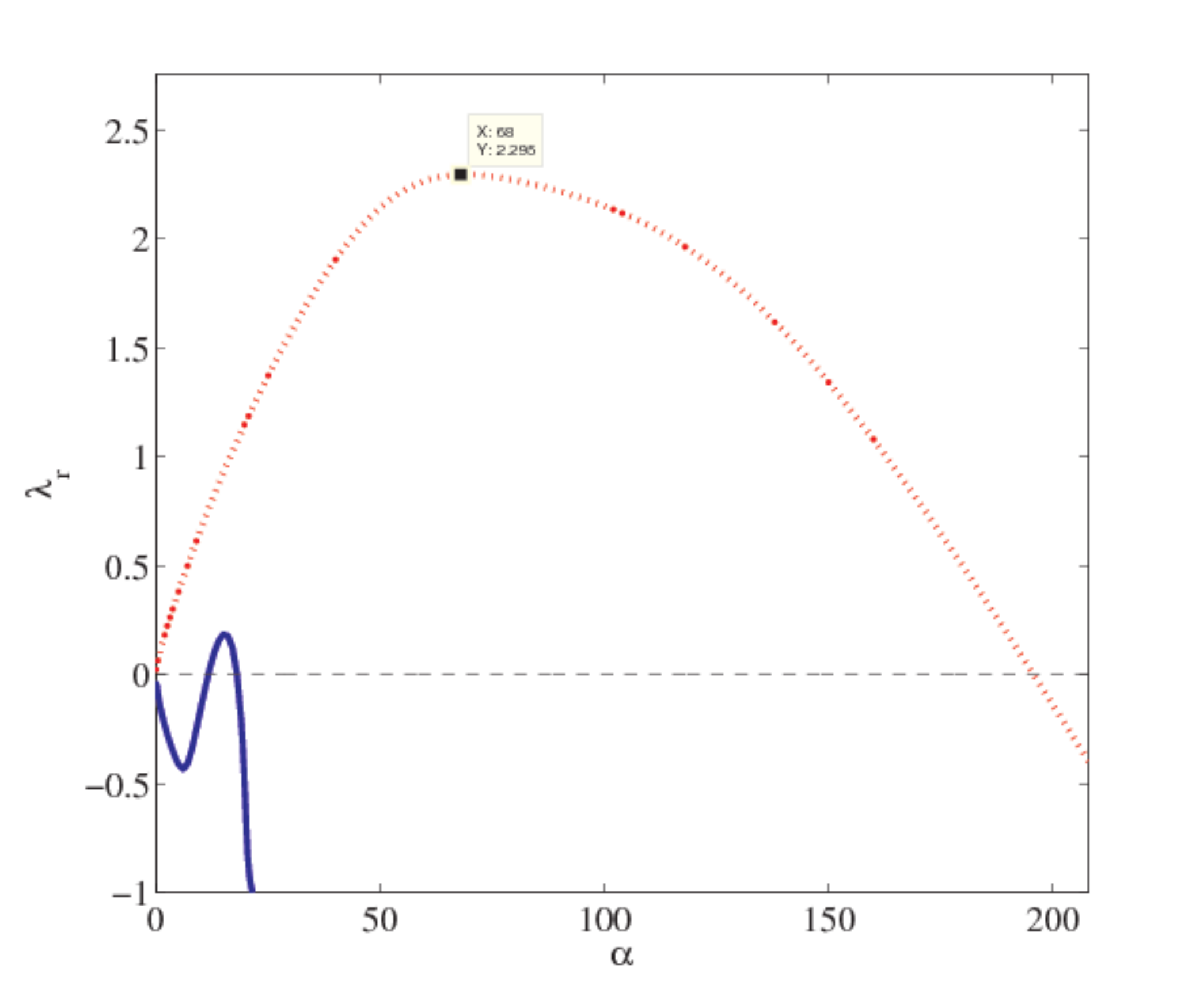}
}
\subfigure[]{
\includegraphics[width=0.3\textwidth]{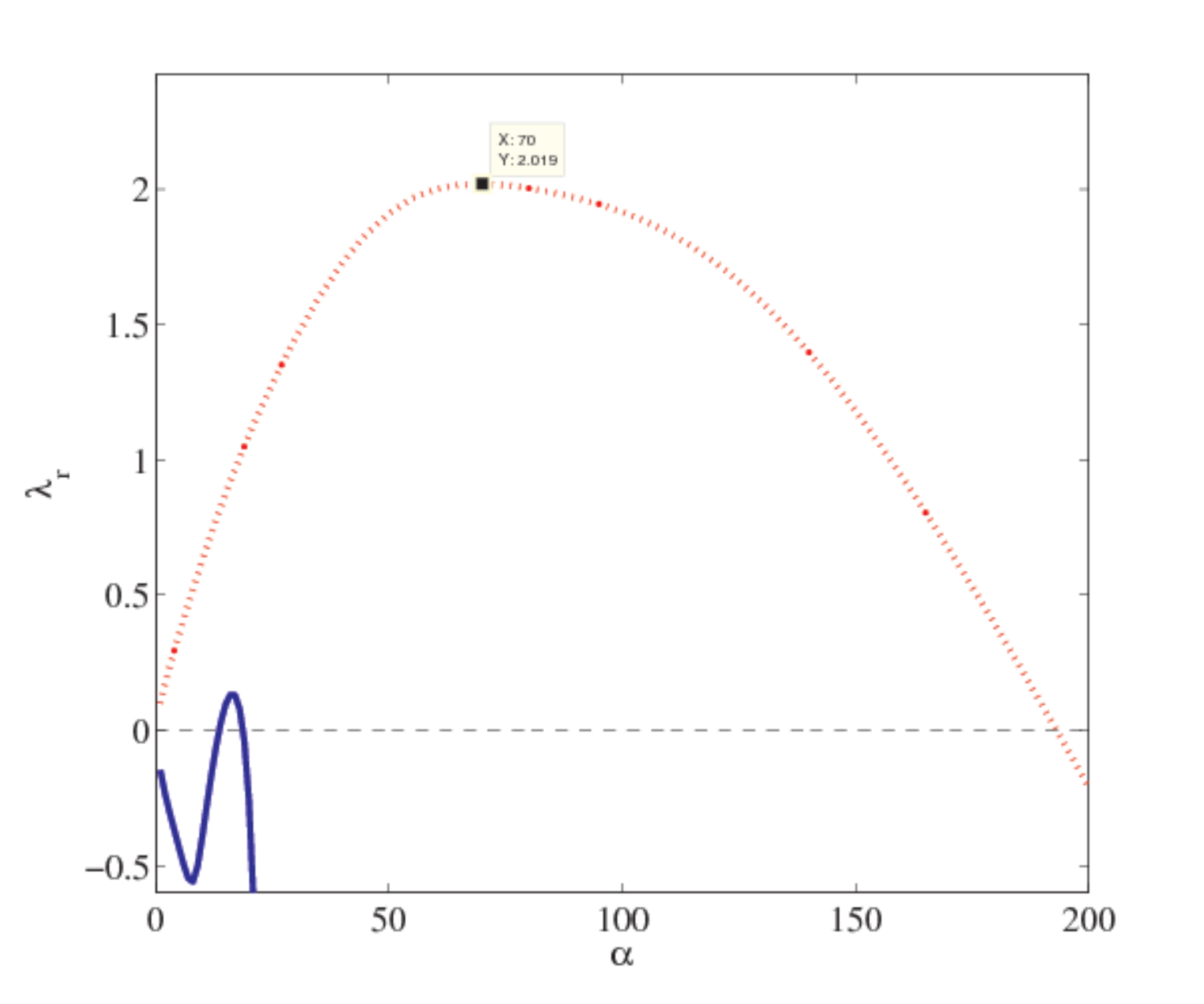}
}
\subfigure[]{
\includegraphics[width=0.3\textwidth]{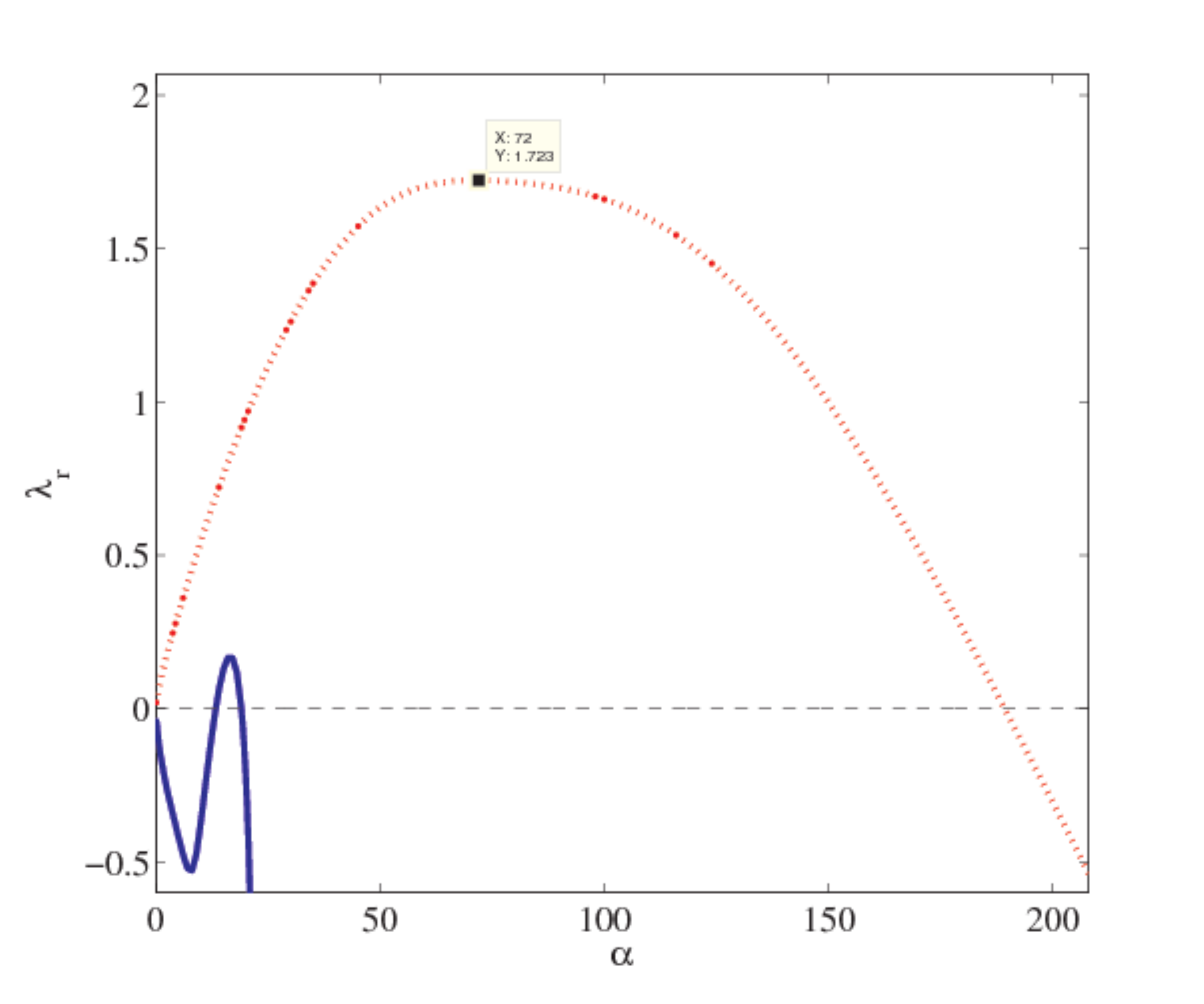}
}
\end{center}
\caption{Dependence of the growth rate on the depth of the interfacial viscous sublayer, $s=\left(5,2,1\right)d_*$, where $d_*=\nu_G/U_{*\mathrm{i}}$.
 By comparing (a) and (b), we see that decreasing the viscous-sublayer thickness
 decreases the maximum growth rate
 of both the internal and the interfacial modes.  The viscous sublayer cannot be reduced further than the value implied by $A_\mathrm{vd}=0$.  Thus, any further increase in $s$ beyond a certain small value $s\approx d_*$ has no effect on the base-state profile, and the growth rate is thereafter unaffected by changes in $s$.
  Here we have set $\left(m,r,\delta,Re_0\right)=\left(55,1000,0.05,4000\right)$.}
\label{fig:competition1}
\end{figure}
 case, we use the smooth-interface model, with a reduced viscous sublayer
 thickness.  Such an approach has been used before~\citep{Morland1993}, where
 it was observed that the reduced viscous sublayer produces a reduced wave
 growth rate.  The second model we use is a modified version of that of \citet{Biberg2007},
 where the eddy-viscosity contains an explicit roughness parameter $K=\ell_{\mathrm{i}}/\left(\kappa
 h\right)$, where $\ell_{\mathrm{i}}$ is the mean height of the roughness
 elements.  We compare these two approaches in this section.

Fig.~\ref{fig:competition1} shows the effect of the viscous-sublayer thickness
on the stability.  The growth rate of the interfacial and internal modes
is shown for viscous-sublayer thicknesses $5d_*$, $2d_*$, and $d_*$.  These different values are obtained from the base-state model by
changing the Van Driest coefficient $A_{\mathrm{vd},\mathrm{i}}$.  Here,
$d_*=\nu_G/U_{*\mathrm{i}}$ is the wall unit based on the interfacial friction
velocity $U_{*\mathrm{i}}$.  From the figure, we see that decreasing the
viscous sublayer thickness decreases the maximum growth rate of both modes. 
%
%
%

Next, we turn to the Biberg model of interfacial roughness.
\begin{figure}
\begin{center}
\subfigure[]{
\includegraphics[width=0.3\textwidth]{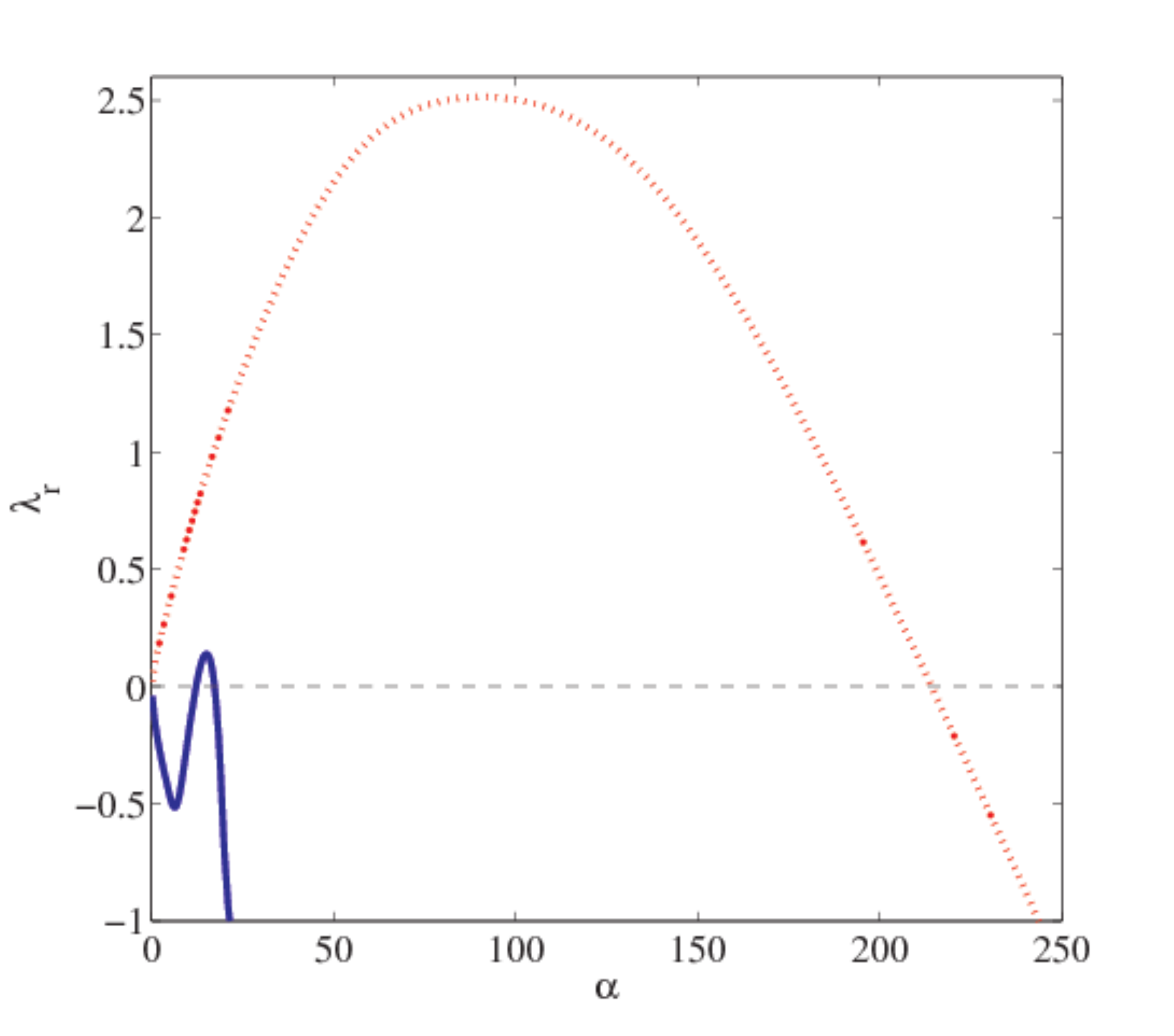}
}
\subfigure[]{
\includegraphics[width=0.3\textwidth]{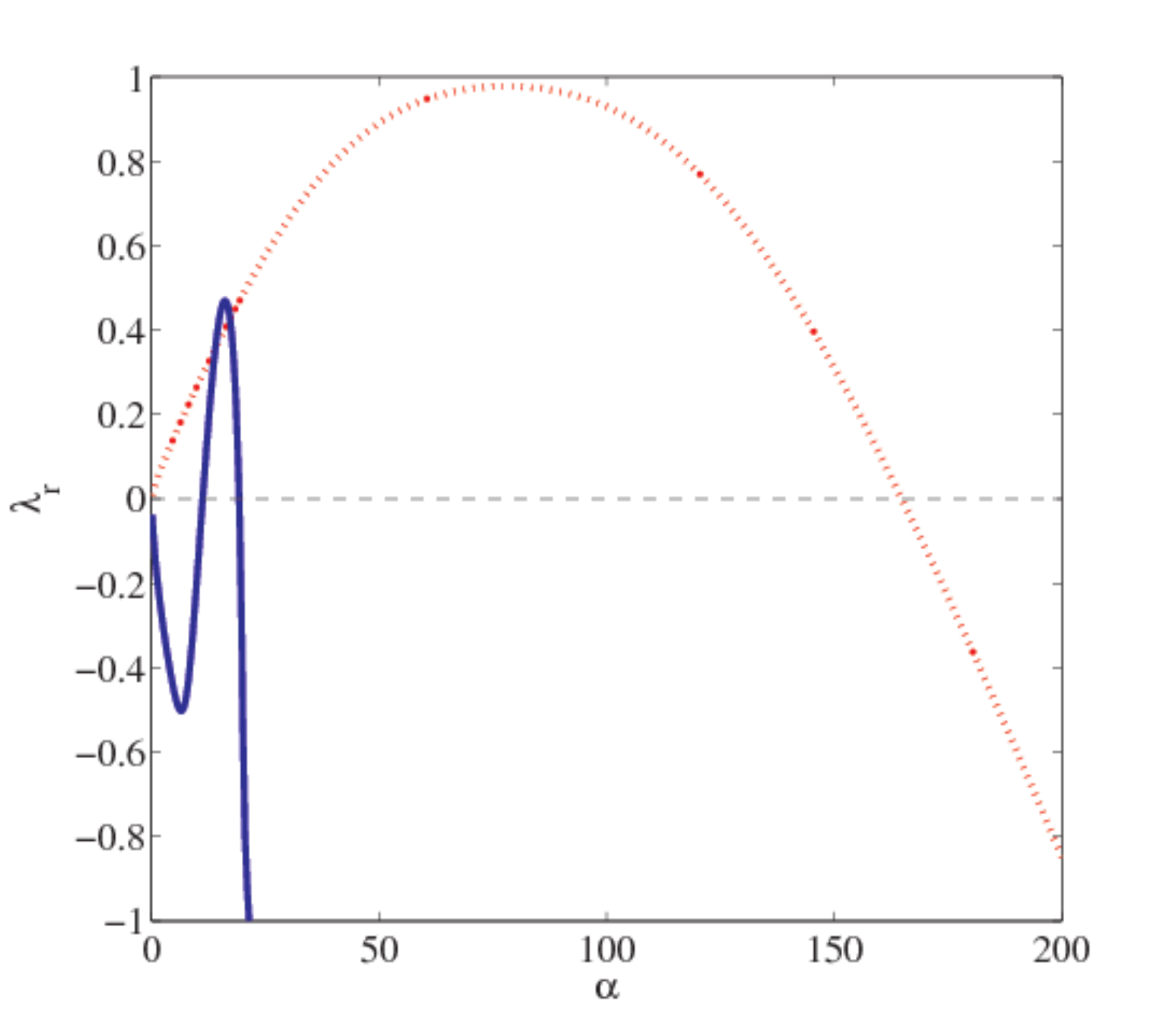}
}
\subfigure[]{
\includegraphics[width=0.3\textwidth]{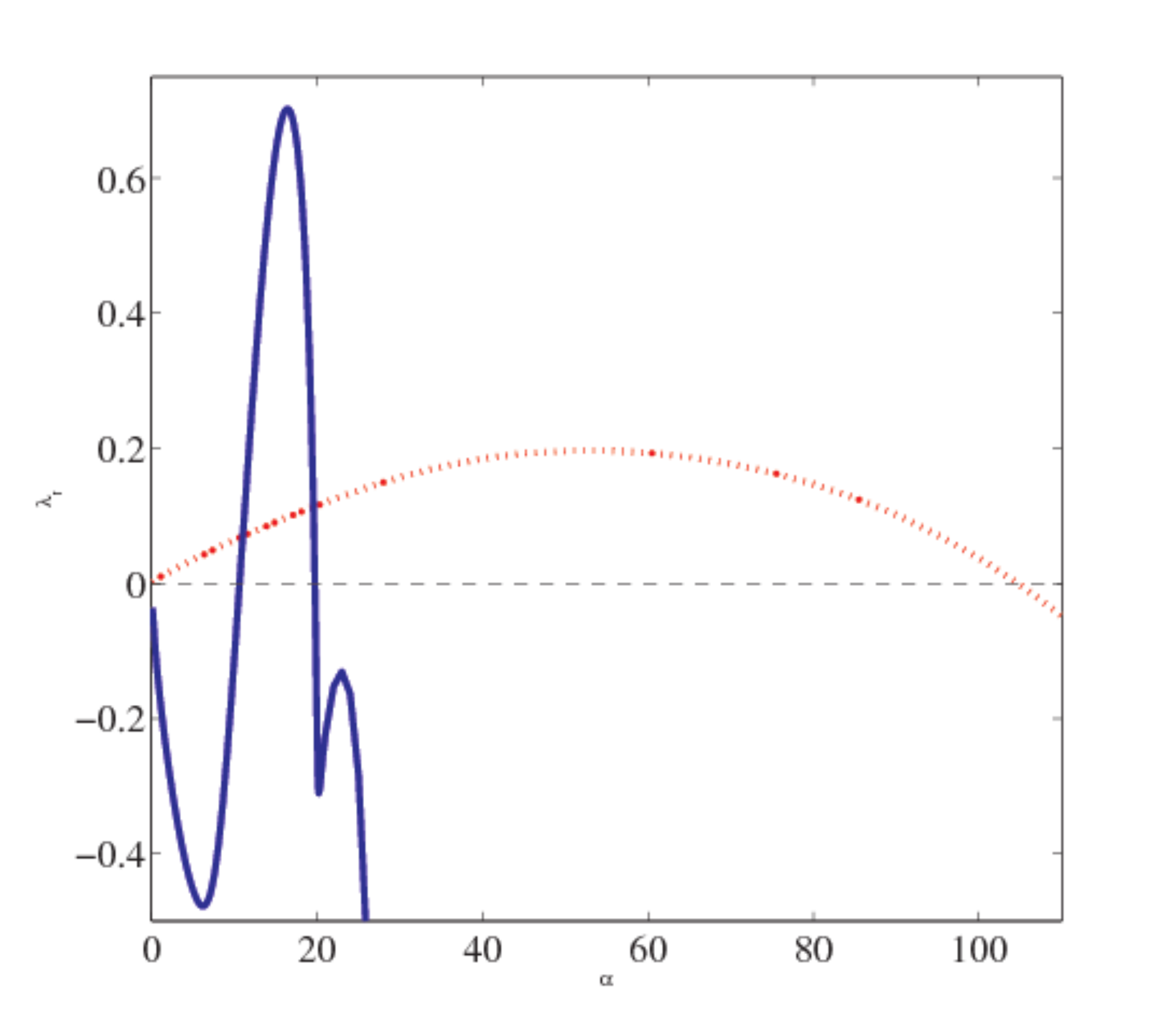}
}
\end{center}
\caption{Dependence of the growth rate on the surface-roughness parameter,
for $K=0,0.001,0.005$ respectively.  Increasing $K$ decreases the maximum
growth rate of the interfacial mode, and increases the maximum growth tate
of the internal mode, to such an extent that the most dangerous mode is internal.
 This crossover happens for $K\apprge0.001$, as in (b).   Here we have set $\left(m,r,\delta,Re_0\right)=\left(55,1000,0.05,4000\right)$.}
\label{fig:competition2}
\end{figure}
Fig.~\ref{fig:competition2} shows the growth rate as a function of the roughness
parameter $K$.  As $K$ increases, the maximum growth rate of the interfacial
mode shrinks dramatically, while the maximum growth rate of the internal
mode increases slightly.  This change is sufficient to promote the maximum
wavenumber-growth rate pair on the internal branch, $\left(\alpha_{\mathrm{max},\mathrm{int}},\lambda_{\mathrm{max},\mathrm{int}}\right)$,
to the status of most dangerous mode.  This crossover occurs for $K\apprge0.001$,
as shown in Fig.~\ref{fig:competition}~(b).
\begin{figure}
\begin{center}
\subfigure[]{\includegraphics[width=0.3\textwidth]{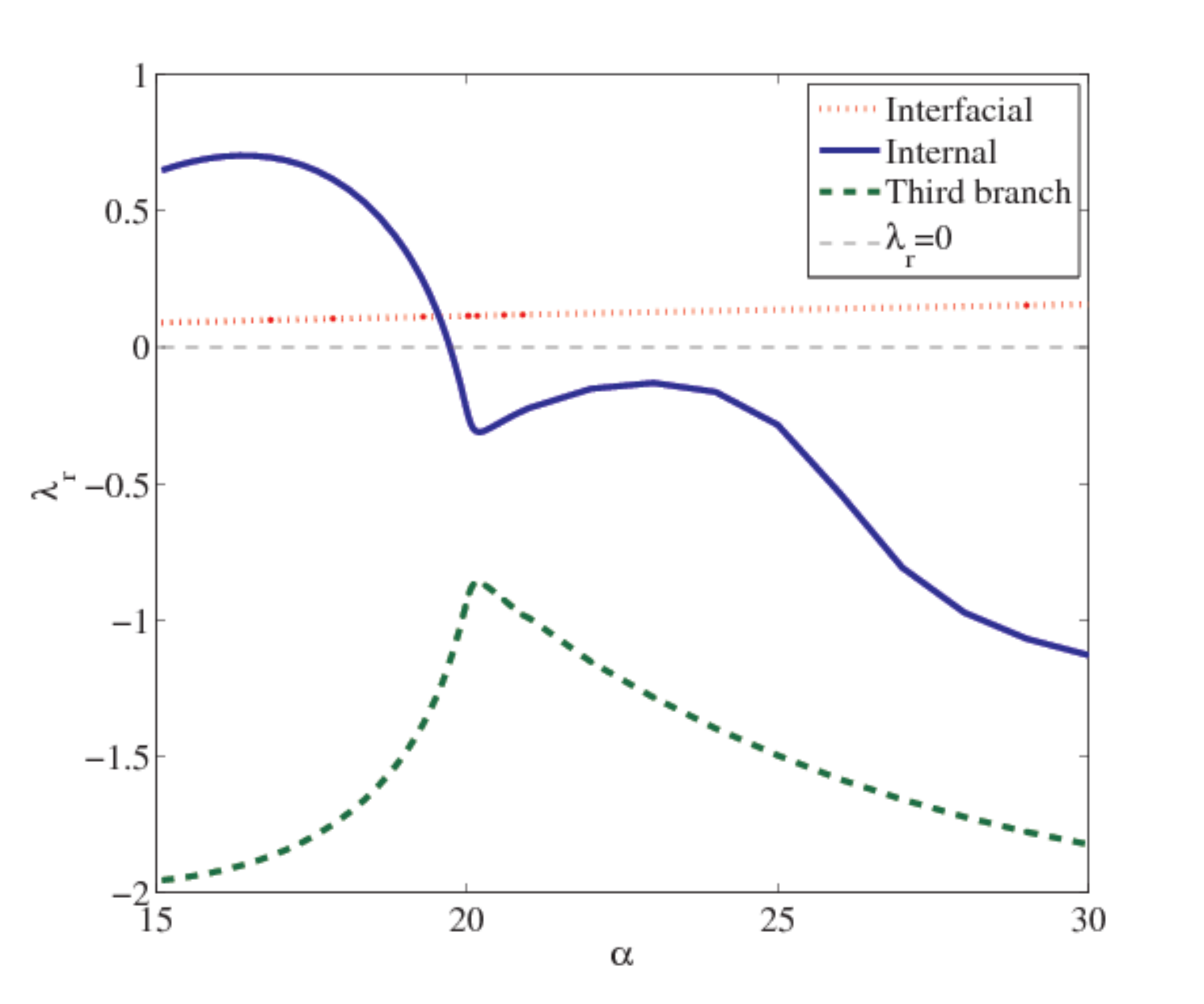}}
\subfigure[]{\includegraphics[width=0.3\textwidth]{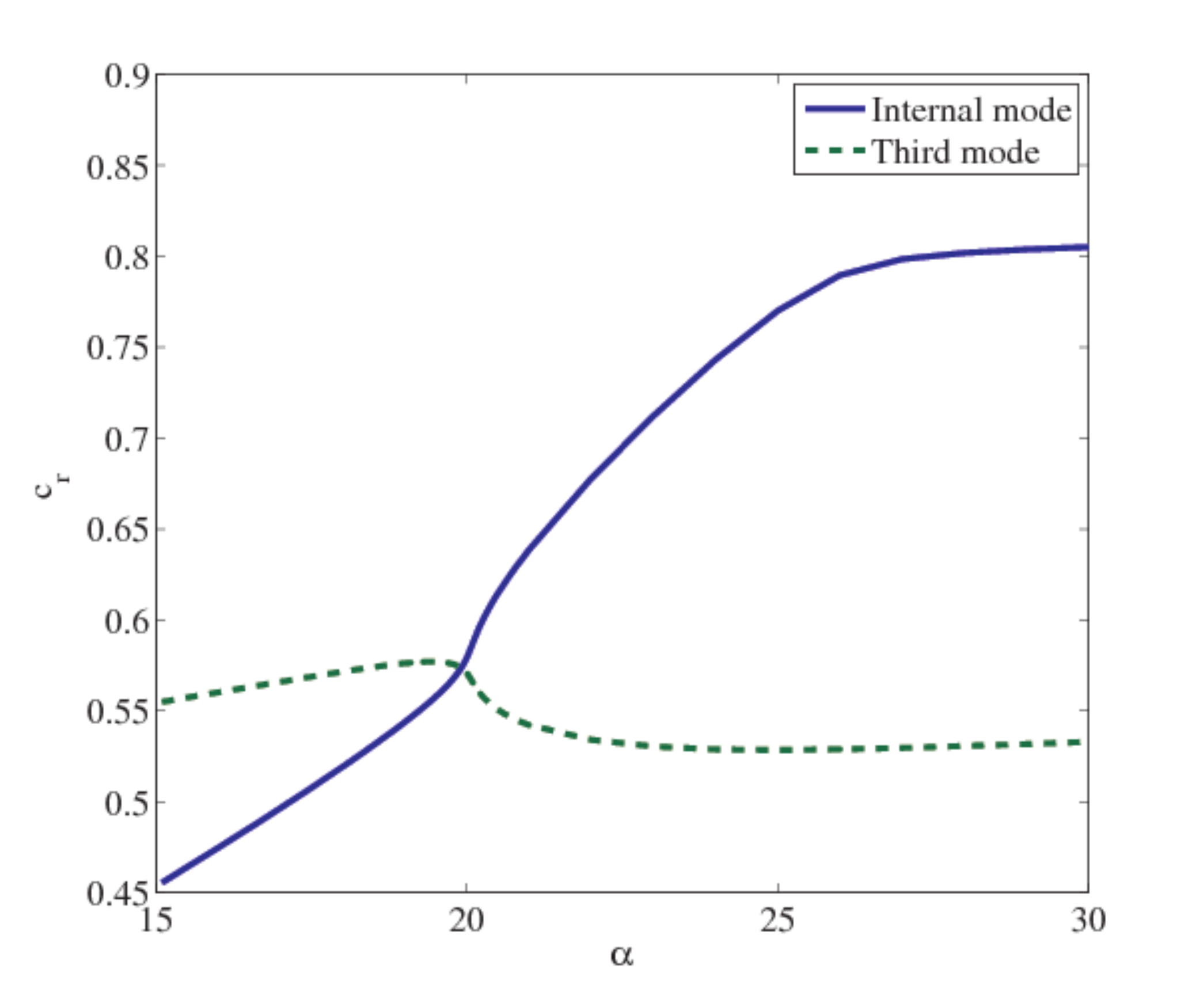}}
\subfigure[]{\includegraphics[width=0.3\textwidth]{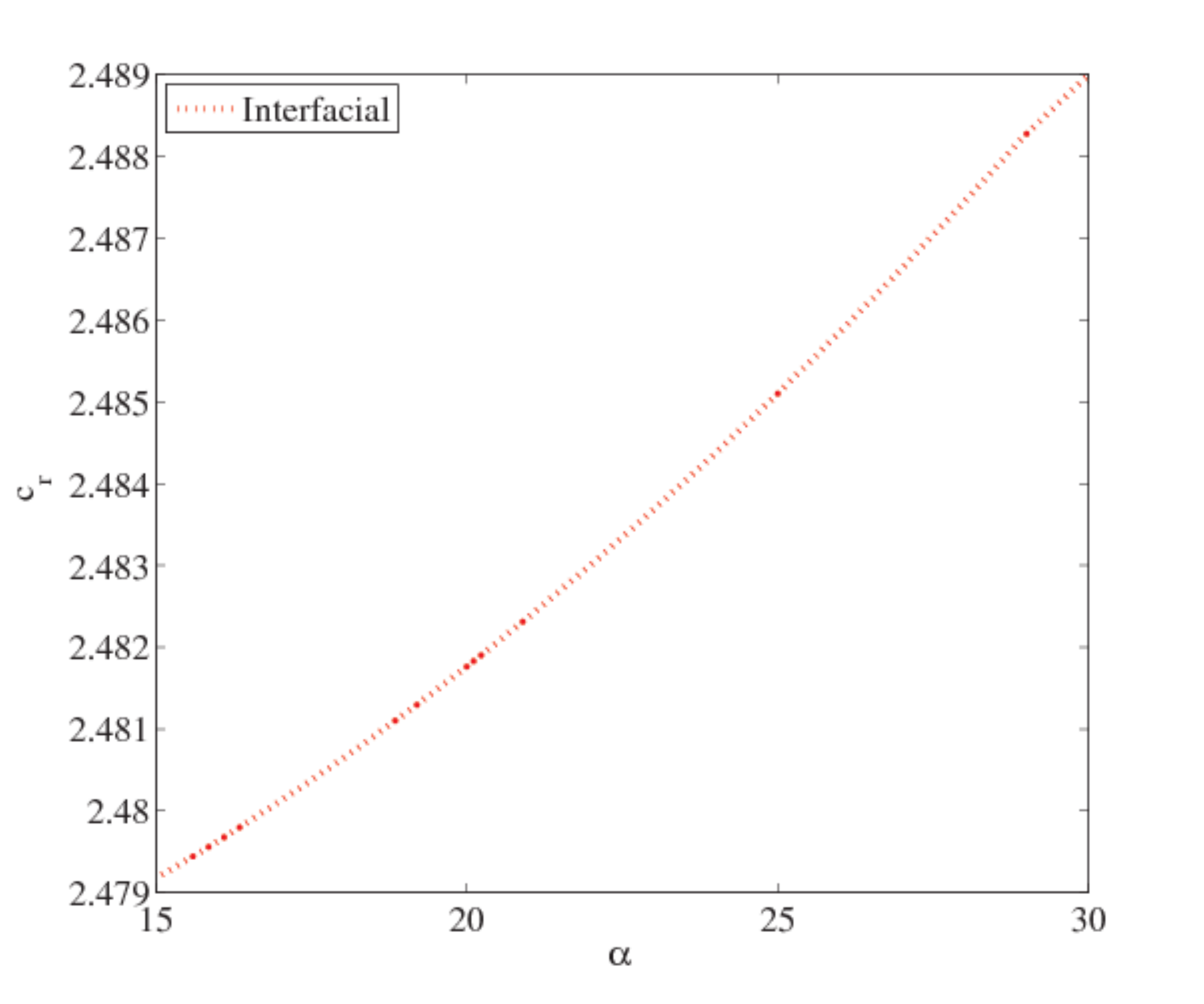}}
\end{center}
\caption{
%
Companion to Fig.~\ref{fig:competition2}~(c).  (a)  The growth rate of the three most dangerous modes at $Re_0=4000$ and $K=0.005$, being an enlarged view of the $\alpha=15$--$30$ region of Fig.~\ref{fig:competition2}~(c).  The sharp kink or extreme point in the growth rate of the internal mode is thus genuine, and not the consequence of mode crossover.  (b) However, by examining the wave speed, we see that the wave speed of the internal and the third most dangerous modes do in fact cross close to the corresponding extreme point of the growth-rate curves.  Finally, (c) shows the wave speed of the interfacial mode.
%
}
\label{fig:competition3}
\end{figure}
In Fig.~\ref{fig:competition2}, the dispersion curve of the internal mode possesses a local minimum near $\alpha\approx20$, similar to Fig.~\ref{fig:competition}~(c).  To verify that this is not due
to a crossover between the second and third modes, we have plotted the three
least negative modes for $K=0.005$ in Fig.~\ref{fig:competition3}~(a).  The second
and third least negative modes are well separated and a crossover effect
is thus ruled out.  Figs.~\ref{fig:competition3}~(b) and~(c) are plots of the wave speed
for the internal and interfacial modes: the continuity of these curves confirms
that no crossover effect is taking place.  Note, however, that the wave speeds of the second and third most dangerous modes intersect close to the point where the internal mode has its local minimum.  Such phenomena often occur in modal coalescence~\citep{Shapiro2005}.
Note finally that $c_{\mathrm{r}}-U_\mathrm{int}$
is negative for the internal  mode, which shows
that the critical layer is in the liquid for the internal mode.

Since the results in Figs.~\ref{fig:competition1} and~\ref{fig:competition2}
are not identical, the two models of interfacial roughness discussed here
are obviously inequivalent.  Which, therefore, is the correct description?
 Reducing the depth of the viscous sublayer, and thus enhancing the extent
 of the logarithmic layer, is clearly a crude model for interfacial roughness.
  The level of detail in the Biberg model is superior, and the predictions
  of this model for a base state with finite roughness agree well with experiments, as explained in
  his paper~\citep{Biberg2007}.  Our prejudice is thus towards the latter
  model, and we therefore expect surface roughness to modify the stability properties
  of the system through mode competition.  However, this contention must
  ultimately be confirmed by DNS, and by experiments.  Although these studies
  are beyond the scope of the present work, we are able to test the predictions of the flat-interface model against experiments, which we do in the next section.


\section{Comparison with other work}
\label{sec:experiments}

In this section, we compare our results with some of the
experimental data from the literature, in particular the work of \citet{CohenHanratty}, and \citet{Craik}.  We
also compare our findings with a model that is frequently used  in
practical applications to predict flow-regime transitions, namely
the viscous Kelvin--Helmholtz theory.  To do this, we refer to
Fig.~\ref{fig:vary_Re1}~(a), which highlights the importance of the
Reynolds number in the stability analysis.  In that figure, the
Reynolds number is varied and the other parameters are held fixed.
For sufficiently large values of $Re_0$, the dispersion curve
associated with the interfacial mode is paraboloidal, with critical
wavenumbers at $\alpha_{\mathrm{c},\mathrm{l}}=0$, and
$\alpha_{\mathrm{c},\mathrm{u}}>0$.  As the Reynolds number
decreases, the the growth rate develops an intermediate
 critical wavenumber $\alpha_{\mathrm{c},\mathrm{u}_0}$, where $0< \alpha_{\mathrm{c},\mathrm{u}_0}< \alpha_{\mathrm{c},\mathrm{u}}$.
 At a critical $Re_0$-value, the maximum growth
rate is zero, and thus the intermediate critical wavenumber is simultaneously
a maximum, and a zero, of the function $\lambda_{\mathrm{r}}\left(\alpha\right)$.   Finally, below this critical $Re_0$-value, the growth rate is negative everywhere.
The experimental works we reference involve a similar path through paramter space.

 \subsection{Comparison with experiments}
\label{subsec:experiments}

\citet{CohenHanratty} report critical Reynolds numbers for millimetre-thick liquid films.
 They observe the development of two-dimensional waves above a critical Reynolds
 number.  They call these waves `fast', in the sense they move at a velocity
 that exceeds the interfacial velocity.
\begin{table}
  \begin{center}
  \begin{tabular}{|l|c|c|c|c|c|c|c|}
\hline
&$d_L$ (mm)&$Re_{CH}$&$Re_{CH}$ (exp)&$c_{\mathrm{r}}/\overline{U}_G$&$c_{\mathrm{r}}/\overline{U}_G$
(exp)&$\ell$ (inches)&$\ell$ (inches,exp)\\
\hline
\hline
(1)&1.89&3810&4050&0.13&0.08&1.1&0.9\\
(2)&3.54&2650&2760&0.15&0.15&0.7&1.2\\
(3)&4.91&1930&1980&--&0.19&0.9&--\\
\hline
\end{tabular}
\vskip 0.1in
\begin{tabular}{|c||c|c|c|c|c|c|c|c|}
\hline
$\alpha$&$KIN_G$&$KIN_L$&$REY_L$&$REY_G$&$DISS_L$&$DISS_G$&$NOR$&$TAN$\\
\hline
\hline
5.5&0.99&0.01&4.58&1.91&-21.32&-31.44&-1.55&48.82\\
\hline
\end{tabular}
\caption{Comparison with the work of \citet{CohenHanratty}.  There is excellent
agreement between the theory and the experiments.  One should note however,
that the experiments carry a margin of error of up to $20\%$.  Thus, the
agreement between the critical Reynolds numbers is both indicative of the
correctness of our theory, and possibly a little fortuitous.  The sub-table is a theoretical energy-budget
calculation related to experiment (3).  The instability is viscosity-induced,
although there are contributions from $REY_L$ and $REY_G$ too.}
\label{tab:experiments1}
\end{center}
\end{table}
These waves are, however, in our classification, `slow' (or on the boundary
beween `slow' and `fast'), since the theoretical
values computed are $c_\mathrm{r}/U_0\apprle1$, and thus the viscosity-contrast
instability is expected.  We show a comparison between the theoretical predictions
of our model and the measurements of Cohen and Hanratty in Tab.~\ref{tab:experiments1}.
 Our estimates for the critical Reynolds number $Re_{CH}=\rho_Gh\overline{U}_G/\mu_G$ are in close agreement with
 the experimental values.  We are mindful, however, that the margin of error
\begin{figure}[htb]
\centering\noindent
\subfigure[]{\includegraphics[width=0.45\textwidth]{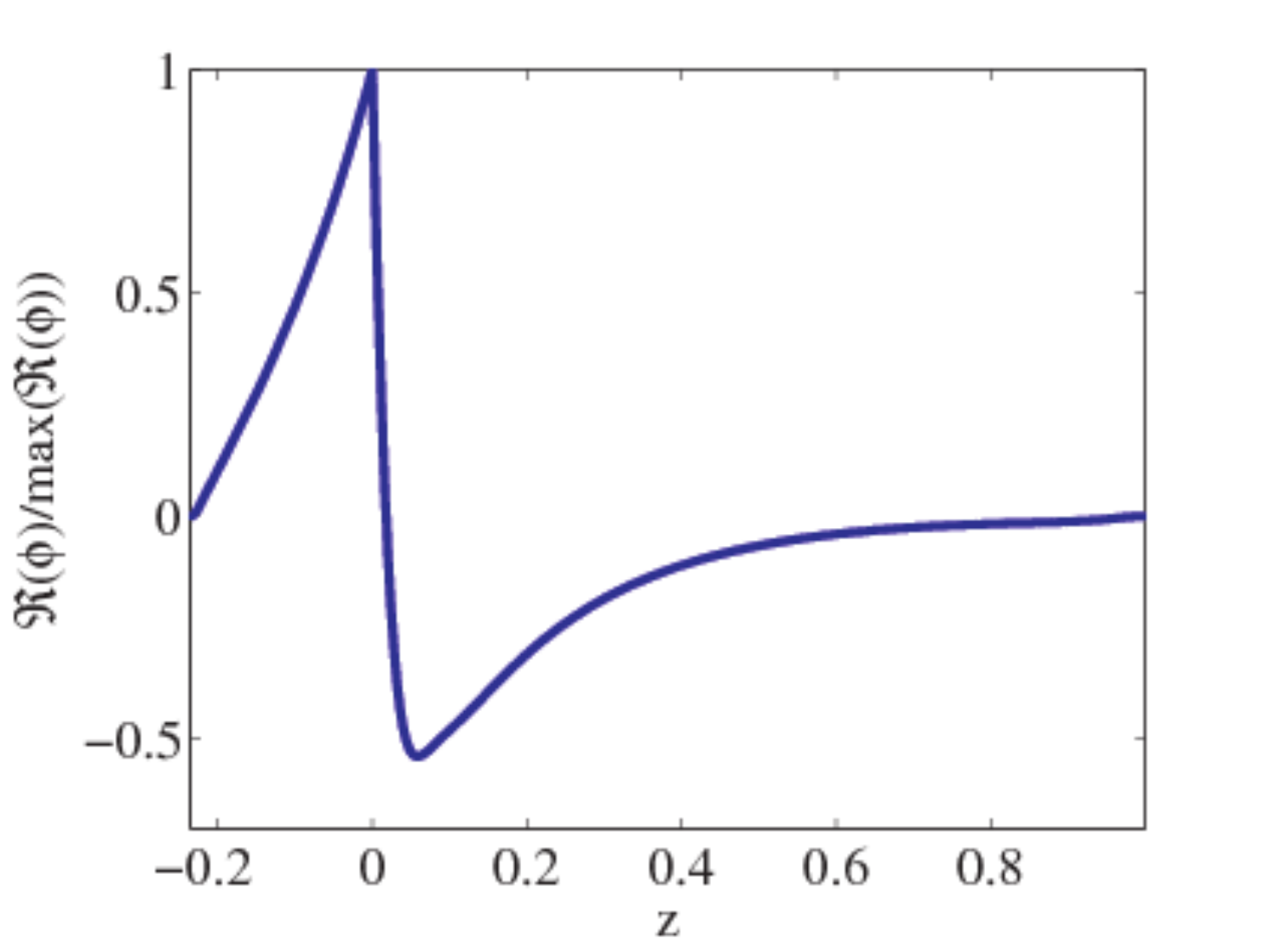}}
\subfigure[]{\includegraphics[width=0.41\textwidth]{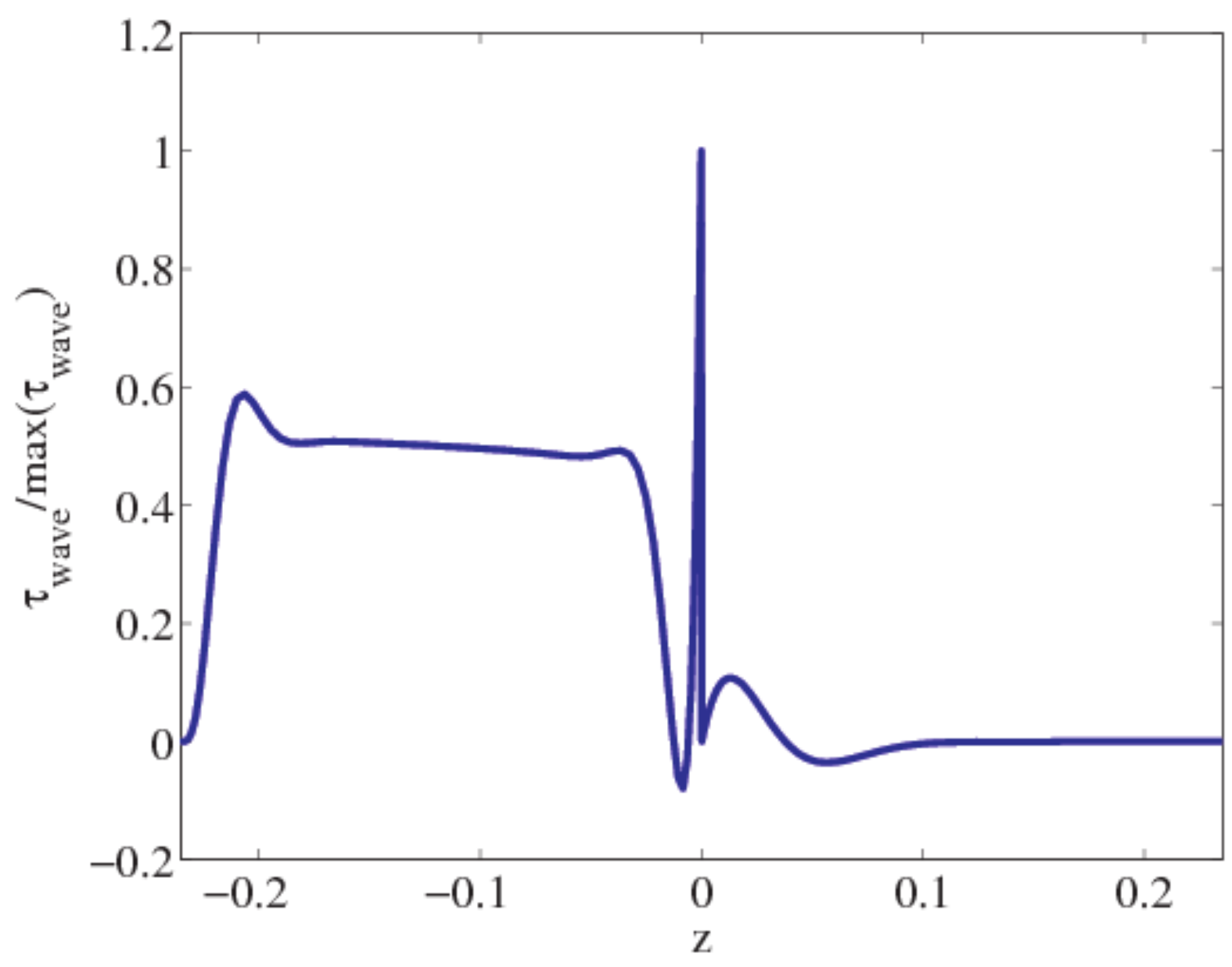}}
\caption{Theoretical calculation based on the parameters in experiment (5)
of Cohen and Hanratty.  (a)  The streamfunction; (b) the wave Reynolds stress.
 In (b) we see a contribution to the instability from transfer terms in the
 liquid and the gas, although the viscosity-contrast across the interface
 gives the most important contribution to the energy of instability.}
\label{fig:streamfunction_cohen}
\end{figure}
 stated in the experiment is between $10\%$ and $20\%$.  We have also compared
 our theoretical model with the measurements of the critical wavelength and
 wave speed.  There is excellent agreement between the theoretical and experimental
 values for the wave speed.
 The spread in values of the
 critical wavelength is larger, although this is acceptable, in view of the
 large error attached to the experimental measurements.  The energy budget
 in Tab.~\ref{tab:experiments1} is based on a theoretical calculation, with
 parameters taken from experiment (3).  The corresponding streamfunction and the wave Reynolds
 stress function are presented in Fig.~\ref{fig:streamfunction_cohen}.  The instability is confirmed to be
 due to the viscosity-contrast mechanism.

\citet{Craik} performs a similar experiment with liquid
films thinner than those found in Cohen and Hanratty.  He reports critical conditions for the generation of unstable waves.
 The trend in the data agrees with that in the theoretical calculations,
 although the quantitative agreement is poor.  Craik explains that waves
\begin{table}
  \begin{center}
  \begin{tabular}{|l|c|c|c|c|c|}
\hline
&$d_L$ (mm)&$Re_{Cr}$&$Re_{Cr}$ (exp)\\
\hline
\hline
(1.1)&0.128&$20$&$30$\\
(1.2)&0.230&$68$&$61$\\
(1.3)&0.218&$66$&$71$\\
(1.4)&0.355&$110$&$140$\\
(1.5)&0.307&$94$&$140$\\
\hline
\end{tabular}
\vskip 0.1in
\begin{tabular}{|l|c|c|c|c|c|c|c|}
\hline
&$d_L$ (mm)&$Re_{Cr}$&$Re_{Cr}$ (exp)&$c_{\mathrm{r}}/U_{\mathrm{int}}$&$c_{\mathrm{r}}/U_{\mathrm{int}}$
(exp)\\
\hline
\hline
%
(2.1)&0.535&$35$&$94$&1.1&1.75\\
(2.2)&0.665&$50$&$89$&1.0&1.9\\
(2.3)&0.820&$56$&$91$&1.1&1.8\\
\hline
\end{tabular}
\caption{Comparison with Table 1 (p. 375) and Table 2 (p. 378) in the work of \citet{Craik}.  There is good agreement between the theory and the experiment in the first case, and only very rough
agreement in the second case.  As explained in the experimental paper, a sharp transition to wavy flows was not observed  in this second case, which explains these quantitative differences.
We do not refer to experiments (2.4)--(2.5), wherein our model predicts laminar gas flow.   }
\label{tab:experiments2}
\end{center}
\end{table}
 are observed for film thickness below that quoted in experiment (1.1), although
 the uniform thin film of liquid is difficult to maintain under these conditions.
  It is possible that the thinness of the film inhibits precision in the
  measurement at film thickness above this lower bound too.  Craik also explains
  that accurate measurements of wave speed were difficult owing to the long
  wavelengths of the observed waves (compared to the channel length).  These
  are sources of error that explain why there is only qualitative agreement
  between the theoretical and experimental data.
\begin{table}
  \begin{center}
\begin{tabular}{|c|c||c|c|c|c|c|c|c|c|}
\hline
&$\alpha$&$KIN_G$&$KIN_L$&$REY_L$&$REY_G$&$DISS_L$&$DISS_G$&$NOR$&$TAN$\\
\hline
\hline
(1.5)&0.3&1.00&0.00&0.00&-165.56&-2.47&-1359.25&-0.09&1528.37\\
(2.1)&0.02&1.00&0.00&0.00&-45.06&-3.23&-701.34&-0.14&750.76\\
\hline
\end{tabular}
\caption{Theoretical energy-budget
calculations related to experiments (1.5) and (2.1) of Craik.  The instability is viscosity-induced,
and there are no other contributions to the instability, unlike in the Cohen
data.}
\label{tab:experiments3}
\end{center}
\end{table}

We also perform theoretical calculations based on the parameters in experiments
(1.5) and (2.3), to examine that character of the unstable waves.  We provide
the energy budgets associated with these calculations in Tab.~\ref{tab:experiments3}.
 The instability is interfacial: the contribution from $REY_L$
 and $REY_G$ present in the Cohen data are absent here.  This makes sense:
 $REY_L$ should be negligible because the liquid layer is so thin and thus
 $\phi_L$ cannot contribute meaningfully to the dynamics, while $REY_G$ is
 unimportant because the waves are slow ($c_\mathrm{r}/U_0$ is $O(10^{-2})$
 or $O(10^{-1})$ for the film thicknesses and Reynolds numbers considered).

 \subsection{Comparison with Viscous Kelvin--Helmholtz theory}
\label{subsec:vkh}

\begin{figure}[htb]
\centering\noindent
\includegraphics[width=0.45\textwidth]{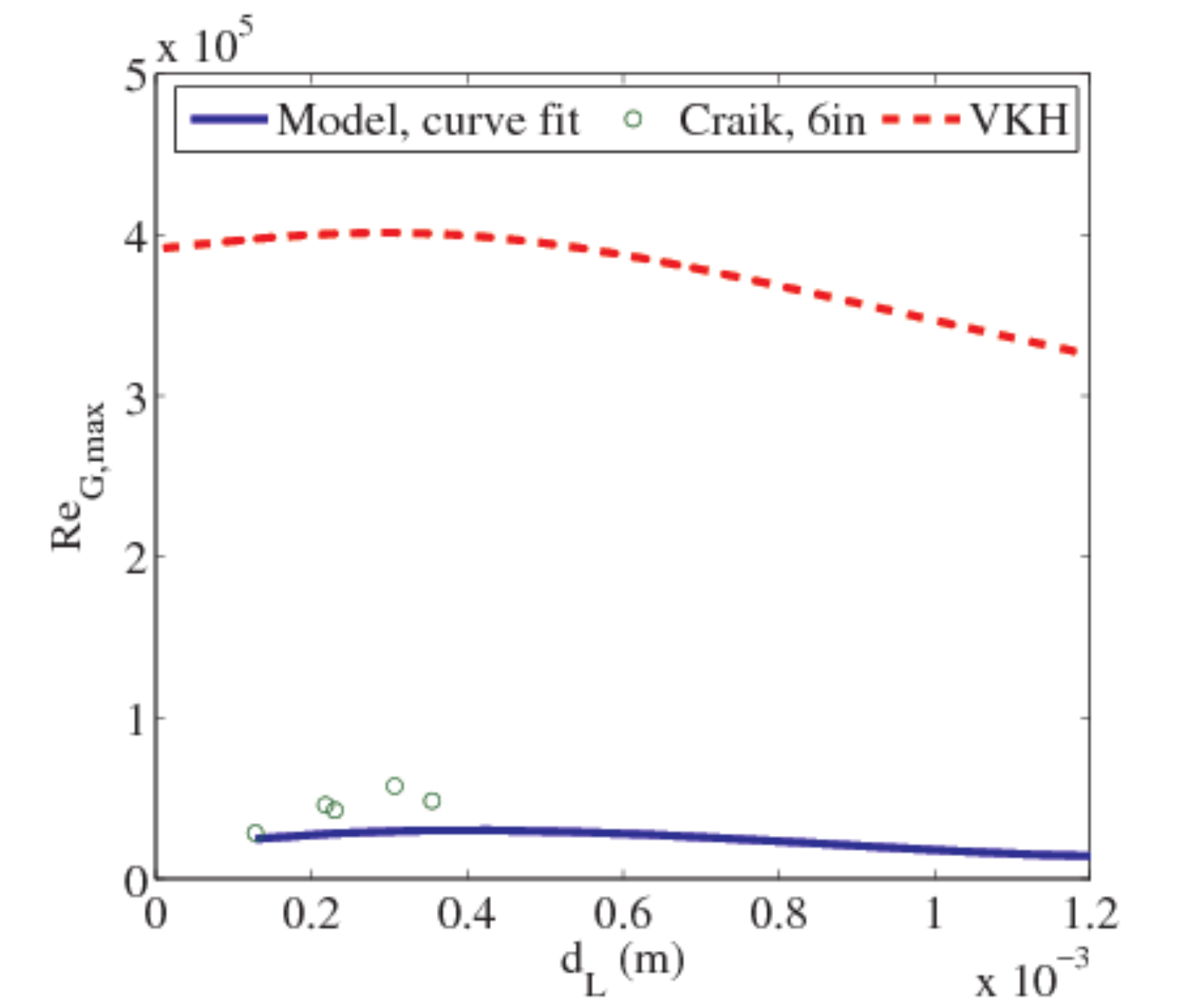}
\caption{Comparison with the viscous Kelvin--Helmholtz model: plots of the liquid film thickness $d_L$ against the Reynolds number $Re_{G\mathrm{max}}=\rho_GU_{G\mathrm{max}}d_G/\mu_G$. The viscous Kelvin--Helmholtz model overpredicts the stability boundary by an order of magnitude.  Better agreement is obtained between the data of Craik  for the $6\,\mathrm{in}$ channel and our theoretical model, although the margin of error in the experimental data is large.
}
\label{fig:vkh_compare}
\end{figure}
In this section, we compare our predictions with those obtained using viscous Kelvin--Helmholtz theory~\citep{barnea1991}.  This is a simplified theory for the interfacial instability of two-phase turbulent flow, and takes account of turbulence in either or both phases.  It is commonly used in one-dimensional models for large-scale stratified and slug-flow predictions.  The velocity field enters only through the liquid- and gas-average values, $u_L$ and $u_G$, while the cross-sectional area fractions $\epsilon_L=d_L/\left(d_L+d_G\right)$ and $\epsilon_G=d_G/\left(d_L+d_G\right)$ also play a role.  Then, the complex frequency $\omega$ is obtainable from a quadratic equation:
\[
\omega^2-2\left(x_0\alpha-x_1i\right)\omega+\left(x_2\alpha^2-x_3\alpha^4-ix_4\alpha\right)=0,
\]
where
\begin{eqnarray*}
\rho_*&=&\frac{\rho_L}{\epsilon_L}+\frac{\rho_G}{\epsilon_G},\\
x_1&=&\frac{1}{\rho_*}\left(\frac{\rho_L u_L}{\epsilon_L}+\frac{\rho_G u_G}{\epsilon_G}\right),\\
x_2&=&-\frac{S_{\mathrm{\mathrm{i}}}}{2\rho_*A}\left(\frac{1}{\epsilon_L}+\frac{1}{\epsilon_G}\right)\left(\frac{1}{\epsilon_G}\frac{\partial\tau_\mathrm{i}}{\partial{u_G}}-\frac{1}{\epsilon_L}\frac{\partial\tau_\mathrm{i}}{\partial{u_L}}+\frac{1}{\epsilon_L}\frac{\partial\tau_{\mathrm{i}L}}{\partial{u_L}}-\frac{1}{\epsilon_G}\frac{\partial\tau_{\mathrm{i}L}}{\partial{u_G}}\right),\\
x_3&=&\frac{1}{\rho_*}\left[\frac{\rho_L u_L^2}{\epsilon_L}+\frac{\rho_G u_G^2}{\epsilon_G}-g\left(\rho_L-\rho_G\right)\left(d_L+d_G\right)\right],\\
x_4&=&\frac{\sigma}{\rho_*}\left(d_L+d_G\right),\\
x_5&=&\frac{S_{\mathrm{\mathrm{\mathrm{i}}}}}{\rho_*A}\left(\frac{\partial\tau_{\mathrm{i}L}}{\partial\epsilon_L}-\frac{\partial\tau_\mathrm{i}}{\partial\epsilon_L}+\frac{u_G}{\epsilon_G}\frac{\partial\tau_{\mathrm{i}L}}{\partial{u_G}}-\frac{u_G}{\epsilon_G}\frac{\partial\tau_\mathrm{i}}{\partial{u_G}}-\frac{u_L}{\epsilon_L}\frac{\partial\tau_{\mathrm{i}L}}{\partial{u_L}}+\frac{u_L}{\epsilon_L}\frac{\partial\tau_\mathrm{i}}{\partial{u_L}}\right),
\end{eqnarray*}
and where the viscous stresses are modelled as
\begin{eqnarray*}
\tau_L&=&\tfrac{1}{2}f_L\rho_L u_L^2,\qquad f_L=C_L\left(\frac{D_L u_L}{\nu_L}\right)^{-n_L},\qquad D_L=4d_L,\\
\tau_G&=&\tfrac{1}{2}f_G\rho_G u_G^2,\qquad f_G=C_G\left(\frac{D_G u_G}{\nu_G}\right)^{-n_G},\qquad D_G=2d_G,\\
\tau_\mathrm{i}&=&\tfrac{1}{2}f_\mathrm{i}\rho_G\left(u_G-u_L\right)|u_G-u_L|,\\
\tau_{\mathrm{i}L}&=&\frac{\tau_L\epsilon_G-\tau_G\epsilon_L}{\epsilon_L+\epsilon_G}.
\end{eqnarray*}
The coefficients $C_G$ and $C_L$ both take the value $0.046$ for turbulent flow and $16$ for laminar flow, while $n_L$ and $n_G$ both take the value $0.2$ for turbulent flow, and $1.0$ for laminar flow.  Finally, the interfacial friction factor $f_\mathrm{i}$ is assumed to be constant and equal to $0.0142$ (see \citet{barnea1991}).
We plot the stability boundary predicted by this theory in Fig.~\ref{fig:vkh_compare}, and compare the results with the Craik data for the $6\,\mathrm{in.}$ channel (Tab.~\ref{tab:experiments2}), and with a curve fit based on a number of points obtained from our theoretical calculations.  The viscous Kelvin--Helmholtz model overpredicts the critical Reynolds number compared with both the data of Craik and our theoretical model by an order of magnitude (Fig.~\ref{fig:vkh_compare}), which casts doubt on the usefulness of such a depth-averaged model.  Our model gives better agreement with the data of Craik, although we are mindful of the uncertainty in these experimental data. Nevertheless, both our theoretical calculations and the experimental data demonstrate the unstable-stable-unstable transition that arises when the film depth is increased, holding the Reynolds number fixed.  This is the statement that our theoretical curve in Fig.~\ref{fig:vkh_compare} is non-monotonic (the non-monotonicity in the model curve is masked somewhat by the large scale necessary to show the VKH results in the same figure).
 In conclusion,
  the qualitative agreement obtained here, together with the good agreement
  obtained relative to the Cohen data, inspires confidence in our model, while the poor agreement between our data and the predictions of the viscous
Kelvin--Helmholtz theory calls into question the validity of this
depth-averaged model, at least for the kind of thin-film waves
studied here.

\section{Conclusions}
\label{sec:conclusions}

In this paper, we have investigated the stability of an interface separating
a thin laminar liquid layer from a turbulent gas in a channel. We have
approached this problem in two steps: first by generalizing the model of \citet{Biberg2007} to describe the interfacial and wall zones in pressure-driven
two-phase channel flow, and second by studying the linear stability of this
state using an Orr--Sommerfeld analysis of the Reynolds-averaged Navier--Stokes
equations.
Our model has enabled us to investigate the stability of the interface as a function
of various parameters.  In general, the interface becomes unstable due
to a mismatch between the viscosities in the liquid and the gas.
In this work, we have taken into account the perturbation turbulent
stresses (PTS) using two distinct models: in both cases, these
stresses have only a quantitative effect on the  stability results
for the thin liquid layers; for the eddy-viscosity model of the PTS,
this effect is particularly small.  These stress contributions are
therefore ignored throughout the work.   We have provided an
explanation for this null result using an analogy with non-Newtonian
fluids.
This work builds upon previous work in the field (notably that of Miesen and co-workers~\citep{Miesen1995,Boomkamp1996,Boomkamp1997}, and \citet{Kuru1995}) by developing an accurate base-state model, validated against numerous experiments and direct-numerical simulations, and by accounting for the PTS.  These efforts result in excellent agreement with the relevant experiments~\citep{Craik,CohenHanratty}.

Our linear stability analyses evince a definition of slow and fast waves.  A slow wave is one
for which the instability is driven by the viscosity contrast across the
interface; a fast wave derives its energy of instability from the critical
layer.  The phase speed $c_{\mathrm{r}}$ of a slow
waves satisfies $c_{\mathrm{r}}/U_0\apprle1$, where $\rho_G U_0^2=h|\partial
p/\partial x|$.  We have carried out a parameter study to find ways of controlling
the wave speed.  The inverse Froude, inverse Weber, and density numbers control
the wave
speed, as suggested by the formula for free-surface waves in a liquid layer.
 However, in all cases considered, the waves are slow.  For certain values
 of the triple $\left(r,Fr,S\right)$, there is a critical-layer contribution
 to the instability, although this is marginal.

For certain parameter values, we also observe a positive growth rate for
the so-called internal mode, associated with instability that is due both
to the shear content of the liquid and to the tangential stress at the
interface.
By a judicious choice of parameters (in particular, for small viscosity ratios),
the internal mode can be made to dominate over
the interfacial mode, and gives rise to mode competition.  To engineer mode
competition in this way, it is necessary to modify the properties of the
two fluids.  However, by increasing the level of turbulence, the flat interface
roughens, and this also has the effect of diminishing the interfacial mode
relative to the internal mode.  \citet{Morland1993} have
explained previously how roughness can reduce the growth rate of the
wave; here we go further and conjecture that this mechanism can, in addition,
engender mode competition.  However, this result is rather conjectural, and
we simply mention it as a route for future experiments.

\subsection*{Acknowledgements}

This work has been undertaken within the Joint Project on Transient
Multiphase Flows and Flow Assurance.  The Authors wish to
acknowledge the contributions made to this project by the UK
Engineering and Physical Sciences Research Council (EPSRC) and the
following: -- Advantica; BP Exploration; CD-adapco; Chevron;
ConocoPhillips; ENI; ExxonMobil; FEESA; IFP; Institutt for
Energiteknikk; PDVSA (INTEVEP); Petrobras; PETRONAS; Scandpower PT;
Shell; SINTEF; StatoilHydro and TOTAL. The Authors wish to express
their sincere gratitude for this support.

%

\renewcommand{\thefootnote}{\ensuremath{\fnsymbol{footnote}}}

\renewcommand{\theequation}{A-\arabic{equation}}
\renewcommand{\thefigure}{A-\arabic{figure}}
\setcounter{equation}{0}  
\setcounter{figure}{0}
\section*{APPENDIX A}  

For further validation of the base state discussed in Sec.~\ref{sec:flat}, we compare our turbulence modelling with experimental data for flow past a wavy wall, obtained from the papers of \citet{Zilker1976}, and \citet{Abrams1985}.
The curvilinear coordinates necessary for this work were previously introduced by \citet{Benjamin1958}$\,\dagger\,$:\makeatletter{\renewcommand*{\@makefnmark}{}
\footnotetext{$\dagger$ We thank S. Kalliadasis and D. Tseluiko for suggesting the application of this coordinate system to the problem.}\makeatother}
\begin{eqnarray}
\xi&=&x-ia\Phi,\nonumber\\
\eta&=&z-a\Phi,\qquad \Phi=e^{-\alpha z}e^{i\alpha x}.
\label{eq:xieta}
\end{eqnarray}
If the streamfunction has the form
\[
\phi=\int_0^\eta U_0\left(s\right)ds+ aF\left(\eta\right)e^{i\alpha\xi}
\]
(where $U_0$ is the single-phase version of the base state in Eq.~\eqref{eq:U_base}),
then the momentum-balance equation for $F$ is
\begin{subequations}
\begin{equation}
i\alpha \left[\left(\partial_\eta^2-\alpha^2\right)F\left(\eta\right)-U_0''\left(\eta\right)F\left(\eta\right)\right]+\mathcal{C}=
\frac{1}{Re_0}\left(\partial_\eta^2-\alpha^2\right)^2F\left(\eta\right)+\mathcal{R},
\label{eq:OSC}
\end{equation}
where $\mathcal{C}$ is the curvature-related term
\begin{equation}
\mathcal{C}=2i\alpha^2 \, U_0'\left(\eta\right)U_0\left(\eta\right)e^{-\alpha\eta}+\frac{1}{Re_0}e^{-\alpha\eta}\left[4\alpha^2U_0''\left(\eta\right)-2\alpha U_0'''\left(\eta\right)\right],
\end{equation}%
and $\mathcal{R}$ is the Reynolds-stress term:
\begin{multline}
{\mathcal{R}}=\left(\partial_\eta^2+\alpha^2\right)\Big\{\mu_T\left[F''\left(\eta\right)+\alpha^2 F\left(\eta\right)+2\alpha U_0'\left(\eta\right)e^{-\alpha\eta}-2\alpha^2U_0\left(\eta\right)e^{-\alpha\eta}\right]\Big\}\\
+2e^{-\alpha\eta}\left[\alpha\tau_0''\left(\eta\right)-\alpha^2\tau_0'\left(\eta\right)\right],\qquad \tau_0\left(\eta\right)=\mu_T\left(\eta\right)U_0'\left(\eta\right).
%
%
%
%
%
%
\end{multline}%
\label{eq:OSC_C2}%
\end{subequations}%
The function $\mu_T\left(\eta\right)$ is the eddy viscosity.  It is set to zero in the quasi-laminar case, and assigned the form of Eq.~\eqref{eq:tss0} if the perturbation turbulent stresses are considered.
We solve Eqs.~\eqref{eq:OSC_C2} subject to the boundary conditions $F=F'=0$ on $\eta=0$ and on $\eta=1$, which are no-slip conditions on the perturbation $F$.  Although the no-slip condition on the upper boundary is at $z=1$, not $\eta=1$, these planes are close to one another:
the physical boundary $z=1$ corresponds to an $\eta$-value $1-ae^{-\alpha}e^{i\alpha x}$, which is close to unity for large $\alpha$-values.   Thus, for simplicity, we impose a boundary condition at $\eta=1$.

\begin{figure}
\begin{center}
\subfigure[]{\includegraphics[width=0.35\textwidth]{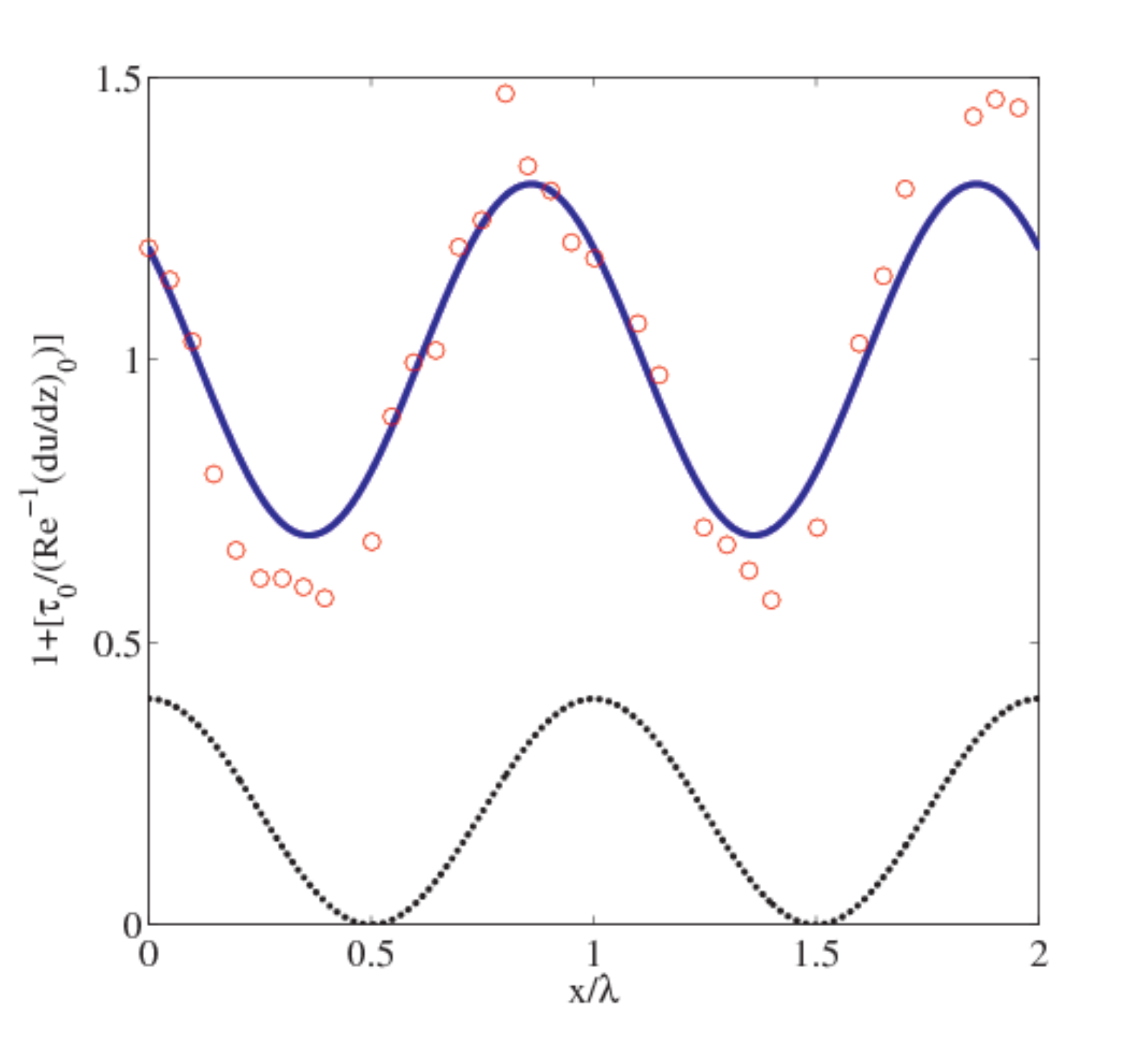}}
\subfigure[]{\includegraphics[width=0.35\textwidth]{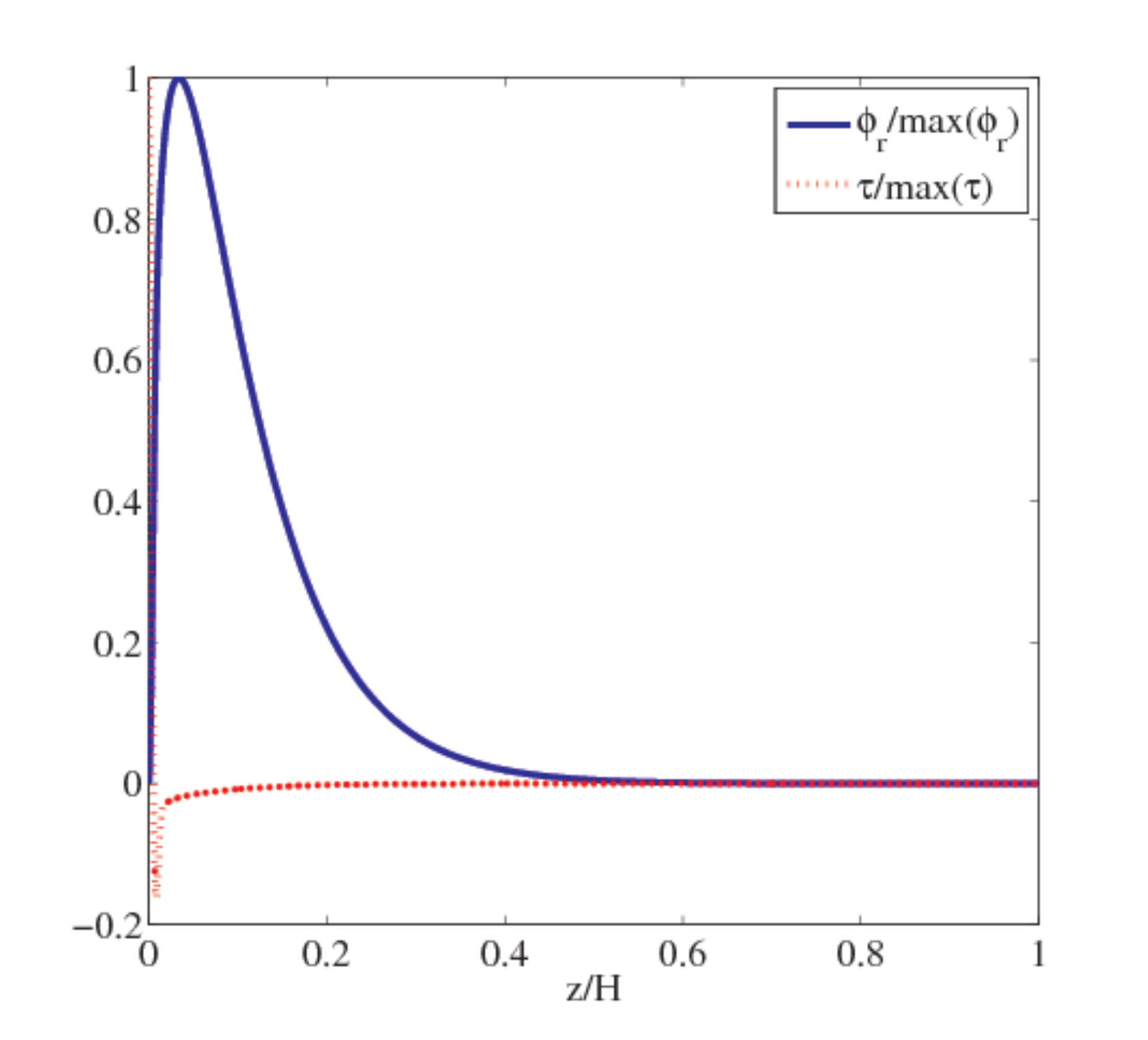}}\\
\subfigure[]{\includegraphics[width=0.39\textwidth]{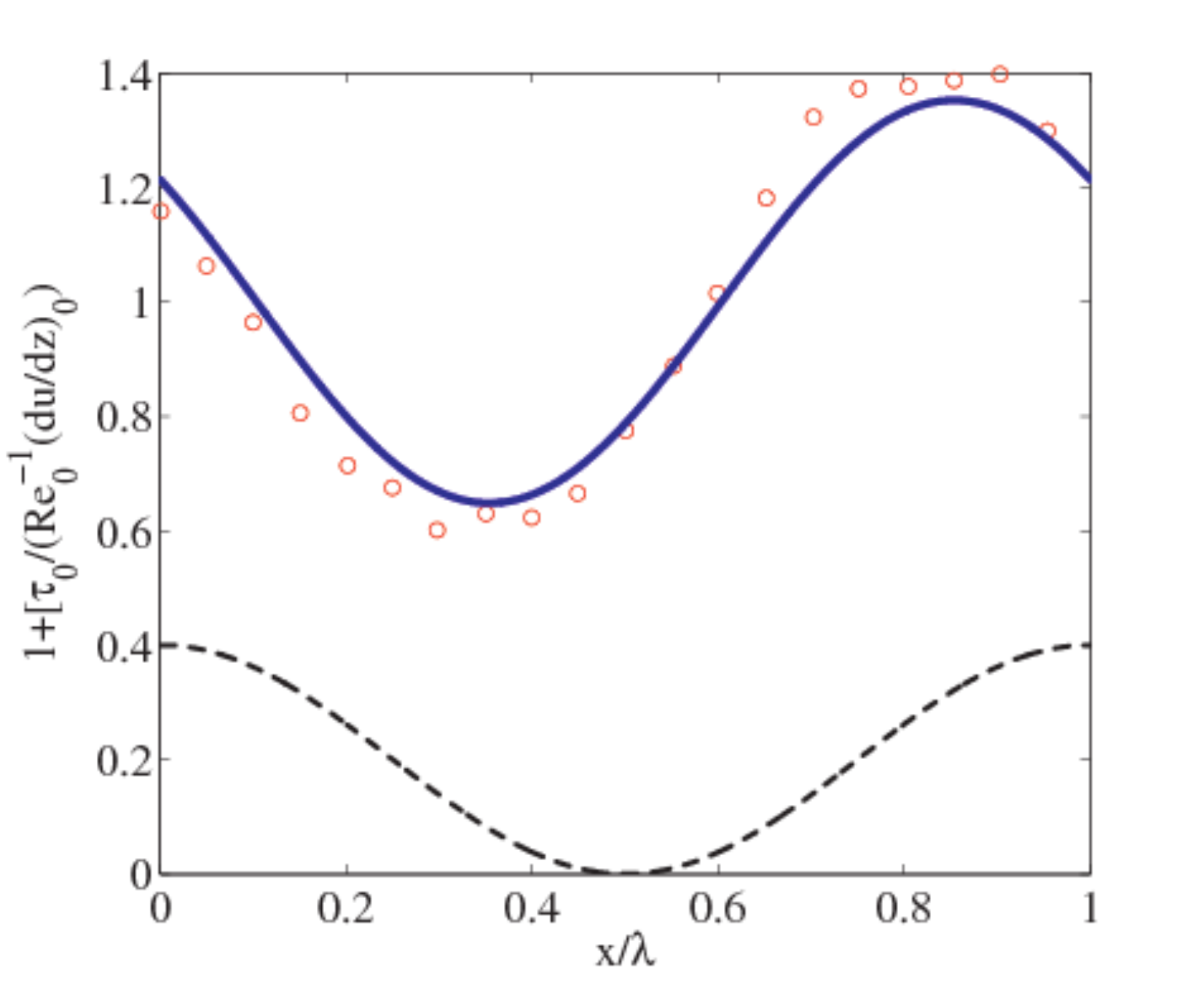}}
\subfigure[]{\includegraphics[width=0.35\textwidth]{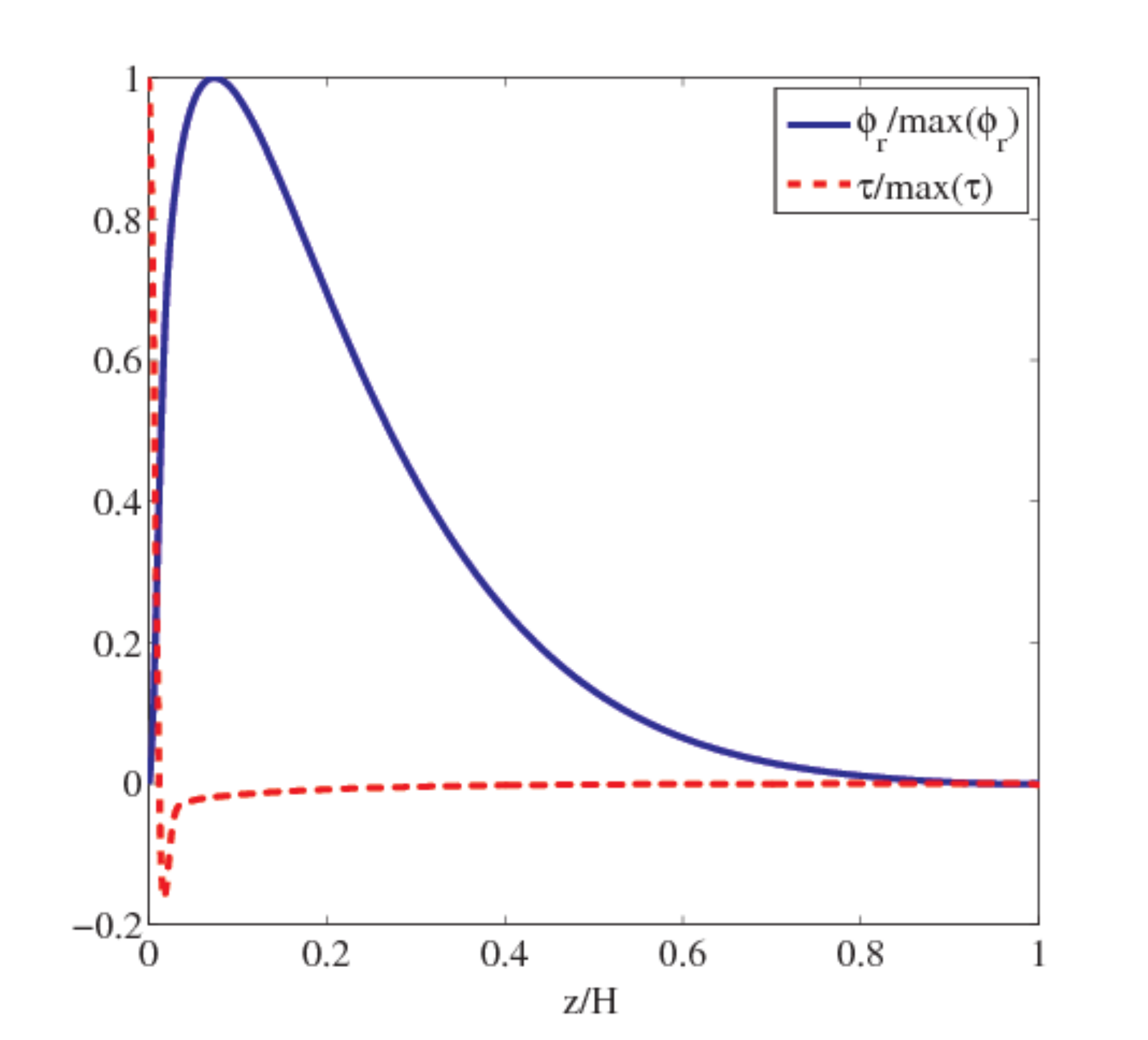}}
\end{center}
\caption{(a) Comparison of the theoretical base state with the work of \citet{Zilker1976}.  We plot the total shear stress at the interface, $\left[Re_0^{-1}U_0'\left(0\right)+\tau_0\left(x\right)\right]/\left[Re_0^{-1}U_0'\left(0\right)\right]$, and compare the theoretical curve with the data from the experiment.  Reasonable agrement is obtained for the amplitude. Excellent agreement is obtained for the phase shift of the shear stress relative to the wavy wall: the data predicts a phase shift of approximately $50^{\mathrm{o}}$, while our model predicts a phase shift $52.6^{\mathrm{o}}$.  The wavy wall is shown in the figure for comparison, albeit with an exaggerated amplitude.  The parameter $\alpha/Re_*$ has the value $0.0059$, while $a\alpha=0.04$;
(b)  Theoretical curves for the shape of the streamfunction and the stress distribution $\tau=Re_0^{-1}\left[F''\left(\eta\right)+\alpha^2 F\left(\eta\right)+2\alpha U_0'\left(\eta\right)e^{-\alpha\eta}-2\alpha^2U_0\left(\eta\right)e^{-\alpha\eta}\right]$;
(c) Comparison of the theoretical base state with the work of \citet{Abrams1985}.   The parameter $\alpha/Re_*$ has the value $2\pi/1110=0.0057$, while $a\alpha=0.04$; (d) Theoretical curves for the shape of the streamfunction and the stress distribution $\tau$.  The streamfunction vanishes slowly as $z\rightarrow H$, in comparison with~(b).
}
\label{fig:zilker}
\end{figure}

The solution of the boundary-value problem facilitates a comparison with experimental data.  In this comparison, we use the \textit{quasi-laminar assumption}: the eddy-viscosity terms $\mathcal{R}$ are set to zero, and turbulence enters only through the shape of the base state $U_0$.
To make an accurate comparison between the experiments and Eqs.~\eqref{eq:OSC_C2}  we study the shear stress at the interface:
\begin{equation}
\tau_0=Re_0^{-1}\left(\phi_{zz}-\phi_{xx}\right)_{\eta=0}
=\frac{ae^{i\alpha\xi}}{Re_0}\left[F''\left(0\right)+\alpha^2 F\left(0\right)+2\alpha\left(Re_*^2/Re_0\right)\right].
\label{eq:tau0}
\end{equation}
We also study the phase shift between this stress function and the wave surface $a\Re\left[e^{i\alpha\xi}\right]$.  We examine the situation described by Fig.~(5) in the work of \citet{Zilker1976}, for which
\[
a/H=0.003,\qquad \alpha H=13.3,\qquad Re_*= 2270,
\]
where $H=5.08\,\mathrm{cm}$ is the channel depth.  We also look at Fig.~(4) in the work of Abrams and Hanratty, where
\[
a/H=0.007,\qquad \alpha H=2\pi,\qquad Re_*= 1110,
\]
where $H$ is the same is in the Zilker experiment.  A comparison between theory and experiment is shown in Fig.~\ref{fig:zilker}, where good agreement is obtained.
%
%
%

%
%
%

Note that in several crucial respects, the stability calculations performed in the main part of paper and the wavy-wall calculations performed in this Appendix are different.
In the case of small-amplitude waves on an interface, the growth of waves is understood through a linear stability analysis.  Thus, the amplitude of the initial interfacial disturbance is assumed to be small, and functions as a small parameter in a linear stability analysis.  In contrast, the amplitude of the wavy wall is not infinitesimally small, and it is the finiteness of this amplitude that gives rise to the curvature terms in the equations~\eqref{eq:OSC_C2}, which in turn affects the distribution of stress at the interface.  Furthermore, the wave speed in the wavy-wall calculation is a known parameter.  In the linear stability analysis, it is determined as the solution of an eigenvalue problem.  It might be tempting to guess the wave speed in the case of interfacial waves by recourse to the free-surface formula~\eqref{eq:c_grav} but our linear stability analysis shows that the $\alpha$-range of maximal wave growth is precisely that range where this formula is least reliable (See Fig.~\ref{fig:test_numerics}).  Thus, a key difference between these two calculations is that in the wavy-call case, $c$ is a parameter; in the wavy-interface case, it must be determined from other parameters.
One similarity between the calculations is the shape of the streamfunction.  In the two-phase calculation, the shape of the streamfunction in the upper layer is similar to that obtained from the wavy-wall calculation (Figs.~\ref{fig:zilker}~(b) and~\ref{fig:zilker}~(d)).  This is a consequence of the boundary conditions, which impose severe limitations on the shape of the streamfunction.
%
%
%
%
%
In conclusion, the wavy-wall calculation, because it assumes that the wave speed is a parameter, is an incomplete model for two-phase wavy interface calculations.
It is, however, a testbed for verifying the turbulence modelling of the basic state.

\bibliographystyle{plainnat}

\end{document}